\documentclass[useAMS,usenatbib]{mn2e}
\usepackage{setspace}
\usepackage{multirow}
\usepackage{graphicx}
\graphicspath{{MainImageFiles/}}
\usepackage[fleqn]{amsmath}
\usepackage{amssymb}
\usepackage{lscape}
\usepackage{url}
\usepackage{color}

\def\araa{ARA\&A}				
\def\apj{ApJ}					
\def\apjl{ApJL}					
\def\aap{A\&A}					
\def\aaps{A\&AS}				
\def\mnras{MNRAS}				
\def\pasp{PASP}					
\def\nat{Nature}				

\title{Alignment of the Angular Momentum Vectors of Planetary Nebulae in the Galactic Bulge}

\author[B. Rees]{B. Rees and A. A. Zijlstra \\Jodrell Bank Centre for Astrophysics, The Alan Turing Building, School of Physics and Astronomy,\\The University of Manchester, Oxford Road, Manchester M13 9PL, UK}

\begin{document}
\setcounter{secnumdepth}{3}
\maketitle
\begin{abstract}
	We use high-resolution H~$\alpha$ images of 130 planetary nebulae (PNe) to investigate whether there is a preferred orientation for PNe within the Galactic Bulge. The orientations of the full sample have an uniform distribution.  However, at a significance level of 0.01, there is evidence for a non-uniform distribution for those planetary nebulae  with evident bipolar morphology. If we assume that the bipolar PNe have an unimodal distribution of the polar axis in Galactic coordinates, the mean Galactic position angle is consistent with $90\degr$, i.e. along the Galactic plane, and the significance level is better than 0.001 (the equivalent of a $3.7\sigma$ significance level for a Gaussian distribution).

The shapes of PNe are related to angular momentum of the original star or stellar system, where the long axis of the nebula measures the angular momentum vector. In old, low-mass stars, the angular momentum is largely in binary orbital motion.  Consequently, the alignment of bipolar nebulae that we have found indicates that the orbital planes of the binary systems are oriented perpendicular to the Galactic plane. We propose that strong magnetic fields aligned along the Galactic plane acted during the original star formation process to slow the contraction of the star forming cloud in the direction perpendicular to the plane. This would have produced a propensity for wider binaries with higher angular momentum with orbital axes parallel to the Galactic plane. Our findings provide the first indication of a strong, organized magnetic field along the Galactic plane that impacted on the angular momentum vectors of the resulting stellar population.

\end{abstract}

\section{Introduction}
\label{intro_fyp}
Planetary nebulae (PNe) are the ionized ejecta from evolved stars. They form
when a low to intermediate-mass star, ascending the Asymptotic Giant Branch,
ejects its envelope in a phase of pulsational instability. The remaining
stellar core evolves to high temperatures and ionizes the expanding ejecta,
before its nuclear burning ceases and the star becomes a white dwarf
 \citep{2005ARA&A..43..435H,2003ARA&A..41..391V}. 

PNe show a variety of morphologies; they range from round to strongly elliptical, together with bipolar and (infrequently) irregular shapes \citep{1995A&A...293..871C,2002ARA&A..40..439B,2006MNRAS.373...79P}. The origin of those morphologies is
disputed. The main contenders are binarity and magnetic fields \citep{2009PASP..121..316D,2009IAUS..259...35B}. Magnetic
fields have been detected in PNe \citep{2007MNRAS.376..378S,2008A&A...488..619V,2009ApJ...695..930G} but may not be strong enough to affect the outflows. \citet{2006PASP..118..260S} has argued
that magnetic fields also require binary companions for their origin. If so, ultimately, in
either case, the morphology is related to angular momentum in the stellar
system and the major axis of the nebula traces the direction of the angular momentum vector. 

\citet{2008PASP..120..380W} have reported that PNe in the direction of the
Galactic centre may have a preferential orientation, in terms of an excess
population at one particular orientation. An alignment between adjacent but
unrelated PNe is unexpected. However, if the orientation of a PN traces the
angular momentum vector of the progenitor system, such an effect could
originate from the formation of the stellar population. But the orientations
could also be influenced by some external factor, for instance the Galactic
magnetic field. This has been argued to be important for supernova remnants
 \citep{1998ApJ...493..781G}.

Studies of the orientation of PNe have been carried out for over thirty years
with conflicting results \citep{2008ApJ...675..380S}. There has been no
convincing evidence for an alignment among nearby Galactic PNe, although there
have been some claims for a preferential alignment of the polar axis with the
Galactic plane \citep{1975MNRAS.171..441M}. The evidence of
 \citet{2008PASP..120..380W} is based on a sample of 440 PNe that were either
bipolar, or elliptical with a major to minor axis ratio greater than $1.2:1$
of which 262 were in the direction of the Galactic centre.  Within that group
they found an excess of PNe with galactic position angle (GPA)
$\sim100\degr\ $.

In this paper we use a morphological survey of 130 PNe likely to be in the
Galactic Bulge to test this claim. As only 19 objects are in common with  
\citet{2008PASP..120..380W}, this provides an independent test of the
potential alignment. The whole of the area containing our sample of PNe lies within one of the four regions (the Galactic Centre) studied by \citet{2008PASP..120..380W}. A number of our objects would not be classified as elongated by those authors and the higher angular resolution of our  sample enables us to use objects with a smaller angular size than is permitted in their sample.

\section{The Observations}
\label{obs}

The initial sample of PNe images consisted of 161 objects observed in 2003
using the EMMI instrument of the European Southern Observatory (ESO) 3.5~m New
Technology Telescope (NTT) and 37 objects observed in 2002 and 2003 using the
Planetary Camera of the WFPC2 of the \textit{HST} (programmes 71.D-0448(A) and 9356 respectively).  After removing foreground objects as described below, the final
Bulge sample used in this paper contained 96 NTT objects and 34 \textit{HST} objects. 

The \textit{HST} targets were selected as a random subset of compact PNe in
the direction of the Bulge. We selected all PNe listed in the Strasbourg-ESO
catalog of Galactic planetary nebulae \citep{1992secg.book.....A}, with a
listed diameter of 4 arcsec or less, or no known diameter. Every second object
from this list was selected for an \textit{HST} SNAPshot program. Of the 60
objects selected, just over half were observed. The observations utilized the F656N filter (2.2~nm wide at 656.4~nm)\footnote{WFPC2 Filter Wavelengths in the HST Archive, Koekemoer, A and Brammer, G; Space Telescope Science Institute;  http://www.stsci.edu/hst/wfpc2/documents/\\wfpc2\_filters\_archive.html; accessed 07/05/10}. The exposure times were about 120 seconds. The images have a pixel size of 0.046 arcsec. 

The NTT sample was selected from the larger nebulae in and towards the Galactic Bulge. The NTT observations
utilized the \#654 H~$\alpha$ filter (3.3~nm wide at 655.4~nm). The images have a pixel size of 0.33 arcsec \citep{EMMI}. The seeing was
typically better than 1.5 arcsec and the exposure times were mostly 60 s but a few were of 30, 120, 180, 240 and 300 seconds.

In the region studied, 80 per cent of the selected objects are
expected to belong to the Galactic Bulge \citep{1990A&A...234..387Z} leaving
few foreground nebulae in the sample. However, we removed PNe from the sample if they were likely foreground (or background) objects, not related to the Bulge, on the basis of the following criteria: 
\begin{enumerate}
\renewcommand{\theenumi}{\arabic{enumi}}
\renewcommand{\labelenumi}{\theenumi}
\item The PN lies beyond 10\degr\ of the Galactic centre in terms of Galactic longitude and latitude. 
\item The PN has a measured component of length of more than 35~arcsec; see \citet{2006IAUS..234..355A}.
\item Where a radio flux at 5~GHz  was available for the PN, for that flux to lie outside the interval $\left(4.2\rmn{~mJy},\: 59.1\rmn{~mJy}\right)$ \citep{S&T2001,1992secg.book.....A}.
\end{enumerate}
It should be noted that PNe remained in the sample unless there was data that
could be used to exclude them. Where two catalogues gave conflicting radio
fluxes the PN was retained in the sample if one of those sets of data
permitted it. 

The NTT images were deconvolved using the {\sevensize LUCY/WAVE} algorithm in
the {\sevensize ESO/MIDAS} package before determining the PNe morphologies and
orientations. The \textit{HST} images did not require deconvolution. 

\begin{figure*}
    \begin{center}

$
    \begin{array}{cc}
    \includegraphics[width=7.5cm]{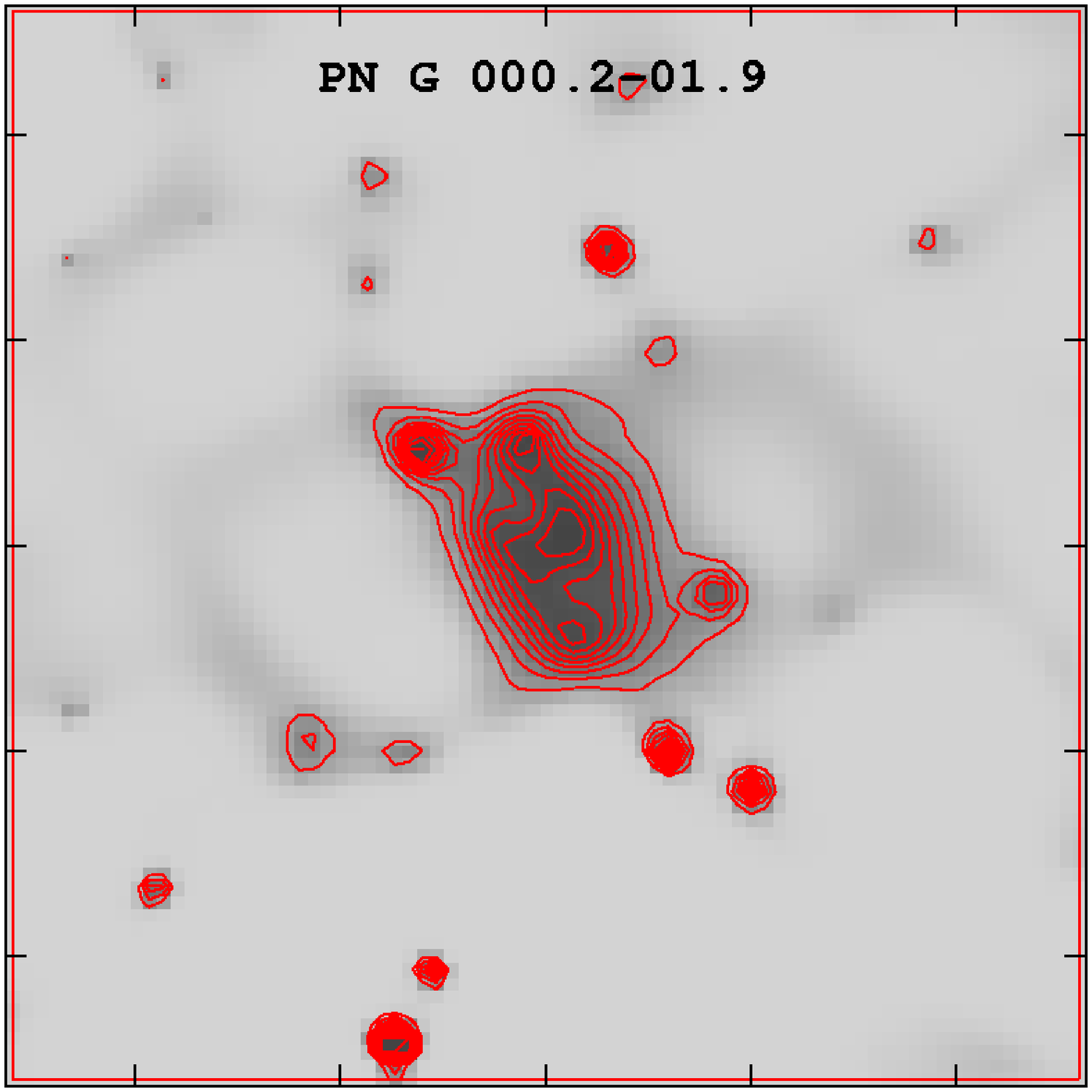} &
    \includegraphics[width=7.5cm]{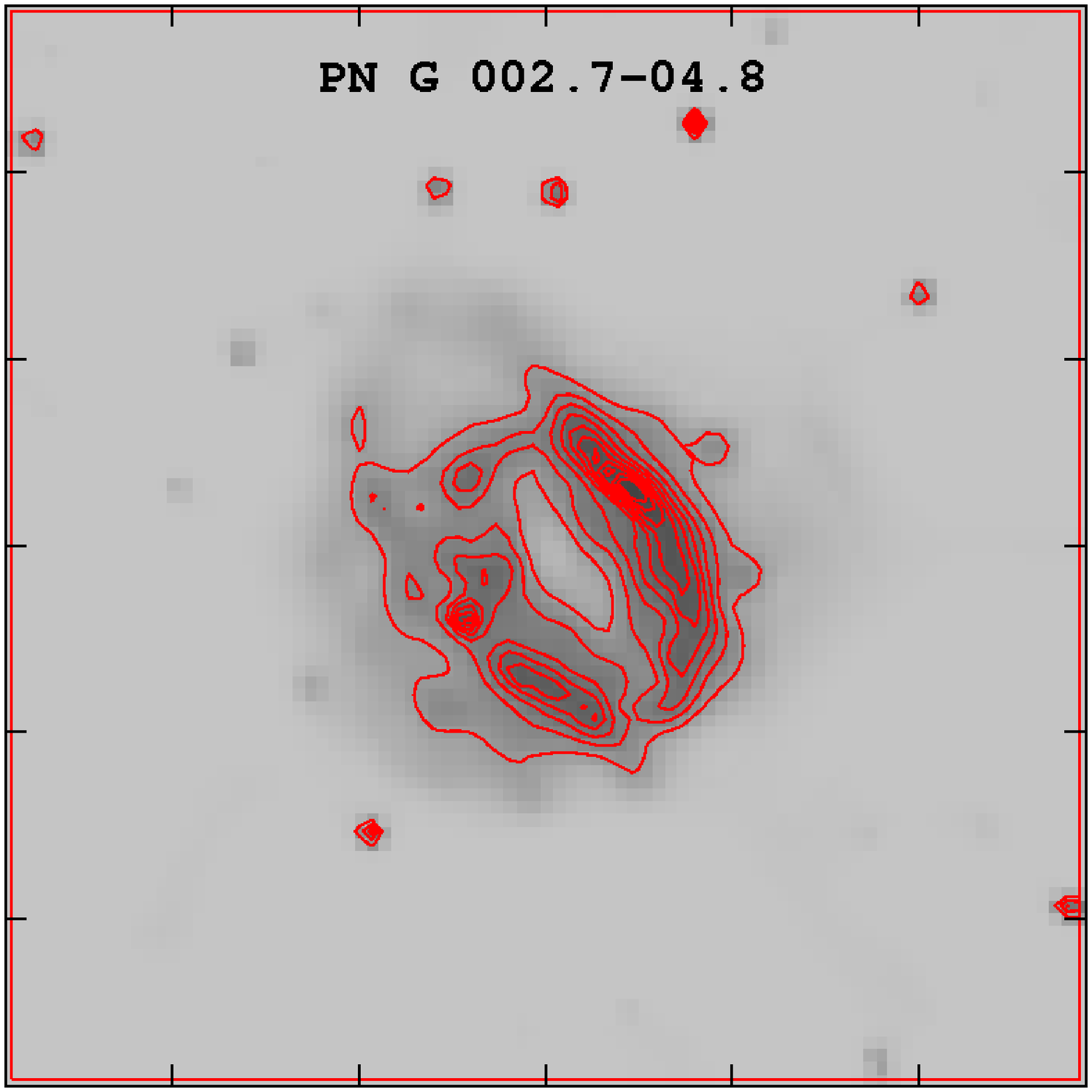}
\\
    \includegraphics[width=7.5cm]{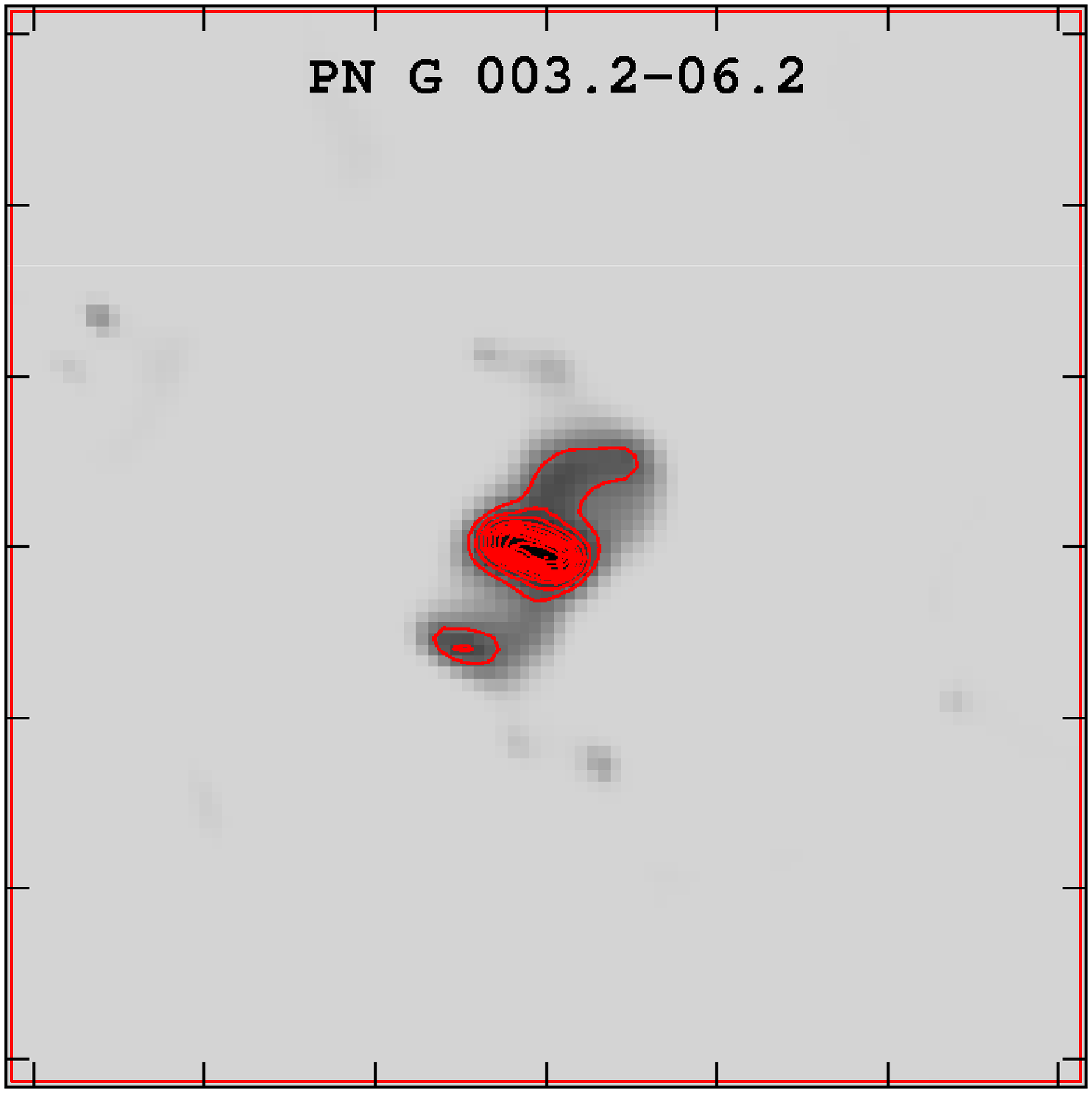} &
    \includegraphics[width=7.5cm]{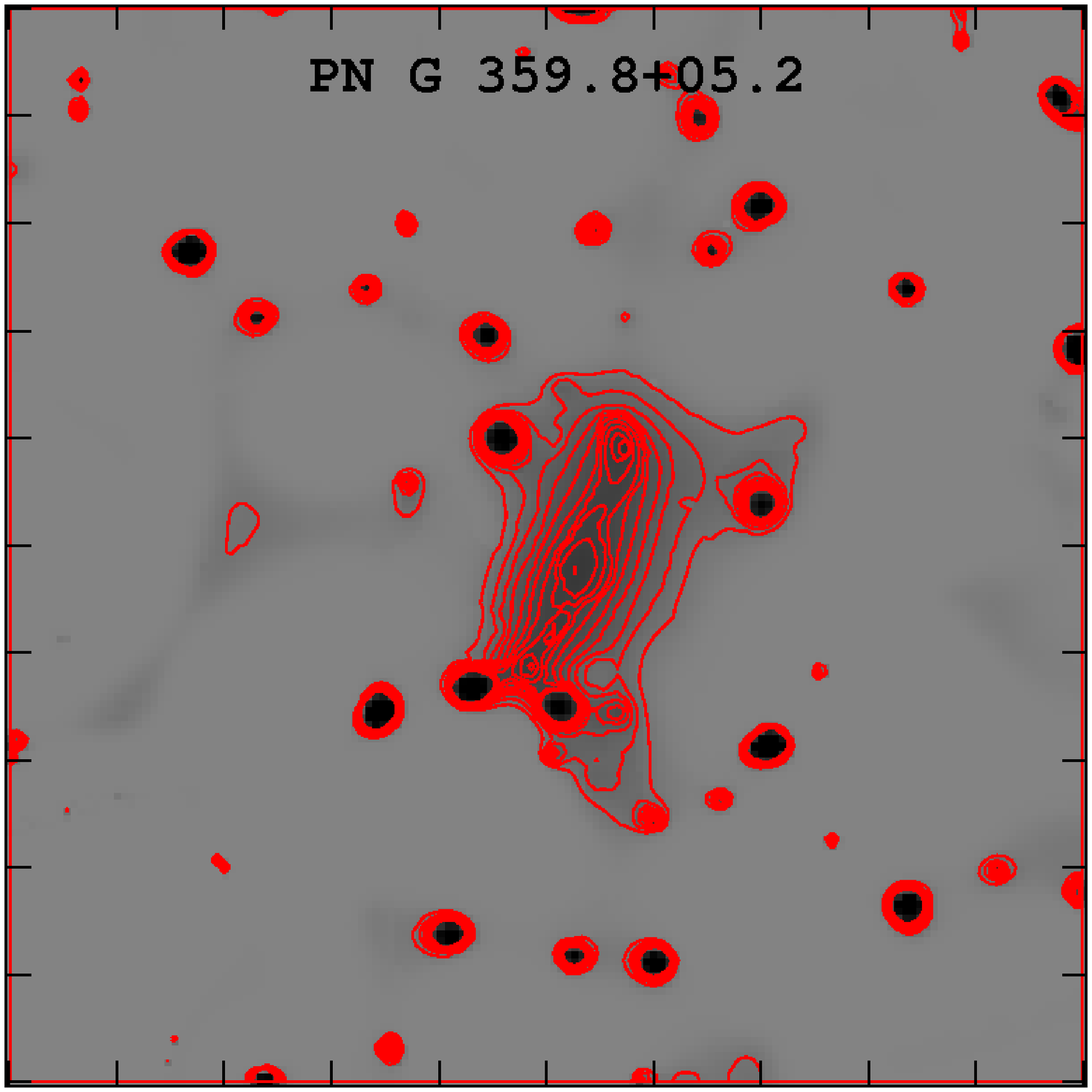}
\\
    \includegraphics[width=7.5cm]{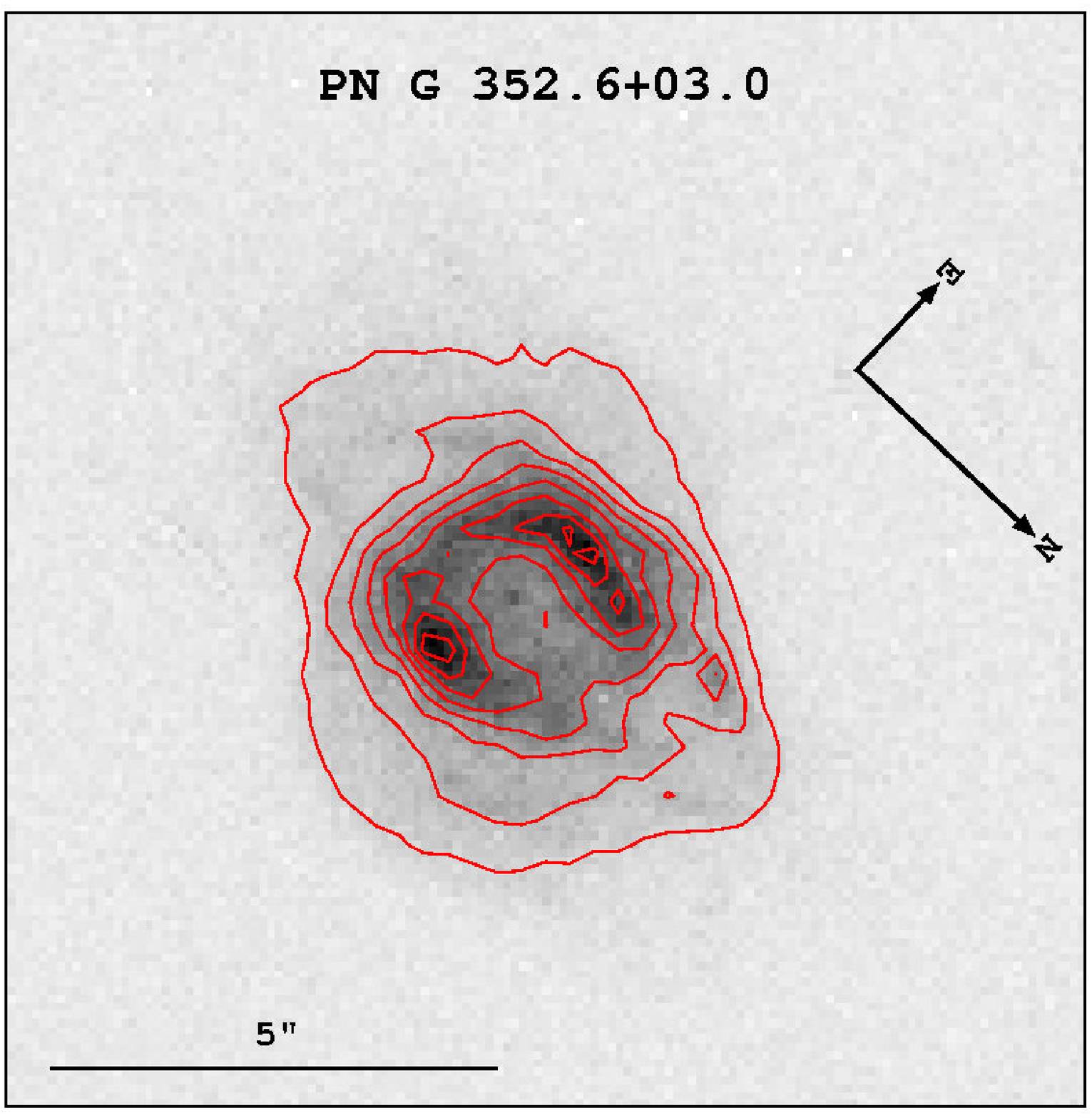} &
    \includegraphics[width=7.5cm]{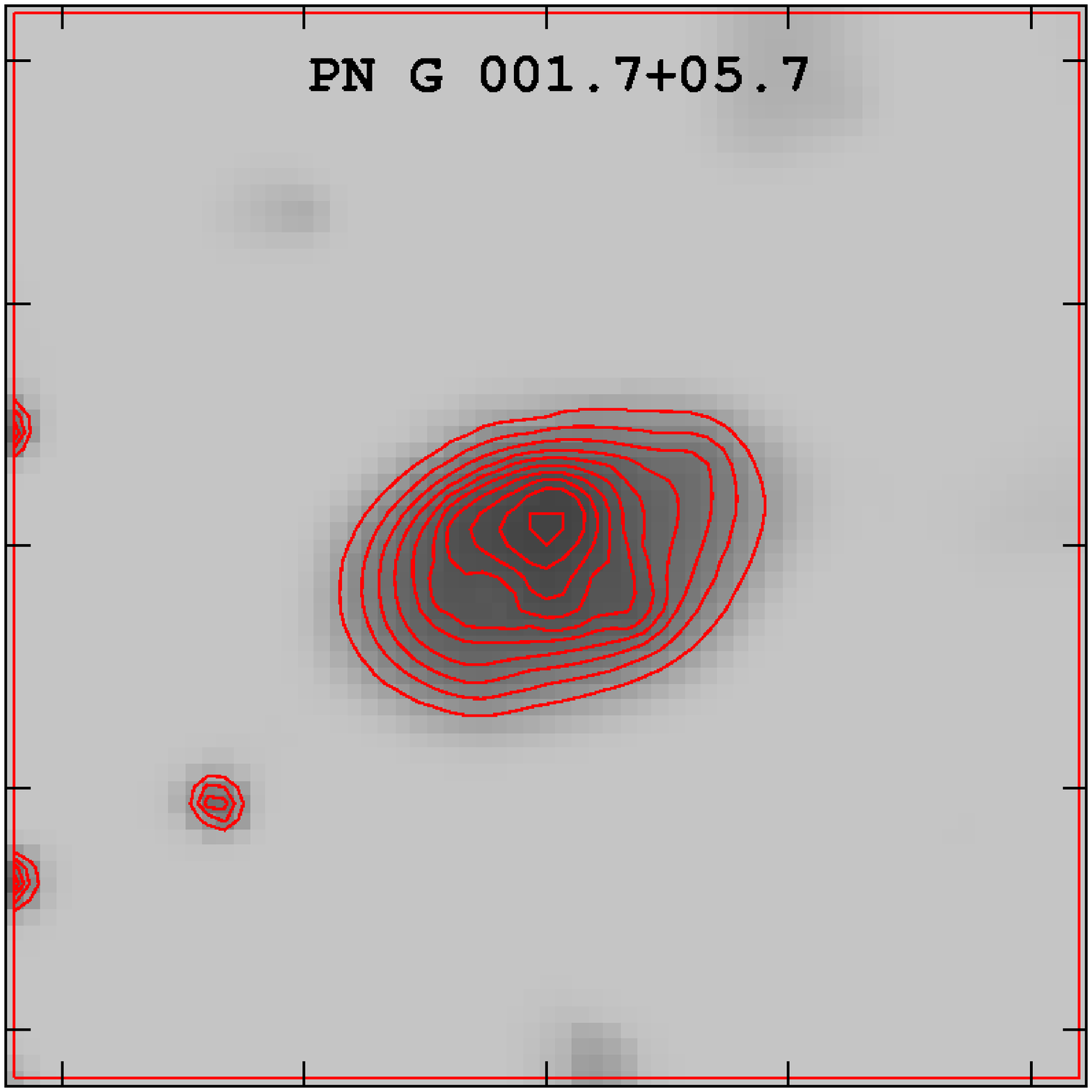}

    \end{array}
$
    \end{center}
    \caption{Examples of the classifications of the PNe. PN G 000.2-01.9 was classified as bipolar and PN G 002.7-04.8 was classified as multilobed bipolar. PN G 003.2-06.2 was also classified as bipolar, i.e point symmetry does not form a class of its own, as was PN G 359.8+05.2, a bipolar remnant, i.e. our bipolar category includes those PNe deemed to be the remnants of bilobed  or multilobed PNe. PN G 352.6+03.0 was classified as polar as it has no lobes and the contours suggest  less material in the north and south directions. PN G 001.7-05.7 shows no internal structure and was classified as non-polar. The greyscale has black as high flux density. The isophotes are shown in red and are based on 10 per cent intervals of the peak PN flux density. The tick marks on the images without a scale bar indicate separations of 5 arcsec and those images have north to the top and east to the left.}
\label{classexamp}
\end{figure*}

We define three morphological categories: bipolar, polar, and non-polar. We had no irregular PNe for which we could fix an orientation. PNe with one or more lobes, or the remnants of lobes, generally at the 1 per cent intensity level were classified as having a bipolar morphology. The bipolar class conforms in the main to that usually used, see \citet{1995A&A...293..871C}. We also followed those authors in not using a separate round class and classifying what might be considered as round PNe as elliptical. Due to the high resolution of the images we were able to split the elliptical category into two distinct classes.  A PN was categorized as having a polar morphology if its image showed no lobes but the intensity distribution above 10 per cent of the peak flux level showed internal structure thus assisting in the  identification of the polar axis. The remaining PNe were classified as non-polar. Fig.~\ref{classexamp} shows examples from each of the classes.

The PN orientation was taken to be that of the polar or major axis of the PN as measured from North to East in the image, and converted to Galactic coordinates. The angle was estimated by eye over the range $\left[0\degr\ ,180\degr\ \right)$ with the assistance of the ruler tool in the NASA HEASARC fv system. The polar angle of a bipolar PN was determined by the alignment of the lower intensity interior of its central shell and the direction of its lobes. The polar angle of a non-polar PN was taken to be that of the major axis at the 10\% level, basically the main symmetry axis of the nebula as suggested in \citet{1998MNRAS.297..617C} and \citet{2008PASP..120..380W}. For the polar PN, the direction of a central cavity and/or low intensity regions suggesting the direction of mass loss was used to assist in the determination of the polar angle. We used the alignments in the high intensity structures to determine the polar angles of the bipolar remnants, except that for PN G 359.8+05.2 we took the alignment to be perpendicular to that of the high intensity structure. Isophote plots of the bipolar PNe, including the bipolar remnants, overlaid with lines indicating the PN  orientation and the uncertainty in that orientation are provided online. Only PNe whose orientation could readily be determined were included in the sample. Note that projection effects were not taken into account, i.e. the angles were measured in the plane of the sky, as we have no information of structure along the line of sight. The orientations for the polar and non-polar PNe were taken to have random uncertainties of 2\degr. Initially we also used that value for the uncertainties in the orientations of the bipolar PNe. However, to ensure that the evidence of alignment that we obtained in the statistical tests for those PNe was not down to using uncertainties that were too small we repeated the tests on the bipolar PNe using more carefully measured uncertainties. For the bipolar PNe we used an uncertainty that is the maximum of: 
\begin{enumerate}
\item the estimated population standard deviation of repeated measurements of the orientation of the PN;
\item the arctangent of the reciprocal of the PN's tip-to-tip length in pixels (to allow for a minimum of 1 pixel variation to account for the image resolution); and
\item 2\degr. 
\end{enumerate}

Overall, 68 of the observed PNe were rejected due to the selection criteria,
imaging problems or uncertain orientation. The data set that remained
consisted of 130 PNe split into the subsamples described above. 

It should be noted that 19 of the PNe, just under one sixth of our final
sample, are also in the analysis presented in \citet{2008PASP..120..380W}. According to our criteria, 5 of those 19 are
bipolar, 10 polar and 4 non-polar. 

\begin{figure*}
    \begin{center}

$
    \begin{array}{cc}
    \includegraphics[width=7.8cm]{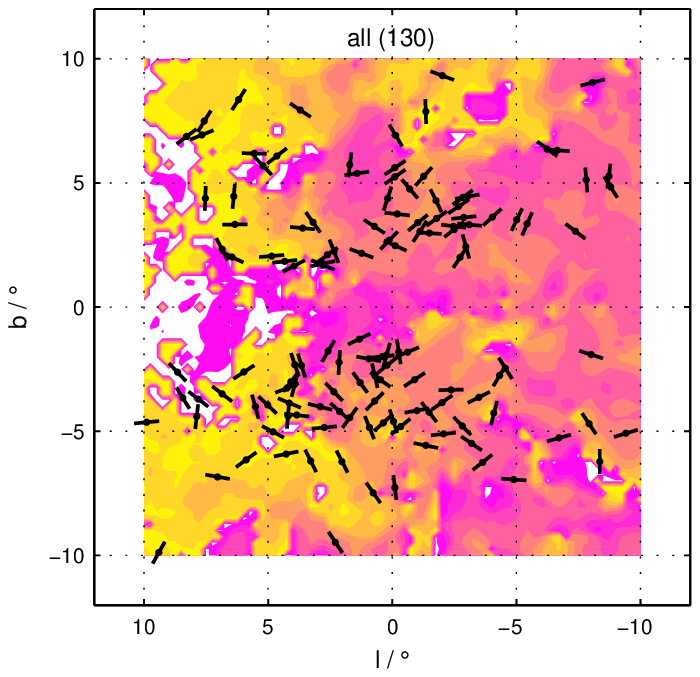} &
    \includegraphics[width=7.8cm]{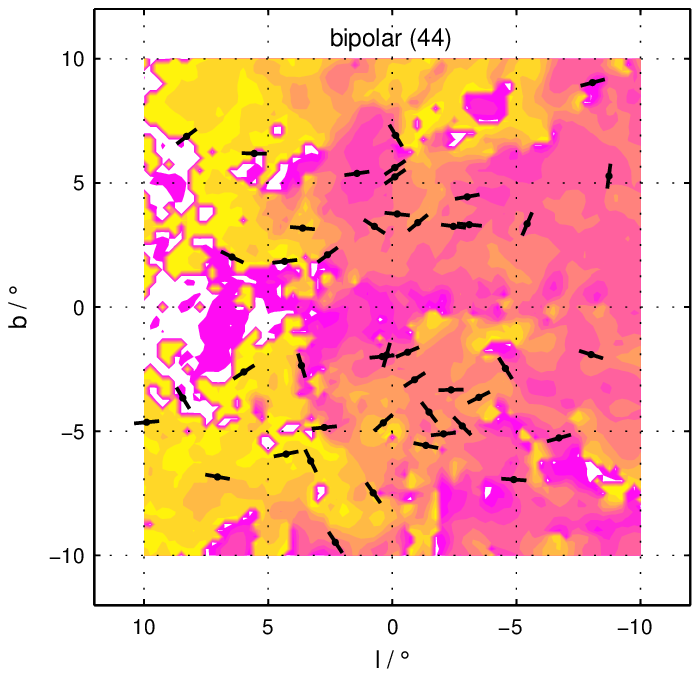}
\\
    \includegraphics[width=7.8cm]{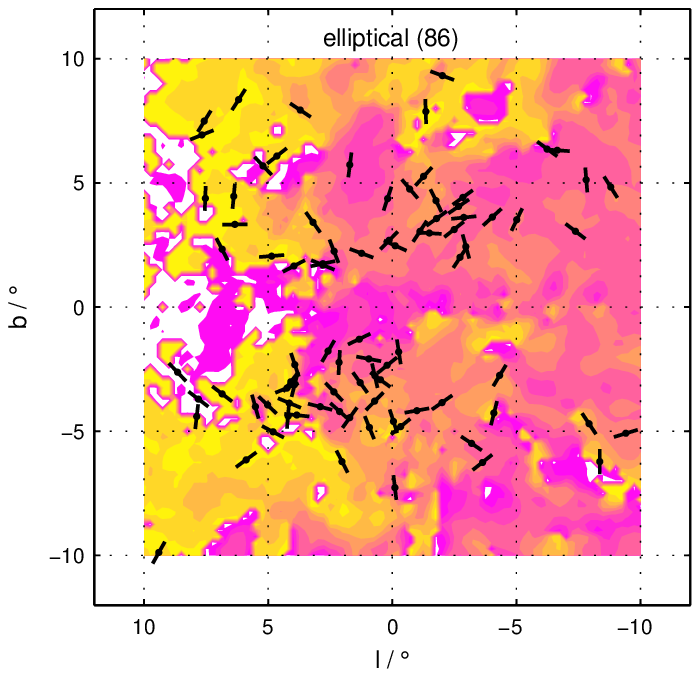} &
    \includegraphics[width=7.8cm]{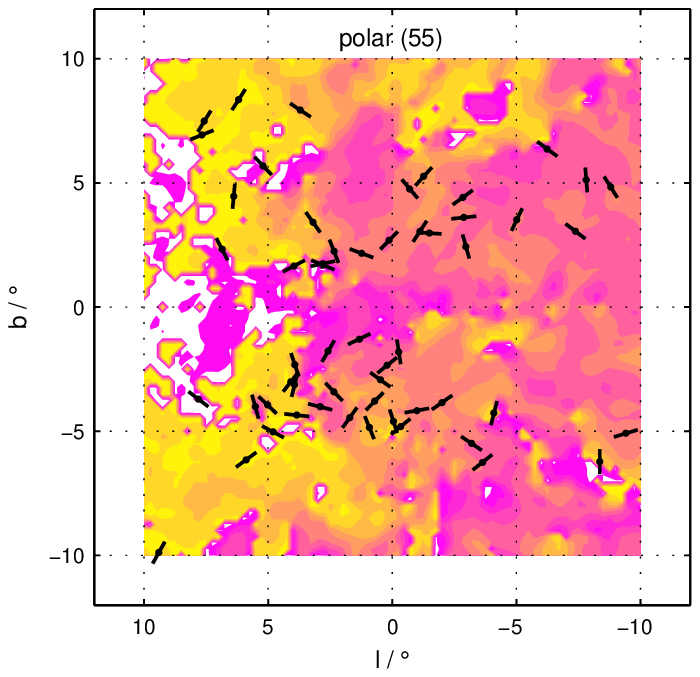}
\\
    \includegraphics[width=7.8cm]{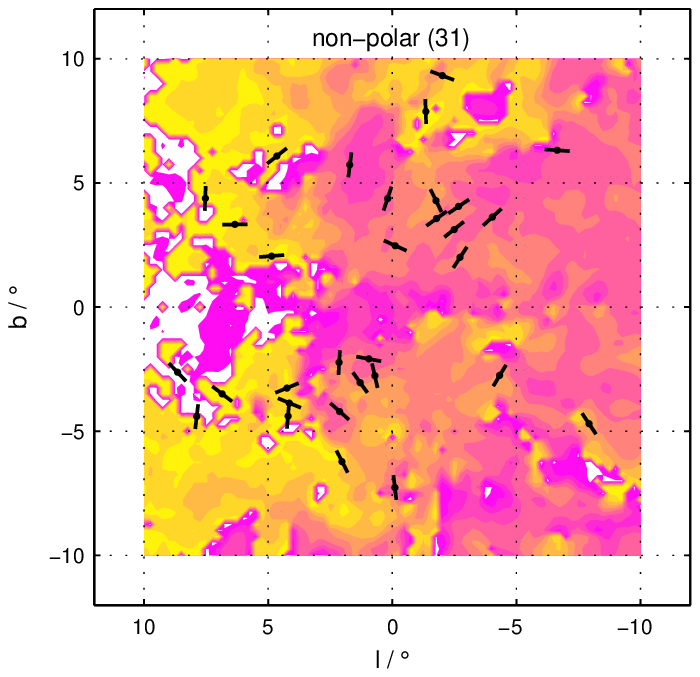} &
    \includegraphics[height=7.8cm]{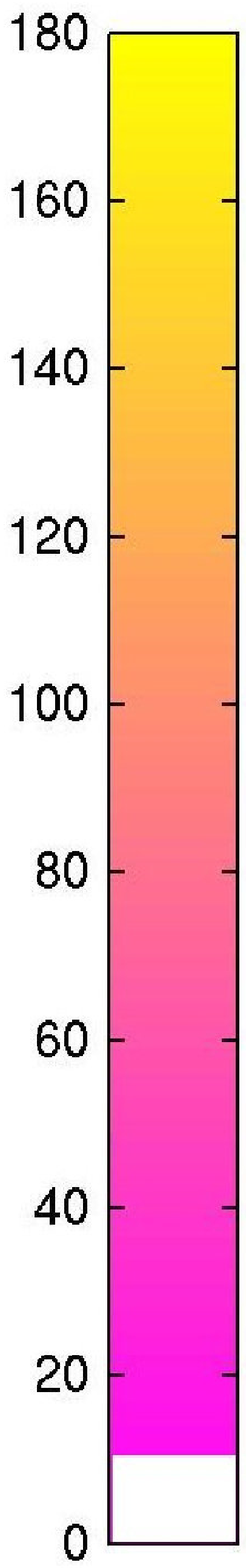} 
    \end{array}
$
    \end{center}
    \caption{The GPA of the PNe in the samples. They are plotted on a background of contour plots of the 21~cm radio continuum polarization angles in the Galactic Bulge which are based on data \citep{2008A&A...484..733T} obtained from Max-Planck-Institut f\"{u}r Radioastronomie Survey Sampler (available from: http://www.mpifr-bonn.mpg.de/survey.html). The polarization angles are with respect to the Galactic North Pole and their magnitudes are indicated by the scale bar at the bottom right. We discuss the use of that polarization data in \S\ref{results}.}
    \label{orienttype}
\end{figure*}

The formula presented in \citet{1998MNRAS.297..617C} and the galactic
coordinates quoted in SIMBAD for each PN were used to convert the measured
polar angles (PA) of the PNe to galactic position angles (GPA). Both sets of
values are shown in Table~\ref{PNorient} and plots of the orientations of the PNe in the sample and main subsamples are shown in Fig.~\ref{orienttype}.

\begin{table*}
\centering
\caption{The orientations of the PNe. The measurement of the position angles and the derivation of their uncertainties and of the GPA are described in Section \ref{obs}. The uncertainties in the PA and GPA are identical.}
\label{PNorient}
\small
\begin{tabular}{|c|l|r|r@{$\pm$}l|c|c|l|r|r@{$\pm$}l|c|}
\hline
PNG & Morphology & PA/\degr & \multicolumn{2}{|c|}{GPA/\degr} & Telescope & PNG & Morphology & PA/\degr & \multicolumn{2}{|c|}{GPA/\degr} & Telescope \\ 
\hline
000.1+02.6 & polar & 78 & 135&2 & NTT & 007.8\,--\,03.7 & polar & 171 & 53&2 & NTT \\ 
000.1+04.3 & non-polar & 105 & 162&2 & NTT & 007.8\,--\,04.4 & non-polar & 111 & 173&2 & NTT \\ 
000.1\,--\,02.3 & polar & 70 & 129&2 & NTT & 008.2+06.8 & bipolar & 68 & 127&2  & \textit{HST} \\ 
000.2\,--\,01.9 & bipolar & 105 & 165&4 & NTT & 008.4\,--\,03.6 & bipolar & 149 & 31&2 & NTT \\ 
000.3\,--\,04.6 & bipolar & 71 & 132&2 & NTT & 008.6\,--\,02.6 & non-polar & 162 & 44&2  & \textit{HST} \\ 
000.4\,--\,01.9 & bipolar & 35 & 95&5 & NTT & 009.4\,--\,09.8 & polar & 86 & 150&2 & NTT \\ 
000.4\,--\,02.9 & polar & 174 & 54&2 & NTT & 009.8\,--\,04.6 & bipolar & 33 & 95&9 & NTT \\ 
000.7+03.2 & bipolar & 179 & 57&2 & NTT & 350.5\,--\,05.0 & polar & 48 & 107&2 & NTT \\ 
000.7\,--\,02.7 & non-polar & 133 & 13&2 & NTT & 351.1+04.8 & polar & 159 & 32&2  & \textit{HST} \\ 
000.7\,--\,03.7 & polar & 75 & 136&2 & NTT & 351.2+05.2 & bipolar & 120 & 172&5 & NTT \\ 
000.7\,--\,07.4 & bipolar & 152 & 35&4 & NTT & 351.6\,--\,06.2 & polar & 121 & 0&2 & NTT \\ 
000.9\,--\,02.0 & non-polar & 19 & 79&2 & NTT & 351.9+09.0 & bipolar & 54 & 105&6 & NTT \\ 
000.9\,--\,04.8 & polar & 141 & 22&2 & NTT & 351.9\,--\,01.9 & bipolar & 15 & 72&2  & \textit{HST} \\ 
001.2+02.1 & polar & 12 & 70&2  & \textit{HST} & 352.0\,--\,04.6 & non-polar & 155 & 34&2 & NTT \\ 
001.2\,--\,03.0 & non-polar & 155 & 35&2 & NTT & 352.1+05.1 & polar & 131 & 5&2 & NTT \\ 
001.3\,--\,01.2 & polar & 56 & 116&2 & NTT & 352.6+03.0 & polar & 178 & 53&2  & \textit{HST} \\ 
001.4+05.3 & bipolar & 42 & 99&3 & NTT & 353.2\,--\,05.2 & bipolar & 46 & 105&6 & NTT \\ 
001.7+05.7 & non-polar & 116 & 173&2 & NTT & 353.3+06.3 & non-polar & 33 & 86&2 & NTT \\ 
001.7\,--\,04.4 & polar & 84 & 145&2  & \textit{HST} & 353.7+06.3 & polar & 0 & 53&2 & NTT \\ 
002.0\,--\,06.2 & non-polar & 145 & 27&2 & NTT & 354.5+03.3 & bipolar & 102 & 157&2  & \textit{HST} \\ 
002.1\,--\,02.2 & non-polar & 117 & 177&2 & NTT & 354.9+03.5 & polar & 99 & 154&2  & \textit{HST} \\ 
002.1\,--\,04.2 & non-polar & 167 & 48&2 & NTT & 355.1\,--\,06.9 & bipolar & 25 & 86&8 & NTT \\ 
002.2\,--\,09.4 & bipolar & 149 & 33&4 & NTT & 355.4\,--\,02.4 & bipolar & 154 & 32&2  & \textit{HST} \\ 
002.3+02.2 & polar & 143 & 21&2 & NTT & 355.6\,--\,02.7 & non-polar & 90 & 149&2 & NTT \\ 
002.3\,--\,03.4 & polar & 163 & 44&2  & \textit{HST} & 355.9+03.6 & non-polar & 77 & 133&2  & \textit{HST} \\ 
002.5\,--\,01.7 & polar & 90 & 150&2 & NTT & 355.9\,--\,04.2 & polar & 106 & 165&2 & NTT \\ 
002.6+02.1 & bipolar & 68 & 127&5 & NTT & 356.3\,--\,06.2 & polar & 67 & 128&2 & NTT \\ 
002.7\,--\,04.8 & bipolar & 33 & 95&4 & NTT & 356.5\,--\,03.6 & bipolar & 57 & 117&3  & \textit{HST} \\ 
002.8+01.7 & polar & 9 & 68&2  & \textit{HST} & 356.8+03.3 & bipolar & 28 & 84&8  & \textit{HST} \\ 
002.8+01.8 & polar & 43 & 102&2 & NTT & 356.8\,--\,05.4 & polar & 175 & 56&2 & NTT \\ 
002.9\,--\,03.9 & polar & 15 & 76&2  & \textit{HST} & 356.9+04.4 & bipolar & 45 & 100&6  & \textit{HST} \\ 
003.1+03.4 & polar & 157 & 35&2  & \textit{HST} & 357.0+02.4 & polar & 139 & 16&2 & NTT \\ 
003.2\,--\,06.2 & bipolar & 144 & 26&6 & NTT & 357.1+03.6 & polar & 40 & 96&2 & NTT \\ 
003.6+03.1 & bipolar & 26 & 85&2  & \textit{HST} & 357.1+04.4 & polar & 70 & 126&2 & NTT \\ 
003.6\,--\,02.3 & bipolar & 136 & 17&5 & NTT & 357.1\,--\,04.7 & bipolar & 164 & 44&3  & \textit{HST} \\ 
003.7+07.9 & polar & 0 & 57&2 & NTT & 357.2+02.0 & non-polar & 90 & 147&2  & \textit{HST} \\ 
003.8\,--\,04.3 & polar & 23 & 85&2 & NTT & 357.3+04.0 & non-polar & 68 & 124&2 & NTT \\ 
003.9+01.6 & polar & 59 & 118&2 & NTT & 357.5+03.1 & non-polar & 72 & 128&2 & NTT \\ 
003.9\,--\,02.3 & polar & 138 & 19&2 & NTT & 357.5+03.2 & bipolar & 26 & 83&5 & NTT \\ 
003.9\,--\,03.1 & polar & 105 & 166&2  & \textit{HST} & 357.6\,--\,03.3 & bipolar & 32 & 91&7 & NTT \\ 
004.0\,--\,03.0 & polar & 81 & 142&2  & \textit{HST} & 357.9\,--\,03.8 & polar & 65 & 125&2 & NTT \\ 
004.1\,--\,03.8 & non-polar & 6 & 67&2  & \textit{HST} & 357.9\,--\,05.1 & bipolar & 35 & 96&5 & NTT \\ 
004.2\,--\,03.2 & non-polar & 51 & 112&2 & NTT & 358.0+09.3 & non-polar & 16 & 70&2 & NTT \\ 
004.2\,--\,04.3 & non-polar & 114 & 176&2 & NTT & 358.2+03.5 & non-polar & 69 & 125&2 & NTT \\ 
004.2\,--\,05.9 & bipolar & 39 & 102&7 & NTT & 358.2+04.2 & non-polar & 150 & 26&2 & NTT \\ 
004.3+01.8 & bipolar & 37 & 96&2 & NTT & 358.5+02.9 & polar & 30 & 87&2  & \textit{HST} \\ 
004.6+06.0 & non-polar & 71 & 129&2 & NTT & 358.5\,--\,04.2 & bipolar & 157 & 37&4  & \textit{HST} \\ 
004.8+02.0 & non-polar & 35 & 95&2  & \textit{HST} & 358.6+07.8 & non-polar & 127 & 2&2 & NTT \\ 
004.8\,--\,05.0 & polar & 0 & 62&2 & NTT & 358.6\,--\,05.5 & bipolar & 18 & 79&5 & NTT \\ 
005.0\,--\,03.9 & polar & 165 & 47&2 & NTT & 358.7+05.2 & polar & 82 & 138&2  & \textit{HST} \\ 
005.2+05.6 & polar & 166 & 44&2 & NTT & 358.8+03.0 & polar & 90 & 147&2 & NTT \\ 
005.5+06.1 & bipolar & 31 & 88&2 & NTT & 358.9+03.4 & bipolar & 73 & 129&2  & \textit{HST} \\ 
005.5\,--\,04.0 & polar & 134 & 16&2 & NTT & 359.0\,--\,04.1 & polar & 40 & 100&2 & NTT \\ 
005.8\,--\,06.1 & polar & 63 & 126&2 & NTT & 359.1\,--\,02.9 & bipolar & 63 & 123&4 & NTT \\ 
005.9\,--\,02.6 & bipolar & 62 & 123&3 & NTT & 359.2+04.7 & polar & 164 & 40&2  & \textit{HST} \\ 
006.1+08.3 & polar & 90 & 147&2  & \textit{HST} & 359.3\,--\,01.8 & bipolar & 54 & 113&5 & NTT \\ 
006.3+03.3 & non-polar & 32 & 91&2 & NTT & 359.6\,--\,04.8 & polar & 68 & 129&2 & NTT \\ 
006.3+04.4 & polar & 116 & 175&2  & \textit{HST} & 359.7\,--\,01.8 & polar & 129 & 8&2 & NTT \\ 
006.4+02.0 & bipolar & 4 & 63&2  & \textit{HST} & 359.8+02.4 & non-polar & 9 & 66&2 & NTT \\ 
006.8+02.3 & polar & 148 & 28&2 & NTT & 359.8+03.7 & bipolar & 27 & 84&4 & NTT \\ 
006.8\,--\,03.4 & non-polar & 171 & 53&2 & NTT & 359.8+05.2 & bipolar & 68 & 125&5 & NTT \\ 
007.0\,--\,06.8 & bipolar & 20 & 80&10 & NTT & 359.8+05.6 & bipolar & 68 & 124&5 & NTT \\ 
007.5+04.3 & non-polar & 119 & 178&2  & \textit{HST} & 359.8+06.9 & bipolar & 156 & 31&4 & NTT \\ 
007.5+07.4 & polar & 90 & 148&2 & NTT & 359.8\,--\,07.2 & non-polar & 124 & 6&2 & NTT \\ 
007.6+06.9 & polar & 54 & 112&2 & NTT & 359.9\,--\,04.5 & polar & 137 & 18&2 & NTT \\ 
\hline
\end{tabular}
\normalsize
\end{table*}

\section{The Analysis}
\label{analysis}

The analysis was performed using Matlab\textsuperscript{\textregistered}, Mathematica\textsuperscript{\textregistered} and CircStat, the circular statistics toolbox for Matlab, \citep{RePEc:jss:jstsof:31:i10}. The full sample and the subsamples of bipolar, elliptical, polar and non-polar PNe were analysed as were subsamples of them based on a split north and south, and east and west of the Galactic Centre.

The orientation angles occupied almost the whole semicircular range so an angle of 1\degr\  modulo 180\degr\  is closer to an angle of 175\degr\  than it is to one of 10\degr.  Consequently the circular nature of the data could not be ignored. Circular statistics techniques were used to investigate its distribution and the statistical tests needed to be invariant under a rotation of the axes.  

As the orientations are axial rather than vector, the angles range only over
180 degrees. These were doubled to bring them to modulo 360\degr\ for much of the 
analysis, see \citet{FisherCircStat}.

\begin{figure*}
    \begin{center}

$
    \begin{array}{cc}
    \includegraphics{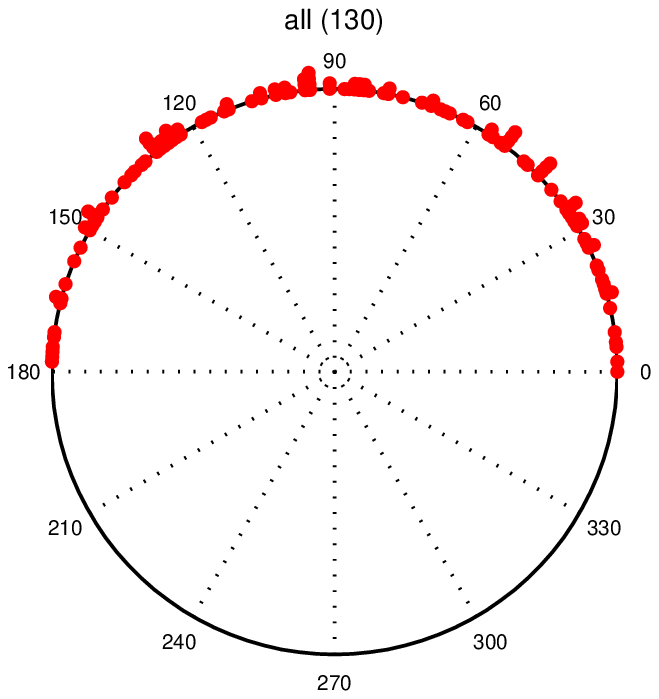} &
    \includegraphics{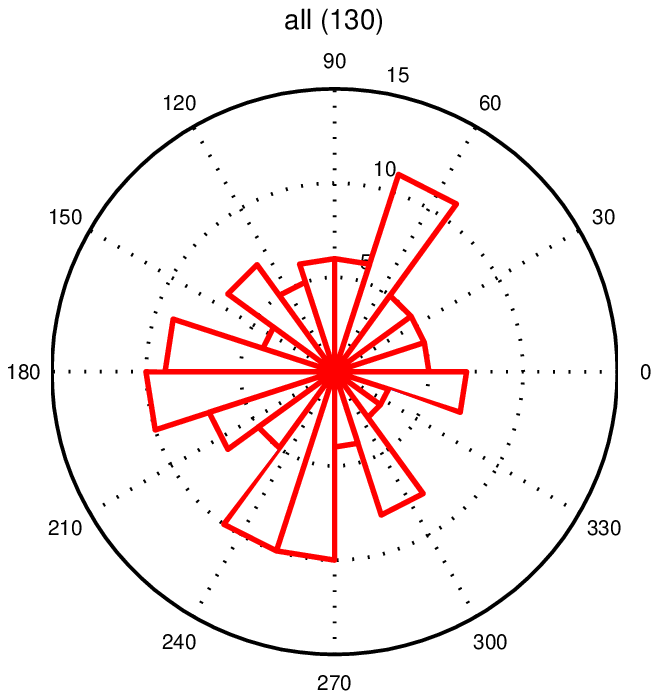}
\\
    \includegraphics{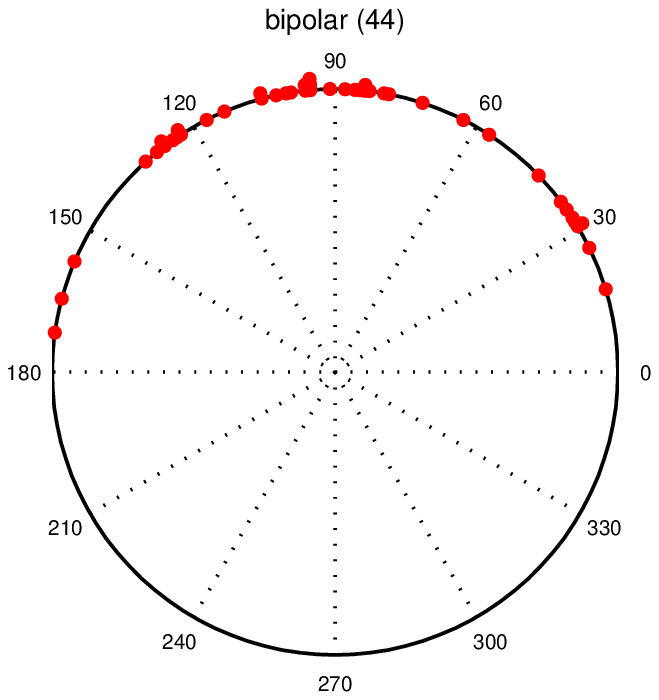} &
    \includegraphics{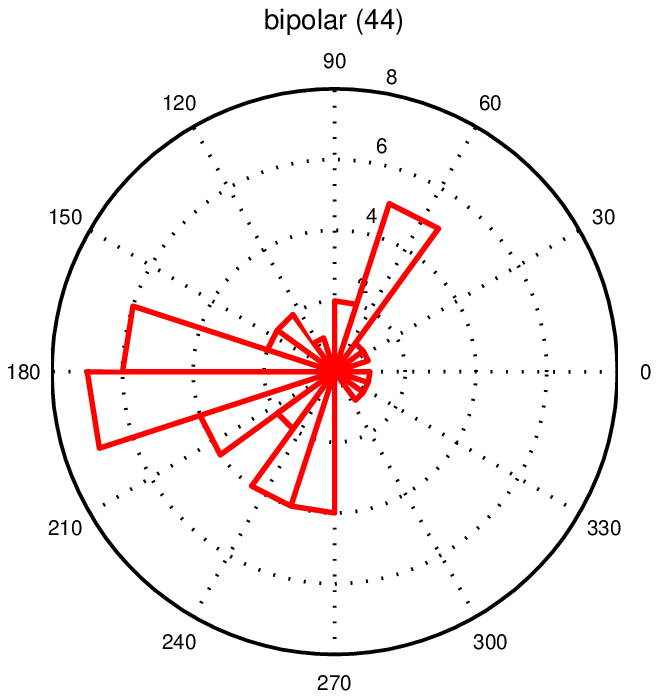}
\\
    \includegraphics{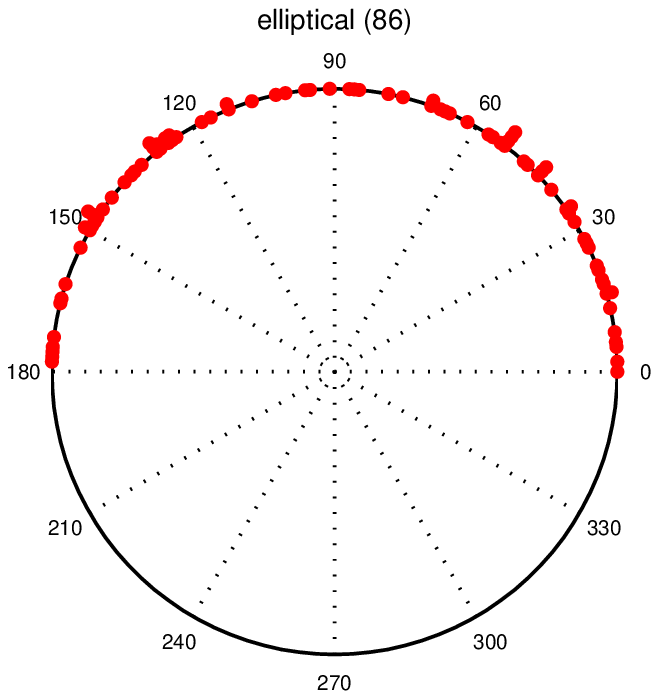} &
    \includegraphics{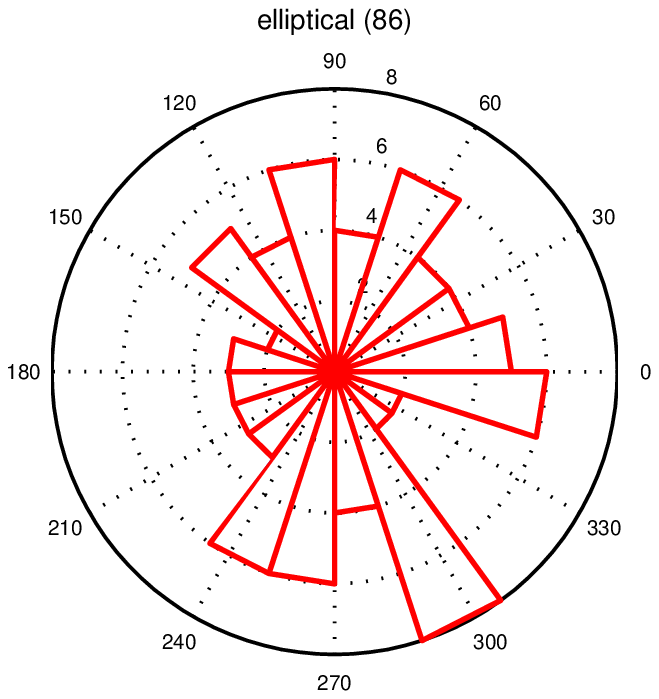}
    \end{array}
$
    \end{center}
    \caption{Finger plots of the GPA and the rose plots of the doubled GPA for the whole sample, the bipolar PNe and the elliptical (polar and non-polar) PNe over the whole area. Each dot on a finger plot represents a GPA in the subsample. The radial scale on the rose plots indicates the number of objects.}
    \label{distcomp1}
\end{figure*}

\begin{figure*}
    \begin{center}

$
    \begin{array}{cc}
    \includegraphics{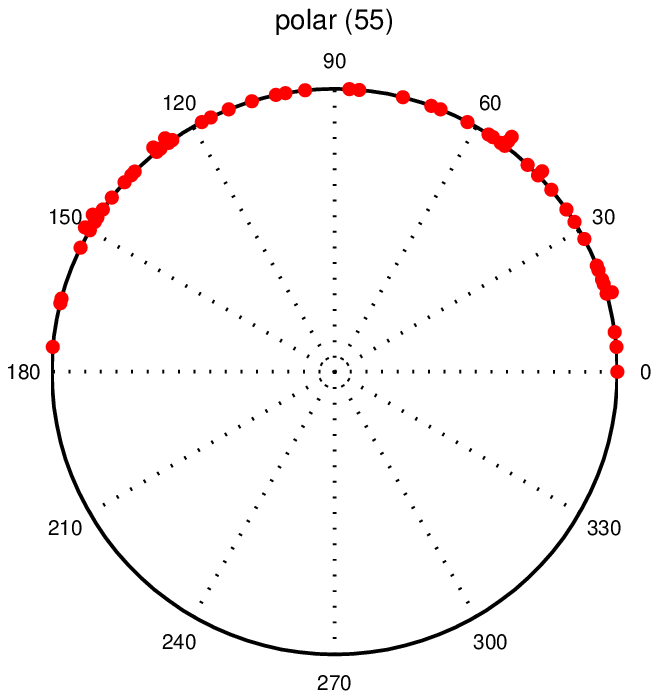} &
    \includegraphics{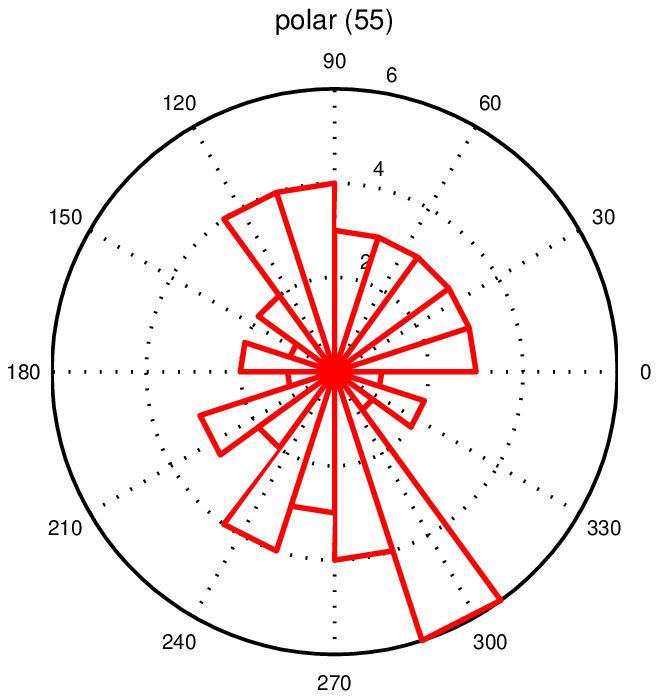}
\\
    \includegraphics{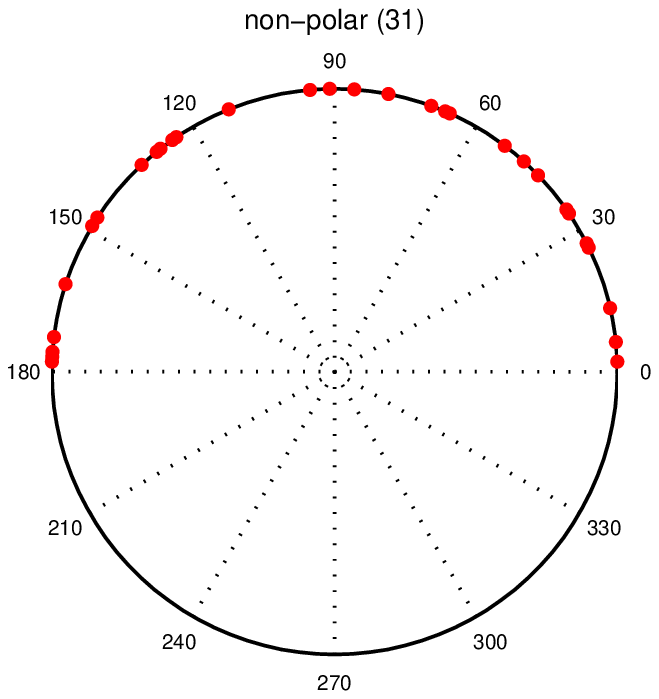} &
    \includegraphics{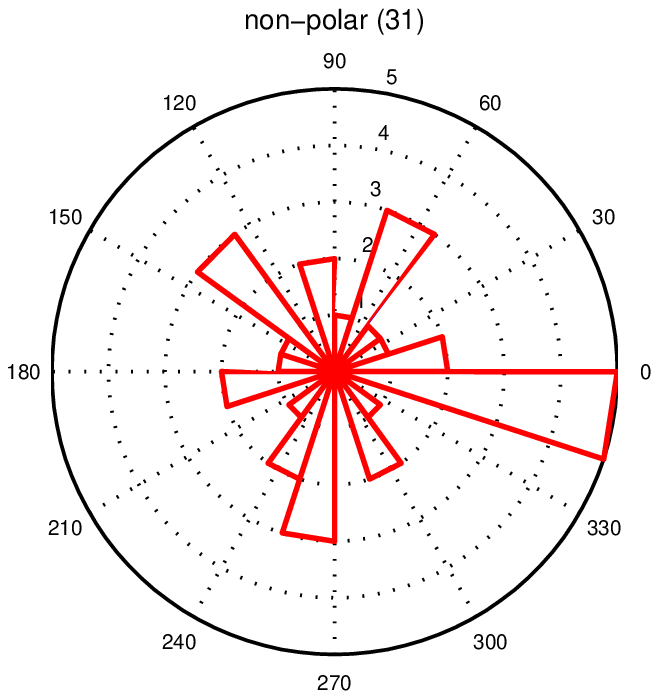}
    \end{array}
$
    \end{center}
    \caption{Finger plots of the GPA and the rose plots of the doubled angles for the polar and non-polar PNe over the whole area. Each dot on a finger plot represents a GPA in the subsample. The radial scale on the rose plots indicates the number of objects.}
    \label{distcomp2}
\end{figure*}

As an initial step, finger plots of the distribution of the GPA and rose plots
(a circular form of bar chart) of the doubled angles were produced for all the
samples. The finger plots provide the fine detail of the information and avoid
the possibly misleading effects of grouping that can occur with any type of
histogram. The plots for the overall sample and the main subsamples are
presented in Figs.~\ref{distcomp1} and \ref{distcomp2}. 

\begin{figure*}
    \begin{center}

$
    \begin{array}{cc}
    \includegraphics[width=8.1cm]{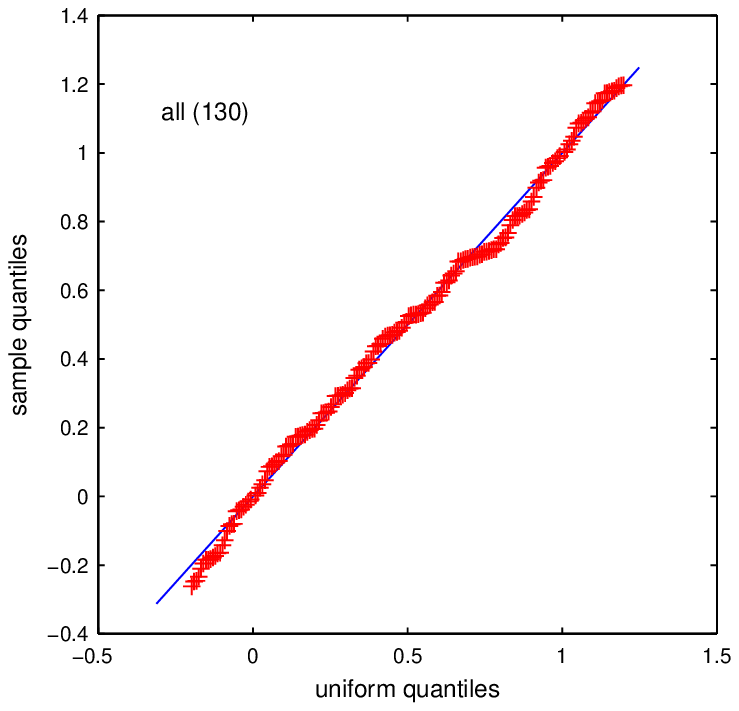} &
    \includegraphics[width=8.1cm]{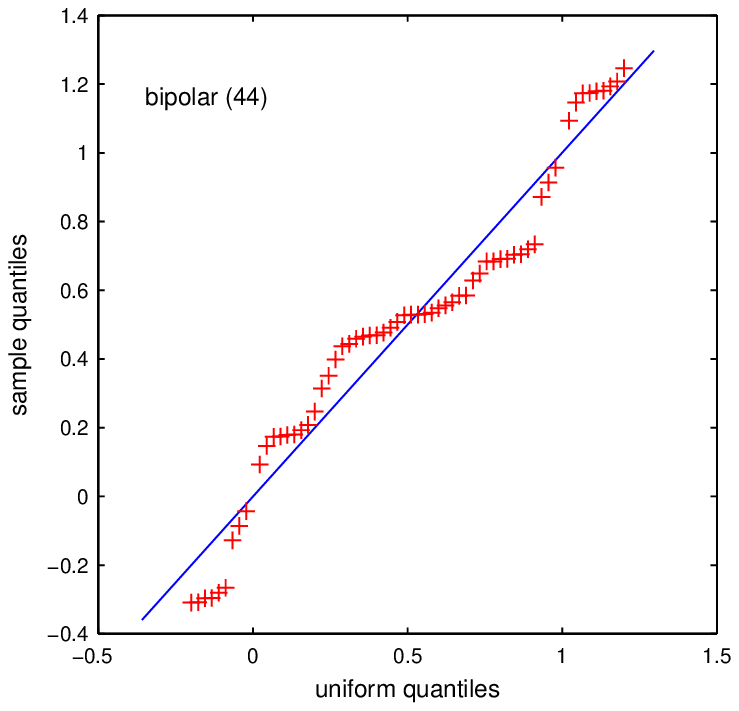}
\\
    \includegraphics[width=8.1cm]{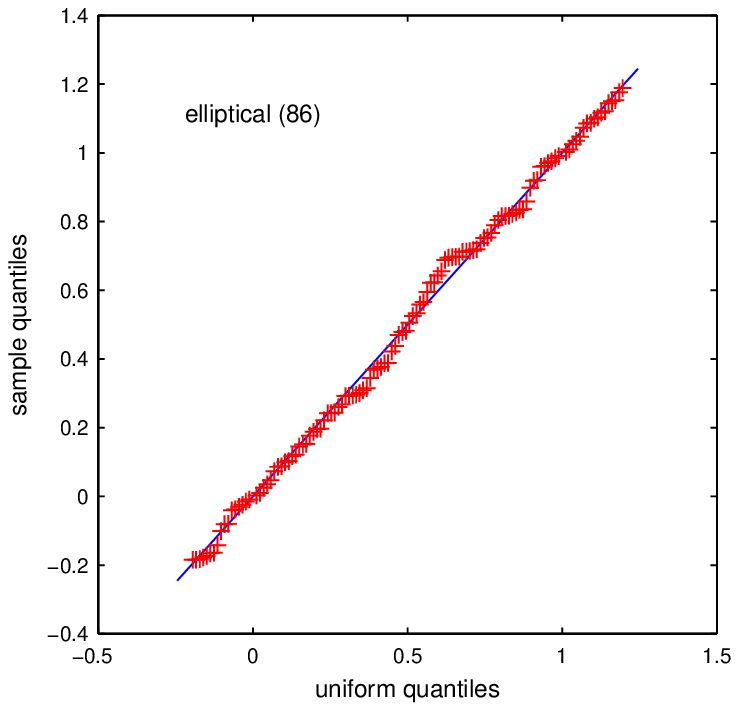} &
    \includegraphics[width=8.1cm]{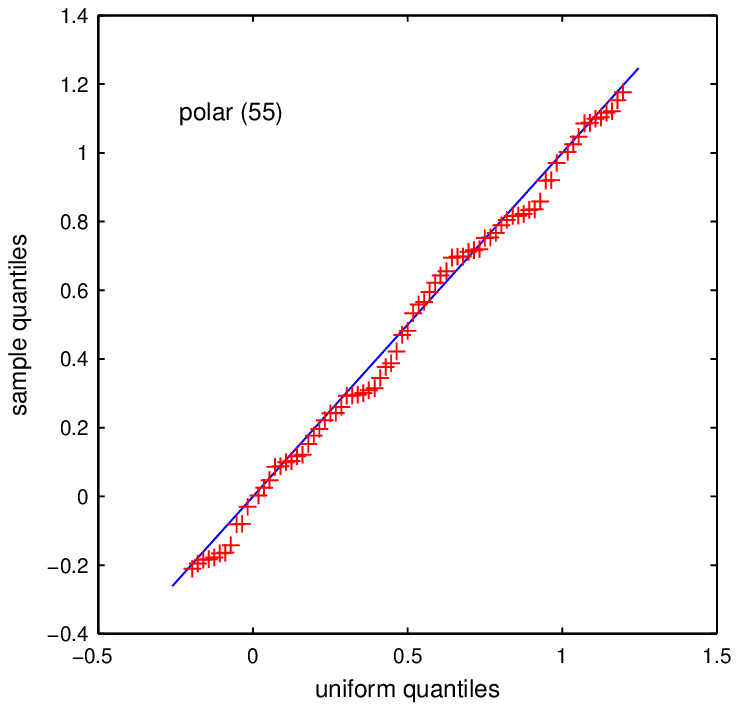}
\\
    \includegraphics[width=8.1cm]{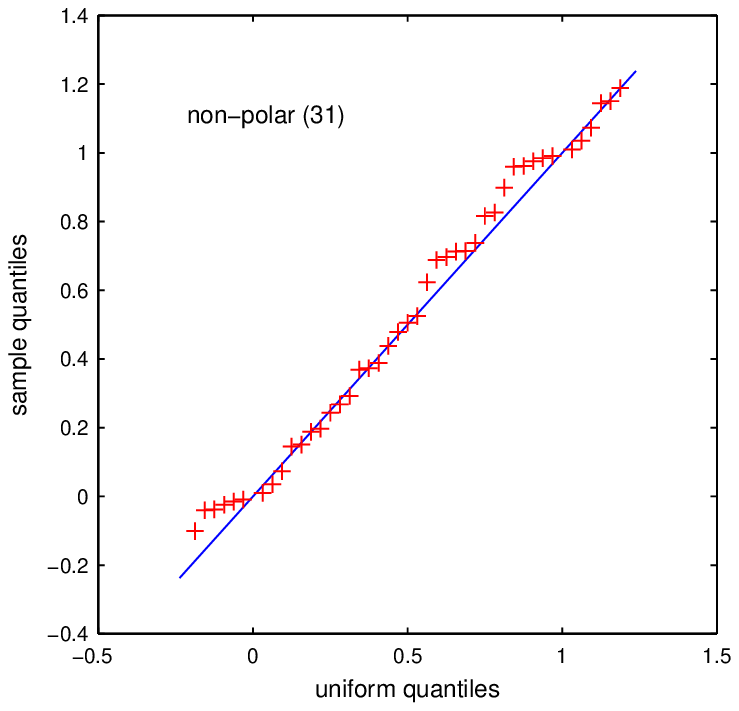} &

    \end{array}
$
    \end{center}
    \caption{The extended quantile-quantile (Q--Q) plots for the full sample and the principal subsamples over the whole area. Q-Q plots for the bipolar PNe in the North, South, East and West subsamples are available online.}
    \label{quantcomp1}
\end{figure*}

Quantile--quantile (Q--Q) plots (a type of probability-probability plot) of the
data for each sample were also prepared \citep{FisherCircStat} and those for
the whole sample and the morphological subsamples are shown in
Fig.~\ref{quantcomp1}. The plots are extended below (0,0) and above (1,1) with
mirrored data from the other end of the plot. This extension is designed to
avoid any misleading effect due to the linear representation of circular
data. Points from an uniform distribution would lie on a 45\degr\  line from
the bottom left to top right of such a plot. Departures from that angle
indicate departures from uniformity. 

The mean GPA values were derived in the usual way for axial data i.e. as a
vector summation over 360\degr\  then halved. 

A Rayleigh test, a Hodges-Ajne test, a Kuiper test and a Watson U$^2$ test for a null hypothesis of uniformity against an alternative hypothesis of non-uniformity were made. The Kuiper test has the advantage that it tests for randomness against any other alternative whereas the Rayleigh test assumes an unimodal alternative \citep{FisherCircStat}. However, if the alternative is a von Mises distribution (also known as the Circular Normal Distribution), the Rayleigh test is the most powerful invariant test for uniformity \citep{springerlink:10.1023/A:1016109603897}. The Hodges-Ajne test counts numbers on arbitrary semi-circles and checks for an excess on one side of the circle. It can also be used for samples from any distribution \citep{TopCircStat,Arach,1985ICRC....3..485P} and is useful when a peak in the number of events is likely to be confined to one half of the angular range. Rayleigh tests for the same null hypothesis against alternative hypotheses of an unimodal distribution with mean angles of 90\degr\ (i.e. along the Galactic Plane) and 100\degr\,, the angle of the excess found by \citet{2008PASP..120..380W}, were also performed. The p--values for the Kuiper and Watson U$^2$ tests were obtained using Mathematica's `DistributionFitTest'. We initially used CircStat \citep{RePEc:jss:jstsof:31:i10} to obtain the p--values for the Hodges-Ajne test but we then modified the procedure to work in Mathematica as using one statistical package for all the statistical tests for uniformity simplified bootstrapping\footnote{Bootstrapping is a process in which new samples of the same size are manufactured by sampling from the original sample with replacement \citep{1979}}. The Rayleigh tests used the formulae presented in \citet{FisherCircStat}. The tests are right tail (one-sided excess) tests, where small p--values would lead us to reject $\mathrm{H}_0$. 

In order to check the effect of the uncertainties in the orientations, the full area tests were repeated after adding randomly selected values from a $\mathrm{N}\left(0,\sigma^2\right)$ distribution, where $\sigma$ is the uncertainty. This process was repeated 100\,000 times. We also used bootstrapping to produce a sampling distribution of the p-values. We repeated this with each of the bootstrapped angles adjusted by a value from its $\mathrm{N}\left(0,\sigma^2\right)$ distribution (a smoothed bootstrap). 

\section{Results}
\label{results}

It can already be seen from the rose plots in Figs.~\ref{distcomp1} and
\ref{distcomp2} that the plot for the bipolar PNe is less balanced
than those for the other classifications. 

Some indications of the uniformity of the distributions of the GPA can be seen
from the Q--Q plots in Fig.~\ref{quantcomp1}. The deviations from the
45\degr\  line tend to increase as the sample size decreases, a situation
that can also be seen in the plots for the area subsamples. However, the bipolar
sample and its area subsamples have greater deviations than do the elliptical
sample and its area subsamples. Note the pronounced deviation from a 45\degr\ 
slope at approximately (0.5,0.5) in the plot for the bipolar data sample.

The p--values obtained for the single sample tests are listed
in Table~\ref{PN_pvals_base} as are the sample sizes. In all five tests the null hypothesis, H$_0$, is
that the distribution of the GPA is uniform (i.e. when the
angles are doubled to cover the full circle it is isotropic).
The alternative hypothesis, H$_1$, for the four tests for
uniformity is simply that the distribution is not uniform.
H$_1$ for the Rayleigh tests against the mean are those of an
unimodal distribution with a mean angle of 90\degr\ (i.e. along the Galactic Plane) or of 100\degr\ (i.e the angle of the excess found by \citet{2008PASP..120..380W}).

\begin{table*} 
\begin{center} 
\caption{The data for the PNe samples and area sub-samples together with the p--values for the single sample tests. The angles are shown rounded to the nearest $1\degr$ and the p--values have been quoted to the first significant digit. The p--values derived for the non-polar south-west subsample have been excluded as the sample size of 3 indicates that they are not
meaningful.  The null hypothesis (H$_0$) for all the tests is that the distribution of the GPA is uniform. East (E), west (W), north (N) and south (S) are with respect to Galactic coordinates on either side of 0\degr longitude and latitude.} 
 \label{PN_pvals_base}
\begin{tabular}{|l|l|r|r|r@{.}l|r@{.}l|r@{.}l|r@{.}l|r@{.}l|r@{.}l|} 
 \hline 
 &  & & & \multicolumn{12}{c|}{p--values}  \\ 
\cline{5-16}
\multicolumn{1}{|c|}{Class} & \multicolumn{1}{|c|}{Area} & \multicolumn{1}{c|}{sample} &  \multicolumn{1}{c|}{mean} &
\multicolumn{8}{c|}{Tests for Uniformity} & \multicolumn{4}{c|}{Rayleigh Test}  \\
 & & \multicolumn{1}{c|}{size} & \multicolumn{1}{c|}{GPA} & \multicolumn{8}{c|}{H$_1$: the distribution is not uniform}  & \multicolumn{4}{c|}{H$_1$: an unimodal mean GPA of:} \\
\cline{5-16} 
 & &  & \multicolumn{1}{c|}{/~$^{\circ}$}  &  \multicolumn{2}{c|}{Rayleigh } &  \multicolumn{2}{c|}{Hodges-Ajne} & \multicolumn{2}{c|}{Kuiper}  & \multicolumn{2}{c|}{Watson U$^2$} & \multicolumn{2}{c|}{90$^{\circ}$} & \multicolumn{2}{c|}{100$^{\circ}$}\\ 
\hline

All & All & 130  & 95  &  0&3  &  0&4  &  0&3  & 0&2 & 0&06 & 0&06  \\
Bipolar & All & 44  & 93  &  0&0008  &  0&02  &  0&0002  & 0&006 & 0&00009 &0&0001  \\
Elliptical & All & 86  & 176  &  0&8  &  0&9  &  0&5  & 0&7 & 0&2 & 0&3  \\
Polar & All & 55  & 158  &  0&9  &  0&9  &  0&6  & 0&7 & 0&4 & 0&4  \\
Non-polar & All & 31  & 8  &  0&7  &  0&9  &  0&9  & 0&9 & 0&2 & 0&2  \\
\\
All & E & 72  & 70  &  0&6  &  0&5  &  0&8  & 0&8 & 0&2 & 0&3  \\
All & W & 58  & 109  &  0&2  &  0&4  &  0&1  & 0&1 & 0&06 & 0&03  \\
All & N & 61  & 107  &  0&1  &  0&3  &  0&1  & 0&1 & 0&04 & 0&02  \\
All & S & 69  & 63  &  0&7  &  0&9  &  0&8  &  0&7 & 0&3 & 0&4  \\
All & NE & 28  & 99  &  0&3  &  0&8  &  0&6  &  0&5 & 0&07 & 0&06  \\
All & NW & 33  & 114  &  0&3  &  0&5  &  0&3  & 0&3  & 0&2 & 0&08  \\
All & SE & 44  & 40  &  0&5  &  0&6  &  0&5  & 0&6  & 0&4 & 0&3  \\
All & SW & 25  & 100  &  0&5  &  0&9  &  0&4  & 0&5  & 0&1 & 0&1  \\
Bipolar & E & 21  & 88  &  0&08  &  0&2  &  0&1  & 0&05  & 0&01 & 0&02  \\
Bipolar & W & 23  & 97  &  0&008  &  0&1  &  0&005  & 0&01  & 0&001 & 0&001  \\
Bipolar & N & 19  & 101  &  0&006  &  0&05  &  0&002  & 0&009  & 0&001 & 0&0006  \\
Bipolar & S & 25  & 84  &  0&05  &  0&1  &  0&04  & 0&02  & 0&009 & 0&02  \\
Bipolar & NE & 8  & 92  &  0&02  &  0&06  &  0&02  &  0&02 & 0&002 & 0&003  \\
Bipolar & NW & 11  & 111  &  0&1  &  0&1  &  0&08  & 0&1  & 0&07 & 0&03  \\
Bipolar & SE & 13  & 75  &  0&7  &  0&5  &  0&6  &  0&4 & 0&2 & 0&3  \\
Bipolar & SW & 12  & 87  &  0&02  &  0&06  &  0&03  & 0&03  & 0&003 & 0&007  \\
Elliptical & E & 51  & 24  &  0&7  &  0&9  &  0&9  & 0&8  & 0&3 & 0&2  \\
Elliptical & W & 35  & 148  &  0&6  &  0&5  &  0&4  & 0&5  & 0&3 & 0&5  \\
Elliptical & N & 42  & 129  &  0&8  &  0&9  &  0&8  & 0&9  & 0&4 & 0&4  \\
Elliptical & S & 44  & 11  &  0&5  &  0&8  &  0&5  &  0&5 & 0&1 & 0&1  \\
Elliptical & NE & 20  & 136  &  0&9  &  0&9  &  1&  & 1&  & 0&5 & 0&4  \\
Elliptical & NW & 22  & 124  &  0&9  &  0&9  &  0&6  & 0&8  & 0&4 & 0&4  \\
Elliptical & SE & 31  & 29  &  0&4  &  0&7  &  0&7  & 0&4  & 0&2 & 0&1  \\
Elliptical & SW & 13  & 159  &  0&4  &  0&5  &  0&6  & 0&5  & 0&2 & 0&3  \\
Polar & E & 33  & 21  &  1&  &  1&  &  0&6  & 0&7  & 0&5 & 0&5  \\
Polar & W & 22  & 155  &  0&8  &  0&9  &  0&8  & 0&9  & 0&3 & 0&4  \\
Polar & N & 26  & 41  &  0&9  &  0&9  &  0&9  & 0&9  & 0&5 & 0&4  \\
Polar & S & 29  & 147  &  0&7  &  0&7  &  0&6  & 0&6  & 0&4 & 0&5  \\
Polar & NE & 14  & 62  &  0&9  &  0&7  &  0&9  & 1&  & 0&4 & 0&5  \\
Polar & NW & 12  & 22  &  0&9  &  1&  &  0&9  & 1&  & 0&4 & 0&3  \\
Polar & SE & 19  & 160  &  1&  &  1&  &  0&7  & 0&7  & 0&4 & 0&4  \\
Polar & SW & 10  & 143  &  0&5  &  0&5  &  0&5  & 0&5  & 0&4 & 0&5  \\
Non-polar & E & 18  & 24  &  0&4  &  0&5  &  0&5  & 0&4  & 0&2 & 0&1  \\
Non-polar & W & 13  & 140  &  0&7  &  0&9  &  0&5  & 0&7  & 0&4 & 0&4  \\
Non-polar & N & 16  & 129  &  0&3  &  0&8  &  0&4  & 0&3 & 0&4 & 0&2  \\
Non-polar & S & 15  & 27  &  0&05  &  0&07  &  0&08  & 0&08  & 0&08 & 0&02  \\
Non-polar & NE & 6  & 143  &  0&5  &  0&2  &  0&3  & 0&4  & 0&4 & 0&5  \\
Non-polar & NW & 10  & 119  &  0&5  &  0&5  &  0&4  & 0&4  & 0&3 & 0&2  \\
Non-polar & SE & 12  & 34  &  0&08  &  0&06  &  0&1  & 0&1  & 0&2 & 0&07  \\
Non-polar & SW & 3  & 4  &  \multicolumn{12}{c|}{ } \\ 
\hline 
 \end{tabular}
\end{center}
 \end{table*} 
\normalsize

The results for the bipolar PNe are shown in Table~\ref{bipolar_pvals}. Bootstrapping increases the mean p--value but reduces the median value. Estimates of the median p-values for the GPA for the whole area samples of the three morphologies and the two combined samples are presented in Table~\ref{PN_pvals}. They were obtained from derived from the bootstrapped plus $\mathrm{N}\left(0,\sigma^2\right)$ uncertainty runs. It can be seen that the results for the bipolar PNe can be considered as significant at, at worst, the 0.005 significance level.

\begin{table*}

\caption{Sampling distributions of the p--values for the GPA of the bipolar PNe sample. There were 100\,000 such samples for each of: the measured value plus a randomly selected $\mathrm{N}\left(0,\sigma^2\right)$ value where $\sigma$ is the uncertainty in the PN orientation, a bootstrapped selection from the measured values with duplicates adjusted by adding multiples of $10^{-8}$ degrees and the bootstrapped selection plus the randomly selected $\mathrm{N}\left(0,\sigma^2\right)$ value. Quoted uncertainties in the statistics are based on the distribution of 100 subsamples of size 1\,000 where that distribution is judged to be Normal using an Anderson--Darling test at a 0.5 significance level.}
\label{bipolar_pvals}
\begin{center}
\scalebox{0.84}{
\begin{tabular}{|l|l|r@{.}l|r@{.}l|r@{.}l|r@{.}l|r@{.}l|} 
 \hline 
\multicolumn{1}{|c|}{test} &\multicolumn{1}{c|}{method} &  \multicolumn{2}{c|}{mean} & \multicolumn{2}{c|}{standard} & \multicolumn{6}{c|}{quantiles} \\ 
\cline{8-12}
 &  &  \multicolumn{2}{c|}{ } & \multicolumn{2}{c|}{deviation} & \multicolumn{2}{c|}{median} & \multicolumn{2}{c|}{16\%} & \multicolumn{2}{c|}{84\%}\\
\hline
Rayleigh & measured + $\mathrm{N}\left(0,\sigma^2\right)$  & 0&00123$\pm$0.00002 & 0&00065$\pm$0.00002 & 0&00109$\pm$0.00002 &  0&00066$\pm$0.00001 & 0&00178$\pm$0.00004 \\
 & bootstrapped  & 0&013$\pm$0.002 & 0&046$\pm$0.007 & 0&00050$\pm$0.00008 &  0&000007$\pm$0.000002 & 0&012$\pm$0.002 \\
 & bootstrapped + $\mathrm{N}\left(0,\sigma^2\right)$ & 0&015$\pm$0.002 & 0&051$\pm$0.007 & 0&0006 &  0&00001 & 0&015$\pm$0.002 \\[2 mm]
Hodges-Ajne & measured + $\mathrm{N}\left(0,\sigma^2\right)$  & 0&02 & 0&0137$\pm$0.0003 & 0&02 &  0&007 & 0&05 \\
 & bootstrapped  & 0&025$\pm$0.002 &  0&058$\pm$0.006 & 0&007 & 0&0001 & 0&05 \\
 & bootstrapped + $\mathrm{N}\left(0,\sigma^2\right)$ & 0&025$\pm$0.002 & 0&062$\pm$0.007 & 0&002 & 0&0001 & 0&05 \\[2 mm]
Kuiper & measured + $\mathrm{N}\left(0,\sigma^2\right)$  & 0&00089$\pm$0.00003 & 0&00095$\pm$0.00007 & 0&00059$\pm$0.00002 & 0&00022$\pm$0.00001 & 0&00149$\pm$0.00006 \\
 & bootstrapped  & 0&0017$\pm$0.0003 & 0&009$\pm$0.003 & 0&00003 & 0&0000002 & 0&0009 \\
 & bootstrapped + $\mathrm{N}\left(0,\sigma^2\right)$ & 0&0026$\pm$0.0005 & 0&01 & 0&00007 & 0&0000009$\pm$0.0000002 & 0&002 \\[2 mm]
Watson U$^2$ & measured + $\mathrm{N}\left(0,\sigma^2\right)$  & 0&00650$\pm$0.00003 & 0&00076$\pm$00002 & 0&00651$\pm$0.00003 & 0&00573$\pm$0.00003 & 0&00725$\pm$0.00004 \\
 & bootstrapped  & 0&0066$\pm$0.0005 & 0&015$\pm$0.003 & 0&0045$\pm$0.0002 & 0&0000006 & 0&0086$\pm$0.0001 \\
 & bootstrapped + $\mathrm{N}\left(0,\sigma^2\right)$  & 0&0082$\pm$0.0007 & 0&020$\pm$0.004 & 0&0051$\pm$0.0002 & 0&00002 & 0&009$\pm$0.0002 \\[2 mm]
Rayleigh & measured + $\mathrm{N}\left(0,\sigma^2\right)$  & 0&000134$\pm$0.000002 &  0&000080$\pm$0.000003 & 0&000120$\pm$0.000003 & 0&000069$\pm$0.000002 & 0&000204$\pm$0.000005 \\
H$_1$: $90^\circ$ & bootstrapped  & 0&0032$\pm$0.0005 & 0&015$\pm$0.003 & 0&00008$\pm$0.00001 & 0&0000012$\pm$0.0000003 & 0&0024$\pm$0.0004 \\
unimodal & bootstrapped + $\mathrm{N}\left(0,\sigma^2\right)$  & 0&0039$\pm$0.0006 & 0&017$\pm$0.004 & 0&0001 & 0&000002 & 0&0031$\pm$0.0005 \\[2 mm]
Rayleigh & measured + $\mathrm{N}\left(0,\sigma^2\right)$  & 0&000200$\pm$0.000004 &  0&000110$\pm$0.000005 & 0&000175$\pm$0.000003 & 0&000105$\pm$0.000002 & 0&000290$\pm$0.000007 \\
H$_1$: $100^\circ$ & bootstrapped  & 0&0054$\pm$0.0007 & 0&023$\pm$0.004 & 0&00012$\pm$0.00002 & 0&000001 & 0&004 \\
unimodal & bootstrapped + $\mathrm{N}\left(0,\sigma^2\right)$  & 0&0062$\pm$0.0008 & 0&025$\pm$0.004 & 0&00016$\pm$0.00003 & 0&000002 & 0&0050$\pm$0.0008 \\
\hline
\end{tabular}
}
\end{center}

\end{table*}

\begin{table*}

\caption{The median p--values for the GPA of the five single sample tests. The values and their uncertainties were obtained by bootstrapping the GPA and adding a randomly selected $\mathrm{N}\left(0,\sigma^2\right)$ value, where $\sigma$ is the uncertainty in the PN orientation for the bipolar PNe and 2\degr\, for the other PNe. The median has been used due to the distributions of the p--values being skewed. There were 100\,000 bootstraps. The samples of 100\,000 were each split into subsamples of 1\,000 and the mean and standard deviation of the median p-value obtained. Where the distribution of the medians was judged to be Normal using an Anderson--Darling test at a 0.5 significance level its standard deviation was used as the uncertainty, otherwise the p--value has been quoted to the first significant digit. The mean GPA quoted is that of the original sample.} 

\label{PN_pvals}
\begin{tabular}{|l|r|r|r@{.}l|r@{.}l|r@{.}l|r@{.}l|r@{.}l|r@{.}l|} 
 \hline 
 &  & & \multicolumn{12}{c|}{median p--values}  \\ 
\cline{4-15}
\multicolumn{1}{|c|}{Class} &  \multicolumn{1}{c|}{sample} &  \multicolumn{1}{c|}{mean} & \multicolumn{8}{c|}{Tests for Uniformity} & \multicolumn{4}{c|}{Rayleigh Test with H$_1$:}  \\
\cline{4-11} 
 & \multicolumn{1}{c|}{size} & \multicolumn{1}{c|}{GPA} & \multicolumn{2}{c|}{\multirow{2}{*}{Rayleigh}} & \multicolumn{2}{c|}{\multirow{2}{*}{Hodges-Ajne}} & \multicolumn{2}{c|}{\multirow{2}{*}{Kuiper}} & \multicolumn{2}{c|}{\multirow{2}{*}{Watson U$^2$}} & \multicolumn{4}{c|}{an unimodal mean GPA of:} \\
\cline{12-15} 
 & &  \multicolumn{1}{c|}{/~$^{\circ}$}  &  \multicolumn{2}{c|}{ } &  \multicolumn{2}{c|}{ } & \multicolumn{2}{c|}{ }  & \multicolumn{2}{c|}{ } & \multicolumn{2}{c|}{90$^{\circ}$} & \multicolumn{2}{c|}{100$^{\circ}$}\\ 
\hline
All &  130  & 95  &  0&17$\pm$0.01  &  0&1  &  0&038$\pm$0.003  & 0&082$\pm$0.005 & 0&063$\pm$0.005 & 0&061$\pm$0.005  \\
Bipolar &  44  & 93  &  0&0006  &  0&002  &  0&00007  & 0&0051$\pm$0.0002 & 0&0001 & 0&00016$\pm$0.00003  \\
Elliptical  & 86  & 176  &  0&40$\pm$0.02  &  0&3  &  0&16$\pm$0.01  & 0&22$\pm$0.01 & 0&205$\pm$0.009  & 0&214$\pm$0.009\\
Polar &  55  & 158  &  0&47$\pm$0.01  &  0&4  &  0&152$\pm$0.008 & 0&194$\pm$0.008 & 0&264$\pm$0.008 & 0&264$\pm$0.008  \\
Non-polar &  31  & 8  &  0&39$\pm$0.02  &  0&5  &  0&24$\pm$0.01  & 0&29$\pm$0.01 & 0&186$\pm$0.008 & 0&183$\pm$0.008  \\

\hline 

\end{tabular}
\end{table*}

Runs were also performed for the GPA of the bipolar sample with the  uncertainties increased by 5, 10 and 20 degrees in order to check the effects of larger uncertainties. As expected, increasing the uncertainties increases the probability that the GPA for the sample come from a randomly orientated population of bipolar Bulge PNe. The results are presented in Table~\ref{uncertainties}. The median p--values for uncertainties increased by no more than 10\degr\ are all below 0.01 with those for the tests against the specified mean orientations below 0.001. With added 20\degr\ uncertainties the Rayleigh tests against the alternative hypotheses of unimodal distributions with mean values of 90\degr\ and 100\degr\ respectively, still produce p--values $\leq 0.01$.  Any increase in the uncertainties above those estimated from the measurements would need to be large in order to change our conclusions.

\begin{table*}

\caption{The effect of uncertainties in the orientations on the sampling distributions of the p--values for the GPA of the bipolar PNe sample. The results indicated to have 0\degr\, additional uncertainty are those shown in the third row for each test in Table~\ref{bipolar_pvals}. They are for 100\,000 bootstrapped selections from the measured values to which were added values randomly selected from a $\mathrm{N}\left(0,\sigma^2\right)$ distribution, where $\sigma$ is the uncertainty in the PN orientation.  Those results are compared to three further sets of 100\,000 for which $\sigma$ was increased by values of 5, 10 and 20\degr\, respectively. Quoted uncertainties in the statistics are based on the distribution of 100 subsamples of size 1\,000 where that distribution is judged to be Normal using an Anderson--Darling test at a 0.5 significance level.}
\label{uncertainties}

\begin{tabular}{|l|l|r@{.}l|r@{.}l|r@{.}l|r@{.}l|r@{.}l|} 
 \hline 
\multicolumn{1}{c|}{test} & \multicolumn{1}{c|}{added} &  \multicolumn{2}{c|}{mean} & \multicolumn{2}{c|}{standard} & \multicolumn{6}{c|}{quantiles} \\ 
\cline{8-12}
 & \multicolumn{1}{c|}{uncertainty/\degr} & \multicolumn{2}{c|}{ } & \multicolumn{2}{c|}{deviation} & \multicolumn{2}{c|}{median} & \multicolumn{2}{c|}{16\%} & \multicolumn{2}{c|}{84\%}\\
\hline
Rayleigh & 0 & 0&015$\pm$0.002 & 0&051$\pm$0.007 & 0&0006 &  0&00001 & 0&015$\pm$0.002 \\
 & 5 & 0&021$\pm$0.002 & 0&067$\pm$0.007 & 0&001 & 0&00002 & 0&025$\pm$0.003 \\
 & 10 & 0&038$\pm$0.003 & 0&100$\pm$0.008 & 0&003 & 0&00005 & 0&018$\pm$0.05 \\
 & 20  & 0&106$\pm$0.006 & 0&190$\pm$0.009 & 0&018$\pm$0.002 & 0&00037$\pm$0.00009 & 0&22$\pm$0.02 \\[2 mm]
Hodges-Ajne & 0 & 0&025$\pm$0.002 & 0&062$\pm$0.007 & 0&002 & 0&0001 & 0&05 \\
 & 5 & 0&029$\pm$0.002 & 0&075 & 0&002 & 0&0001 & 0&05 \\
 & 10 & 0&043$\pm$0.003 & 0&102$\pm$0.008 & 0&007 & 0&0001 & 0&05 \\
 & 20  & 0&103$\pm$0.006 & 0&179$\pm$0.008 & 0&02 & 0&0006 & 0&2 \\[2 mm]
Kuiper & 0 & 0&0026$\pm$0.0005 & 0&01 & 0&00007 & 0&0000009$\pm$0.0000002 & 0&002 \\
 & 5 & 0&0071$\pm$0.0008 & 0&028$\pm$0.005 & 0&00027$\pm$0.00004 & 0&000004 & 0&0066$\pm$0.0009 \\
 & 10 & 0&016$\pm$0.002 & 0&051$\pm$0.006 & 0&0008 & 0&000012$\pm$0.000003 & 0&02 \\
 & 20  & 0&050$\pm$0.004 & 0&111$\pm$0.007 & 0&0052$\pm$0.0007 & 0&00009$\pm$0.00002 & 0&085$\pm$0.009 \\[2 mm]
Watson U$^2$ & 0  & 0&0082$\pm$0.0007 & 0&020$\pm$0.004 & 0&0051$\pm$0.0002 & 0&00002 & 0&009$\pm$0.0002 \\
 & 5 & 0&0132$\pm$0.0009 & 0&03 & 0&0062$\pm$0.0002 & 0&0012$\pm$0.0002 & 0&01 \\
 & 10 & 0&024$\pm$0.002 & 0&060$\pm$0.007 & 0&0074$\pm$0.0002 & 0&0024$\pm$0.0003 & 0&03 \\
 & 20  & 0&067$\pm$0.004 & 0&126$\pm$0.007 & 0&01 & 0&0050$\pm$0.0003& 0&12$\pm$0.01 \\[2 mm]
Rayleigh & 0  & 0&0039$\pm$0.0006 & 0&017$\pm$0.004 & 0&0001 & 0&000002 & 0&0031$\pm$0.0005 \\
H$_1$: $90^\circ$  & 5 & 0&006 & 0&024$\pm$0.004 & 0&0002 & 0&000004 & 0&0056$\pm$0.0007 \\
unimodal  & 10 & 0&013$\pm$0.001 & 0&041$\pm$0.004 & 0&00058$\pm$0.00008 & 0&000009$\pm$0.000002 & 0&014$\pm$0.002 \\
 & 20  & 0&047$\pm$0.003 & 0&093$\pm$0.004 & 0&006 & 0&0001 & 0&090$\pm$0.008 \\[2 mm]
Rayleigh & 0  & 0&0062$\pm$0.0008 & 0&025$\pm$0.004 & 0&00016$\pm$0.00003 & 0&000002 & 0&0050$\pm$0.0008 \\
H$_1$: $100^\circ$  & 5 & 0&0088$\pm$0.0009 & 0&032$\pm$0.004 & 0&00032$\pm$0.00005 & 0&000004 & 0&009$\pm$0.001 \\
unimodal  & 10 & 0&02 & 0&05 & 0&0008$\pm$0.0001 & 0&00001 & 0&019$\pm$0.002 \\
 & 20  & 0&052$\pm$0.003 & 0&097$\pm$0.004 & 0&007 & 0&0001 & 0&104$\pm$0.009 \\
\hline
\end{tabular}


\end{table*}

We note that the NTT images were taken at a fixed equatorial orientation while the HST images have a random orientation.  Instrumental effects could only have introduced Equatorial alignments for the NTT images for which we find no evidence. The conversion from Equatorial to Galactic coordinates gives some rotation of the polar direction  across the area of the Bulge. We therefore repeated the tests for uniformity of the bipolar PNe using the observed (equatorial) PA in order to check for the effect of the conversion on the p--values. Those results are shown in Table~\ref{bipolar_pvals_epa}. For all but the Hodges-Ajne test, the median p--values for the bootstrapped plus $\mathrm{N}\left(0,\sigma^2\right)$ distributions are larger than those for the GPA. The non-random nature of the bipolar orientations is therefore an
attribute of the Galactic coordinates, and is not introduced by the Equatorial coordinates.

\begin{table*}

\caption{Sampling distributions of the p--values for the PA of the bipolar PNe sample. There were 100\,000 such samples for each of: the measured value plus a randomly selected $\mathrm{N}\left(0,\sigma^2\right)$ value where $\sigma$ is the uncertainty in the PN orientation, a bootstrapped selection from the measured values with duplicates adjusted by adding multiples of $10^{-8}$ degrees and the bootstrapped selection plus the randomly selected $\mathrm{N}\left(0,\sigma^2\right)$ value. Quoted uncertainties in the statistics are based on the distribution of 100 subsamples of size 1\,000 where that distribution is judged to be Normal using an Anderson--Darling test at a 0.5 significance level.}

\label{bipolar_pvals_epa}
\scalebox{0.9}{
\begin{tabular}{|l|l|r@{.}l|r@{.}l|r@{.}l|r@{.}l|r@{.}l|} 
 \hline 
\multicolumn{1}{|c|}{test} &\multicolumn{1}{c|}{method} &  \multicolumn{2}{c|}{mean} & \multicolumn{2}{c|}{standard} & \multicolumn{6}{c|}{quantiles} \\ 
\cline{8-12}
 & &   \multicolumn{2}{c|}{ } & \multicolumn{2}{c|}{deviation} & \multicolumn{2}{c|}{median} & \multicolumn{2}{c|}{16\%} & \multicolumn{2}{c|}{84\%}\\
\hline
Rayleigh & measured + $\mathrm{N}\left(0,\sigma^2\right)$ &  0&00228$\pm$0.00004 & 0&00115$\pm$0.00004 & 0&00204$\pm$0.00004 &  0&00126$\pm$0.00003 & 0&00328$\pm$0.00008 \\
 & bootstrapped &  0&019$\pm$0.002 & 0&060$\pm$0.007 & 0&0010$\pm$0.0001 &  0&00002 & 0&022$\pm$0.003 \\
 & bootstrapped + $\mathrm{N}\left(0,\sigma^2\right)$&  0&022$\pm$0.002 & 0&066$\pm$0.007 & 0&0012$\pm$0.0002 &  0&00002 & 0&026$\pm$0.003 \\[2 mm]
Hodges-Ajne & measured + $\mathrm{N}\left(0,\sigma^2\right)$ &  0&0310$\pm$0.0005 & 0&02 & 0&02 &  0&02 & 0&05 \\
 & bootstrapped &  0&031$\pm$0.002 &  0&071$\pm$0.005 & 0&007 & 0&0001 & 0&05 \\
 & bootstrapped + $\mathrm{N}\left(0,\sigma^2\right)$&  0&037$\pm$0.002 & 0&082$\pm$0.006 & 0&007 & 0&0006 & 0&05 \\[2 mm]
Kuiper & measured + $\mathrm{N}\left(0,\sigma^2\right)$ &  0&00117$\pm$0.00005 & 0&001 & 0&00074$\pm$0.00003 & 0&00026$\pm$0.00001 & 0&0020$\pm$0.0001 \\
 & bootstrapped &  0&0024$\pm$0.0004 & 0&012$\pm$0.004 & 0&00006 & 0&0000004 & 0&002 \\
 & bootstrapped + $\mathrm{N}\left(0,\sigma^2\right)$&  0&0038$\pm$0.0006 & 0&017$\pm$0.004 & 0&0001 & 0&000001 & 0&0030$\pm$0.0003 \\[2 mm]
Watson U$^2$ & measured + $\mathrm{N}\left(0,\sigma^2\right)$ &  0&00721$\pm$0.00003 & 0&00076$\pm$0.00002 & 0&00722$\pm$0.00003 & 0&00645$\pm$0.00004 & 0&00796$\pm$0.00003 \\
 & bootstrapped &  0&0087$\pm$0.0006 & 0&020$\pm$0.003 & 0&0053$\pm$0.0002 & 0&0002 & 0&0093$\pm$0.0002 \\
 & bootstrapped + $\mathrm{N}\left(0,\sigma^2\right)$ &  0&0105$\pm$0.0008 & 0&025$\pm$0.004 & 0&0058$\pm$0.0002 & 0&0008$\pm$0.0002 & 0&01 \\
\hline
\end{tabular}
}

\end{table*}

Our statistical analysis having determined that the bipolar PNe have a distribution of their orientations that is unlikely to be random we searched for a possible cause.

We used data from the Max-Planck-Institut f\"{u}r Radioastronomie's Survey Sampler and the circ\_corrcc function in CircStat to check for a relationship between the GPA of the PNe and the Galactic magnetic
field using 21~cm polarization angles in the Galactic Bulge from the Villa
Elisa survey as a tracer for the field \citep{2008A&A...484..733T}. The correlations obtained between the polarization angles
and the GPA of the sample and main subsamples have coefficients that lie in
the interval $\left(-0.13, 0.21\right)$. The low magnitudes of the correlation coefficients suggest that
there is little, if any, relationship between the orientation of PNe and the current Galactic magnetic field. It should, however, be noted that the result does not allow for depolarization due to Faraday rotation \citep{2008A&A...484..733T} and that the wavelength of the Villa Elisa survey is too long to avoid that effect. Moreover the resolution of the survey is too low to match the locations of the PNe with sufficient precision so we are unable to come to any conclusion regarding a possible relationship.
 
\begin{figure*}
    \begin{center}

$
    \begin{array}{cc}
    \includegraphics[width=8.1cm]{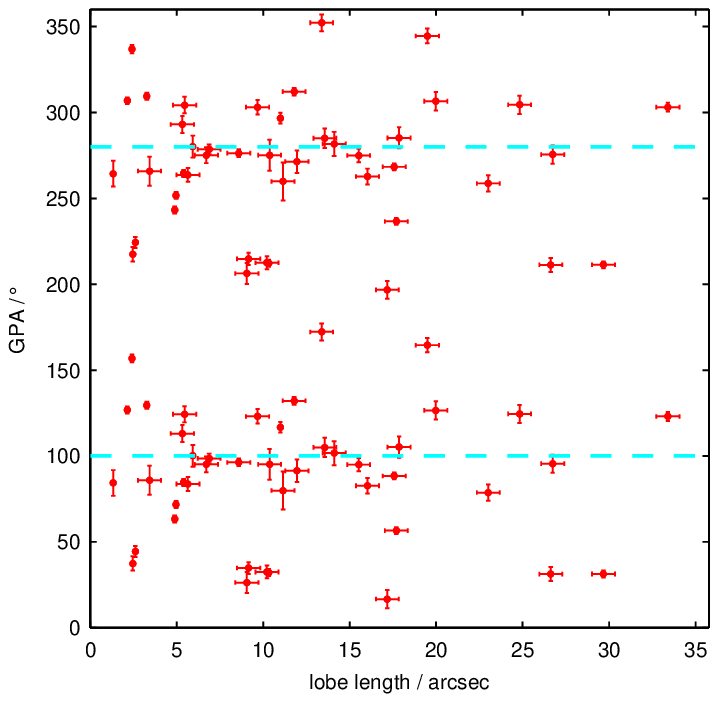} &
   
\\
    \includegraphics[width=8.1cm]{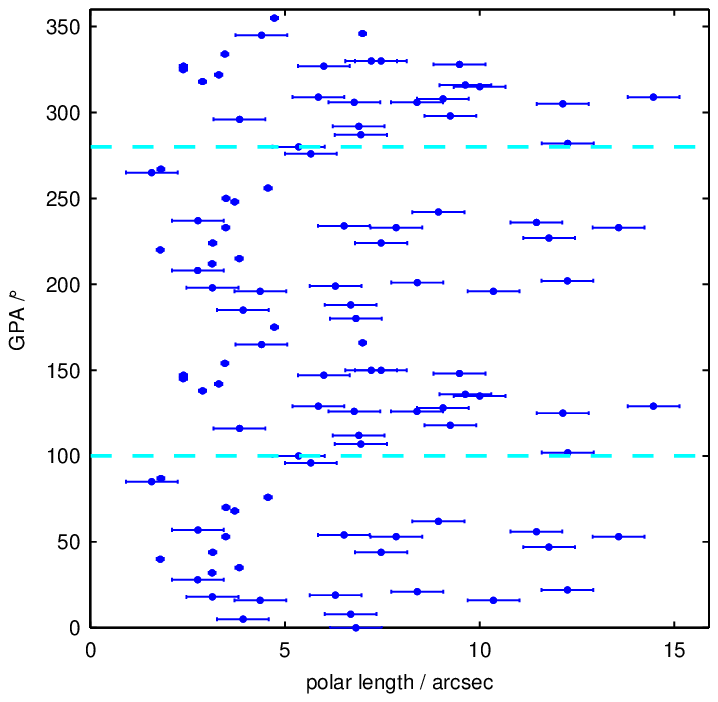} &
    \includegraphics[width=8.1cm]{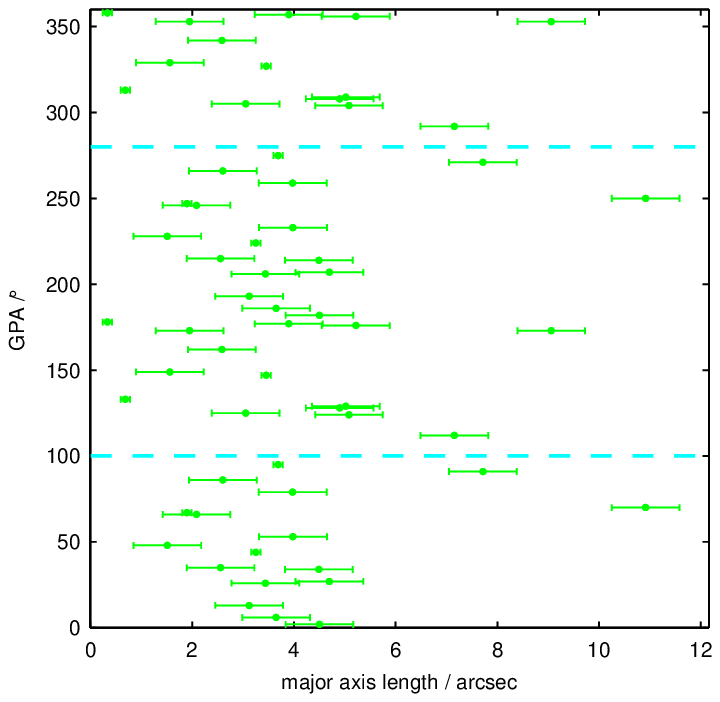}
    \end{array}
$
    \end{center}
    \caption{The GPA of the PNe plotted against the lobe to lobe dimensions of the 44 bipolar PNe (red, top left), the polar lengths of the 55 polar PNe (blue, bottom left) and the lengths along the major axis of the 31 non-polar PNe (green, bottom right). The PNe appear twice in order to emphasize the axial nature of the angular data, once for the GPA and once for the GPA plus 180\degr. The horizontal lines at 100\degr and 280\degr are to allow comparison with the angle of excess found by \citet{2008PASP..120..380W}. The uncertainties forming the error bars for the lengths are dominated by the pixelation and are taken to be 1 s.d.}
    \label{dimang}
\end{figure*}

\begin{figure}
    \begin{center}

    \includegraphics[width=8.1cm]{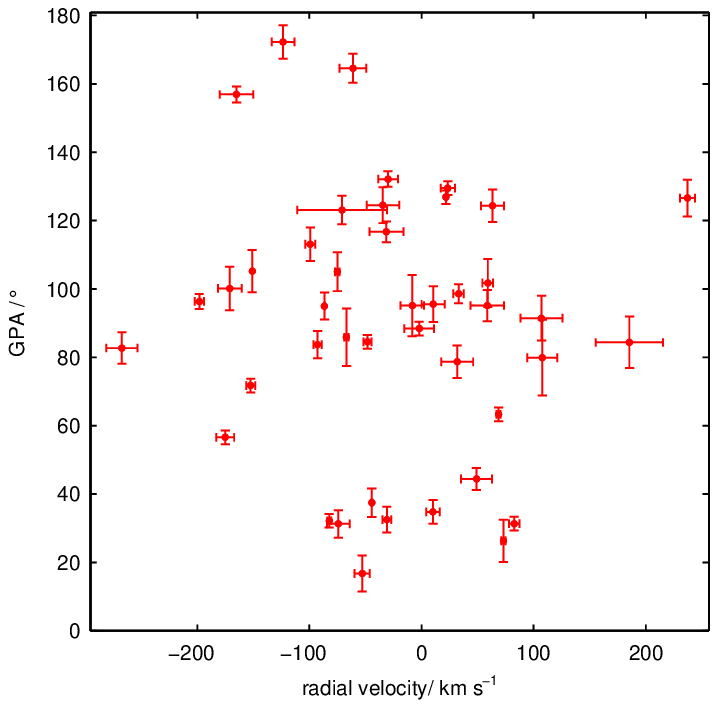}
    \end{center}
    \caption{The GPA of the 44 bipolar PNe plotted against their Radial Velocities.The uncertainties forming the error bars for the radial velocities are taken from the catalogues. The figures for the polar and non-polar PNe are available online.}
    \label{velang}
\end{figure}

As any effect from external magnetic fields is likely to become more pronounced with
the age of a PN we also decided to look for any relationship between the lobe
lengths of the bipolar nebulae and their orientation. We measured the lengths
of the PN from lobe tip to lobe tip at (for the most part) 1~per~cent of the
peak intensity level. Linear regression (the procedure g02ca of The NAG Toolbox for MATLAB) was used to test for any relationship between GPA and that length. Plots of the relationship between PN orientation and PNe size are shown in Fig.~\ref{dimang}. A plot of the orientation of the bipolar PNe against radial velocity is provided as Fig.~\ref{velang} and those for the polar and non-polar PNe are provided online. The Pearson product-moment correlation coefficient from the linear regression of the GPA against length are  -0.009 for the bipolar GPA against lobe length, 0.02 for the polar GPA against polar length and -0.005 for the non-polar GPA against the length along the major axis. We thus find no relationship between nebular age and orientation.

We also used linear regression to check for any relationship between the PN orientation and its radial
or expansion velocity. The radial velocities were obtained from \citet{1998A&AS..132...13D,Malaroda2006,0004-637X-515-2-610,1992secg.book.....A} and the expansion velocities from \citet{2007A&A...467L..29G} and K. Gesicki (private communication). The correlation coefficients for the GPA against the radial velocity lie in the interval (-0.1,0.08) with that for the bipolar PNe being -0.092. The correlation coefficients for the GPA against expansion velocity lie in the interval (-0.2,0.3). Therefore there appears to be no relationship between orientation and either velocity. 

\begin{figure}
    \begin{center}

    \includegraphics[width=8.4cm]{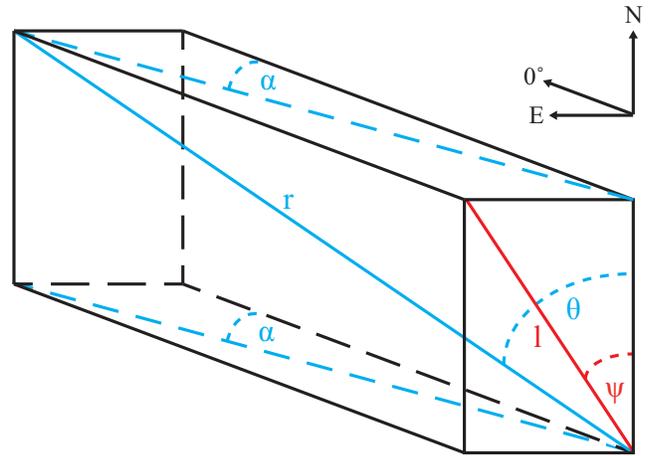}
    \end{center}
    \caption{A schematic of the 3D situation. The axis of the PN is denoted by the blue line, its length by \textit{r} and the angle from Galactic North to the axis by $\theta$. The projection onto the sky has its axis denoted by the red line, its length by \textit{l} and the GPA by $\psi$. The projection onto the Galactic Plane is shown by the dotted blue line and its angle from 0\degr by $\alpha$. That line and its angle are repeated on the top face of the cuboid to clarify the derivation of the angle in the text. Similar schematics are valid for a PN with $\psi > 90\degr$ and/or $\alpha > 90\degr$.}
    \label{cuboid}
\end{figure}

Some caution is required when considering this analysis in that the angles were measured in the plane of the sky and no adjustment to allow for projection effects was attempted. We do not have the additional data to derive the extent of the PNe along the line of sight. The fact that the data is two dimensional has two effects: First, it raises the possibility that some bipolar PNe were observed approximately along the lobes and consequently misclassified. Secondly, if we consider the angles of the PNe taken from the Galactic North Pole, through the lobe-to-lobe axis of PN towards the Galactic Plane the true angles will be more concentrated towards 90\degr. This can readily be seen by considering the GPA as lying on the diagonal of the front (rear) face of a cuboid with the PN stretching from the bottom right of that face to the top left of the rear (front) face see Fig.~\ref{cuboid}. From the figure we can see that:
\begin{equation}
r\cos\theta=l\cos\psi \label{eqnnorth}
\end{equation}
\begin{equation}
r\sin\alpha\sin\theta=l\sin\psi \label{eqnplane}
\end{equation}
and
\begin{equation}
l \leq r \label{eqnseq}
\end{equation}
where $l$ and $r$ are the observed and actual lengths of the PN,  and $\psi$, $\theta$ and $\alpha$ are the GPA, the angle from Galactic North to the axis of the PN and the angle of the axis of the PN projected onto the Galactic Plane from the PN's Galactic longitude, respectively. We also have 
\begin{equation}
r\leq r_{max} \label{rlim}
\end{equation}
where $r_{max}$ is the maximum permitted PN length of 35~arcsec, see \S\ref{obs}. 
From equation~\ref{eqnnorth} we can see that the value of the angle from Galactic North to the PN axis is closer to 90\degr than is the GPA.

Due to the rotation of the Bulge it seems reasonable to assume that the orientations of the projections onto the Galactic Plane are random. We created 100\,000 possible samples based on our bipolar samples where, for each sample, the true length of each PN was estimated based on a random angle in the range [0,180) degrees that was taken as the (axial) orientation of its projection onto the Galactic Plane. Equations \ref{eqnnorth} to \ref{eqnseq}, the PN's GPA and its projected length on the sky, and the maximum permitted PN length were used to ensure the validity of the angle on the Plane. We used the measured PN lengths of Rees, Zijlstra and Gesicki  (in preparation) and the maximum permitted PN length of 35~arcsec (neither the observed nor maximum length changes due to the change in coordinates).  The process was repeated for each PN until a `valid' angle on the Plane was obtained whereupon the value of $\theta$ obtained from equation~\ref{eqnnorth} was used as the angle from Galactic North down to the axis of the PN. The randomness of the angles in the sample was then checked using the same tests as in \S\ref{analysis}. The results are shown in Table~\ref{3D} and suggest that the distribution of the angles from North is not random. Note that this process did not involve bootstrapping. Histograms of the derived angles are provided online.

\begin{table*}

\caption{The p--values for the deprojected angles from Galactic North to the axes of the bipolar PNe sample. They are for 100\,000 samples made up as follows. A random angle in the range [0,180) degrees was produced for each PN and used as the (axial) orientation when projection onto the Galactic Plane. The equations in \S\ref{results} together with the PN's GPA, its projected length  on the sky as measured from the observations, and a maximum permitted `true' PN length of 35~arcsec, see \S\ref{obs}, were used to ensure the validity of the angle on the Plane. The process was repeated until a `valid' angle on the Plane was obtained whereupon the angle from Galactic North down to the axis of the PN obtained from equation~\ref{eqnnorth} was used as the sample value for the PN. The null hypothesis (H$_0$) for all the tests is that the distribution of the angles of the nebular axes from Galactic North is uniform. Quoted uncertainties in the statistics are based on the distribution of 100 subsamples of size 1\,000 where that distribution is judged to be Normal using an Anderson--Darling test at a 0.5 significance level. The mean angle from Galactic North was estimated as $91.10\pm0.04$\degr.}
\label{3D}

\begin{tabular}{|l|r@{.}l|r@{.}l|r@{.}l|r@{.}l|r@{.}l|} 
 \hline 
\multicolumn{1}{c|}{test} & \multicolumn{2}{c|}{mean} & \multicolumn{2}{c|}{standard} & \multicolumn{6}{c|}{quantiles} \\ 
\cline{6-11}
 & \multicolumn{2}{c|}{ } & \multicolumn{2}{c|}{deviation} & \multicolumn{2}{c|}{median} & \multicolumn{2}{c|}{16\%} & \multicolumn{2}{c|}{84\%}\\
\hline
Rayleigh  & \multicolumn{2}{l|}{$3\times 10^{-6}$} & \multicolumn{2}{l|}{$5\times 10^{-6}$} & 8&3$\pm0.6\times 10^{-7}$ &  1&2$\pm0.1\times 10^{-7}$ & 4&3$\pm0.3\times 10^{-6}$ \\
~~H$_1$: non-uniform & \multicolumn{10}{c|}{ } \\
Hodges-Ajne & 1&5$\pm0.1\times 10^{-4}$ & 3&9$\pm0.6\times 10^{-4}$ & \multicolumn{2}{l|}{$3\times 10^{-5}$} & \multicolumn{2}{l|}{$4\times 10^{-6}$} & \multicolumn{2}{l|}{$1\times 10^{-4}$} \\
Kuiper & 7&1$\pm0.4\times 10^{-6}$ & 1&4$\pm0.1\times 10^{-5}$ & 2&3$\pm0.2\times 10^{-6}$ & \multicolumn{2}{l|}{$3\times 10^{-7}$} & 1&22$\pm0.08\times 10^{-5}$ \\
Watson U$^2$  & \multicolumn{2}{l|}{$6\times 10^{-5}$} & 2&5$\pm0.2\times 10^{-4}$ & \multicolumn{2}{l|}{$<10^{-7}$} &  \multicolumn{2}{l|}{$<10^{-7}$} &  \multicolumn{2}{l|}{$<10^{-7}$} \\
Rayleigh  & \multicolumn{2}{l|}{$2\times 10^{-7}$} & \multicolumn{2}{l|}{$4\times 10^{-7}$} & \multicolumn{2}{l|}{$6\times 10^{-8}$} & \multicolumn{2}{l|}{$9\times 10^{-9}$} & 3&3$\pm0.03\times 10^{-7}$ \\
~~H$_1$: unimodal, $90^\circ$ mean & \multicolumn{10}{c|}{ } \\
\hline
\end{tabular}


\end{table*}

\section{Discussion}
\label{disconc}

\subsection{Alignment}

We divide the sample of PNe into three separate morphological categories. The statistics tell us that one of those categories, the bipolar PNe, shows evidence for alignment with the Galactic Plane while the other two appear to have random alignments. Our result for the bipolar PNe appears to be in reasonable conformity with that for the elongated PNe in the Galactic Centre region of \citet{2008PASP..120..380W}.

The GPA of the full sample of 130 PNe has an uniform distribution even with a significance level of 0.01 as used by \citet{2008PASP..120..380W}. This is also true for the samples of 55 polar, 31 non-polar and 86 elliptical (combined polar and non-polar) PNe and is the case for all the statistical tests for uniformity, see Tables~\ref{PN_pvals_base} and \ref{PN_pvals}. 

A different result arises when we consider the bipolar subsample. If we restrict ourself to the measured values with no added uncertainties and no bootstrapping (Table~\ref{PN_pvals_base}), the Kuiper and Rayleigh tests for uniformity against a simple non-uniform alternative allow us to reject the null hypothesis of uniformity of the sample at a 0.001 significance level, the Watson U$^2$ test does so at the 0.01 level. 

The Hodges-Ajne test does not let us reject uniformity. That test counts numbers on arbitrary semi-circles on the full circle and checks for an excess on one side. It has low resolution for a sample size of 44 as an increase or decrease of 1 in the minimum number on a semicircle can cause a large change in the p--value. A minimum of 11 angles on the semicircle produces a p--value of 0.02 whereas values of 10 and 12 produce p--values of 0.01 and 0.05 respectively. In effect there is a problematical step change in the p--values. The test also has higher standard deviations for the p--values than do the other tests for all the categories in Tables~\ref{bipolar_pvals}, \ref{bipolar_pvals_epa} and \ref{uncertainties}. This suggests that it has less discriminatory value than other tests and we  therefore place more weight on the results from them. 

The Rayleigh tests for uniformity against the alternative of an unimodal distribution with mean values of 90\degr\ (i.e. along the Galactic Plane) and 100\degr\ both produce p--values below 0.0002.   \citet{FisherCircStat} points out that tests with a simple non-uniformity alternative may not be very good at detecting something specific like unimodality whereas a test designed to detect that attribute may not be very good at detecting other types of non-uniformity. This difference might explain the difference in the results. The finger plots and rose plots of the bipolar sample indicate that it is reasonable to make an assumption of unimodality for it, which supports the use of the Rayleigh test against unimodal distributions.

The median p--values given in Table~\ref{PN_pvals} confirm the basic result. Two of the tests for uniformity against a simple non-uniform alternative allow us to reject the null hypothesis of uniformity of the bipolar sample at the 0.001 (99.9 per cent) significance level and the other two do so at the 0.01 (99 per cent) significance level.  The  Rayleigh tests for uniformity against unimodal distributions with mean values of 90\degr and 100\degr both allow us to reject the null hypothesis of uniformity of the bipolar sample at the 0.0002 (99.98 per cent) significance level.

We did not pursue the results for any of the area subsamples using bootstrapping. However, the basic results, shown in Table~\ref{PN_pvals_base}, suggest that the non-random nature of the orientations is not due to some localized subset of the bipolar PNe. At the 0.01 significance level, the Rayleigh, Kuiper and Watson U$^2$  tests indicate that the orientations of the bipolar PNe are not random for either the West or North subsamples.  The Rayleigh tests for uniformity against the alternative of an unimodal distribution with mean values of 90\degr\ (i.e. along the Galactic Plane) and 100\degr\  also permit us to reject uniformity for some of the bipolar area subsamples at that significance level. We can do so for the West, North, North East and South West subsamples for both those mean orientations and for the South and East subsamples for the 90\degr\ mean orientation.

\citet{1998MNRAS.297..617C} found little evidence for alignments of PNe within the Galaxy. However, their sample of 209 PNe were spread around the Galaxy in terms of Galactic longitude and contained only 12 Bulge PNe. \citet{2008PASP..120..380W} found an excess of elongated PNe toward the Galactic Centre that are oriented at 100\degr. As our sample and theirs have only 19 objects in common, of which only five are bipolar in our classification, we have an independent test of a potential alignment. We have restricted our study to PNe in the Bulge and, having taken the factors listed above into account, we consider that there is evidence that the GPA of the bipolar PNe in the Galactic Bulge do not have an uniform distribution. That is, that the two-dimensional orientation of the bipolar PNe in the Galactic Bulge is not random.  Moreover, we find evidence of an alignment with a mean approximately along the Galactic Plane. 

PN orientations are expected to be independent of each other because a PN's shape is set by processes internal to the originating star or stellar system. A mechanism is required.

\subsection{Orientation Changes After PN Ejection}

The velocity graphs in Fig.~\ref{velang} together with the small
Pearson product-moment correlation coefficients for the radial and expansion velocity
regressions suggest that there is no relationship between PN orientation and either PN radial velocity or PN
expansion velocity. Similarly, Fig.~\ref{dimang} together with low Pearson product-moment correlation coefficients suggest that there is no relationship between PN orientation and size. In particular, the plot for the bipolar PNe in Fig.~\ref{dimang} together with a Pearson product-moment correlation coefficient of -0.009 from a simple linear
regression of the GPA against the PNe angular size from lobe tip to lobe tip,
demonstrates that the orientation of bipolar PNe is not related to lobe length. The variation in
distance to the PNe is unlikely to be such as to change that conclusion. 

The lobe length of a bipolar PN is related to the age of the PN. The lack of relation between orientation and lobe length suggests that the orientation is set during the youth of the PN or before it
has formed. There is no evidence that the orientation  changes during the PN
evolution. The expansion velocity  tends to be higher for higher-mass central
stars (although this may not be important for the old stellar population of
the Galactic Bulge), or for emission-line central stars \citep{2006A&A...451..925G}.
These aspects are not shown to be related to the orientation. We also found no evidence that orientation is related to radial velocity.

It has been suggested that magnetic fields in the interstellar medium (ISM) elongate stellar wind bubbles and that this is important in the production of bilateral supernova remnants \citep{1998ApJ...493..781G}. In the case of PNe however, that magnetic field can have little effect during the formation of the nebula due to the small size scales involved (the wind acceleration region is located within 1--10 AU from the star) and the high densities in the wind. Moreover, this mechanism would not explain why the alignment is found only for bipolar PNe.

\subsection{Angular momentum}

The currently favoured view for shaping non-spherical PNe is that of binary
interactions \citep{2009PASP..121..316D} . In this model, either an AGB star
provides spherical ejecta and its interactions with a binary companion shapes
those ejecta, or a close binary forms a common envelope leading to aspherical
ejecta \citep{2007BaltA..16...79Z}. Stellar magnetic fields also provide
possible shaping mechanisms, however \citet{2006PASP..118..260S} argues that 
sufficiently strong fields still require a companion to continuously spin up
the envelope. 

The distinction between bipolar and other PNe is likely related to the amount of angular momentum available for shaping. The angular momentum available for shaping is set by a balance between the orbital angular momentum and the efficiency of the transfer of momentum. Bipolar nebulae show dense tori which shape the polar flows. They may arise from intermediate $\left(\sim\!1~\mathrm{yr}\right)$ and short-period $\left(\leq 0.1~\mathrm{yr}\right)$ binaries. Shorter period binaries form common-envelope systems and nebulae ejected by them often show thin rings rather than dense tori \citep{2007BaltA..16...79Z}, possibly related to the lower angular momentum available in close systems. For common envelope systems the momentum transfer  efficiency will be 100\%; the percentage is lower for wider systems. The widest binaries $\left(\gtrsim 10~\mathrm{yr} \right)$ show spiral features but do not form bipolar nebulae \citep{2012Natur.490..232M}.

Although the orbit of a star around the Galaxy may change due to interactions with other bodies, the angular momentum vector of the stellar system tends to remain stable. The Kozai--Lidov mechanism (a three-body interaction) \citep{2013ApJ...766...64S} could prevent the retention of any alignment of binary systems that occurs during the early stages of star formation i.e. it could randomize the alignments. However, any method of alignment of the PNe in the later stages of the lives of the systems that form them would then be required to operate over kiloparsec distances and appears implausible. Consequently, given that we have found an alignment, that alignment must have originated during the formation of the stars. 

\subsection{A proposed mechanism}

The dense molecular clouds within about 4\degr\ of the Galactic centre tend to have their magnetic fields
aligned along the Galactic plane \citep{1998IAUS..184..331M} and these
fields are quite uniform and strong.  If that was also the case for the 
Bulge at the time of its star formation, a propensity for the stars which form the bipolar PNe
to align with the magnetic field of the cloud from which they themselves are
formed could provide an explanation for the non-random distribution of the PNe
orientations.

A strong magnetic field embedded in a star-forming cloud slows the contraction of star-forming cores along the direction perpendicular to the magnetic field \citep{2009ApJ...700..251T}.  The material around the collapsing young stellar objects may be expected to contract more quickly along the magnetic field. If a binary forms in the thus-flattened cloud, the separation will on average be larger perpendicular to the field. As wider binaries carry more angular momentum, this gives a preferential angular momentum vector of the binary population along the magnetic field.  So the stronger the field, the greater is the propensity towards forming wide binaries with orbits perpendicular to that field. Whether this occurs in reality is controversial. Hydrodynamical models indicate that for larger fields $\left(\sim80~\mu\mathrm{G}\right)$, outflows indeed become oriented along the magnetic field lines \citep{2006ApJ...637L.105M}. Also, the 3D MHD simulations of \citet{2008ApJ...687..354N} appear to show that star formation in relatively diffuse clouds containing strong magnetic fields is controlled by those fields, not by turbulence.

The evidence for orientation of stellar rotation axes along interstellar magnetic field lines in nearby-clusters is not conclusive. While the T Tauri stars FU Ori and DR Tau have their stellar axes aligned with the larger magnetic field direction \citep{2005MNRAS.359.1049V}, other studies suggest that the orientation of young stellar objects in a cluster is random \citep{2004A&A...425..973M,2010MNRAS.402.1380J}. However, no studies have been made on possible alignment of binary orbits within young clusters. Binary orbits carry far more angular momentum than does stellar rotation, and the result from the Galactic Bulge PNe could indicate that the alignment survives only in binary systems. We can also not exclude the possibility that magnetic fields within the Bulge during its formation were considerable stronger than those in the Solar
neighbourhood.  \citet{2006ApJ...637L.105M} show that the strength of the field is an important parameter, with alignment occurring above 80~$\mu\mathrm{G}$.

The alignment of bipolar PNe suggests that there was an alignment of stellar systems in the Bulge at the time of the formation of the stars, most clearly seen for those objects with the largest angular momentum. That alignment lies  approximately along the Galactic plane. This implies that the binary systems in the Galactic Bulge have angular momentum vectors that are preferentially aligned along the Galactic plane.  There is strong evidence for cylindrical rotation in the Bulge, with angular momentum perpendicular to the Galactic plane \citep{2009ApJ...702L.153H}. We thus have the counter-intuitive situation that binary systems out of which bipolar PNe are formed have their angular momentum vector perpendicular to that of the Bulge in which they formed.

In conclusion, we have evidence that the orientation of bipolar PNe in the Bulge is not random. The argument that we use is that the angular momentum vectors are aligned. This effect is likely to date from the origin of the central stars of the PNe. Interstellar magnetic fields at the time of the formation of the Bulge provide a possible mechanism.

\section{Summary}
We have classified 130 PNe that we consider to be in the Galactic Bulge into three morphological categories. For each of those PNe, we determined a position angle for the polar direction and converted that to an angle from Galactic north to Galactic east. We then undertook a statistical analysis of those angles and found that for one of the morphological classes, the bipolar class, there is evidence for a non-random distribution of the angles at significance levels equivalent to that between $2.3\sigma$ and $3.7\sigma$ (depending on the test used) for a Gaussian distribution. The mean  orientation is aligned approximately to that of the galactic plane. On the basis that the orientation of the bipolar nebulae is independent of their size we propose a mechanism for the production of that non-random distribution of orientations.

The debate on the main shaping mechanism in PNe has  concentrated on binarity
and magnetic fields. Currently, binary systems are seen as the dominant effect
(this may be of course be influenced by the fact that binaries are easier to
detect than magnetic fields),  with magnetic fields having a secondary
role.  The hypothesis put forward in this paper is that the alignment of bipolar planetary nebulae can be understood if  star formation in the Bulge occurred in the presence of strong magnetic fields oriented along the Galactic plane and if the parameters of the binary systems themselves are in part the product of magnetic fields present at their origin. This reverses again cause and effect. The long debate has shown that both effects are difficult to separate, and are to a significant degree intertwined.The estimated ages of stars in the Bulge vary with estimates from 8 to $\sim\!13$~Gyr \citep{2001ASPC..245..216R,2008IAUS..245..323M,2010A&A...512A..41B}. The claim in this paper is that this interaction between field and binarity dates back to the processes acting during the origin of the stars, a form of archaeomagnetism of the early Universe. 

We are of the view that a similar analysis should be performed on high resolution observations of a much larger sample of bipolar PNe in the Galactic Bulge.

\section*{Acknowledgements}
\label{ack}
We thank Clive Dickinson for discussions on the available polarization data for the Galactic Bulge and Iain McDonald for discussions on the projections.

This research has made use of the SIMBAD database, and the VizieR catalogue access tool, CDS, Strasbourg, France.

\bibliographystyle{mn2e}


\begin{thebibliography}{48}
\expandafter\ifx\csname natexlab\endcsname\relax\def\natexlab#1{#1}\fi

\bibitem[{{Acker} {et~al}\mbox{.}(1992){Acker}, {Marcout}, {Ochsenbein},
  {Stenholm}, \& {Tylenda}}]{1992secg.book.....A}
{Acker} A., {Marcout} J., {Ochsenbein} F., {Stenholm} B., {Tylenda} R., 1992,
  {Strasbourg - ESO catalogue of galactic planetary nebulae. Part 1; Part 2}.
  Garching: European Southern Observatory

\bibitem[{{Acker}, {Peyaud} \& {Parker}(2006){Acker}, {Peyaud}, \&
  {Parker}}]{2006IAUS..234..355A}
{Acker} A., {Peyaud} A.~E.~J., {Parker} Q., 2006, in {Barlow} M.~J., {M{\'e}ndez}
  R.~H., eds., IAU Symposium, Vol. 234,
  Planetary Nebulae in our Galaxy and Beyond,  p. 355

\bibitem[{{Balick} \& {Frank}(2002)}]{2002ARA&A..40..439B}
{Balick} B., {Frank} A., 2002, \araa, 40, 439

\bibitem[{{Beaulieu}, {Dopita} \& {Freeman}(1999){Beaulieu}, {Dopita}, \&
  {Freeman}}]{0004-637X-515-2-610}
{Beaulieu} S.~F., {Dopita} M.~A., {Freeman} K.~C., 1999, \apj, 515, 610

\bibitem[{{Bensby} {et~al}\mbox{.}(2010){Bensby}, {Feltzing}, {Johnson},
  {Gould}, {Ad{\'e}n}, {Asplund}, {Mel{\'e}ndez}, {Gal-Yam}, {Lucatello},
  {Sana}, {Sumi}, {Miyake}, {Suzuki}, {Han}, {Bond}, \&
  {Udalski}}]{2010A&A...512A..41B}
{Bensby} T. {et~al.}, 2010, \aap, 512, A41

\bibitem[{{Berens}(2009)}]{RePEc:jss:jstsof:31:i10}
{Berens} P., 2009, Journal of Statistical Software, 31, 1

\bibitem[{{Blackman}(2009)}]{2009IAUS..259...35B}
{Blackman} E.~G., 2009, in {Strassmeier} K.~G., {Kosovichev} A.~G.,
  {Beckman} J., eds., IAU Symposium, Vol. 259, Cosmic Magnetic Fields:
  from Planets, to Stars and Galaxies,  p. 35

\bibitem[{Bogdan, Bogdan \& Futschik(2002)Bogdan, Bogdan, \&
  Futschik}]{springerlink:10.1023/A:1016109603897}
Bogdan M., Bogdan K., Futschik A., 2002, Annals of the Institute of Statistical
  Mathematics, 54, 29

\bibitem[{{Corradi}, {Aznar} \& {Mampaso}(1998){Corradi}, {Aznar}, \&
  {Mampaso}}]{1998MNRAS.297..617C}
{Corradi} R.~L.~M., {Aznar} R., {Mampaso} A., 1998, \mnras, 297, 617

\bibitem[{{Corradi} \& {Schwarz}(1995)}]{1995A&A...293..871C}
{Corradi} R.~L.~M., {Schwarz} H.~E., 1995, \aap, 293, 871

\bibitem[{{de Marco}(2009)}]{2009PASP..121..316D}
{de Marco} O., 2009, \pasp, 121, 316

\bibitem[{{Durand}, {Acker} \& {Zijlstra}(1998){Durand}, {Acker}, \&
  {Zijlstra}}]{1998A&AS..132...13D}
{Durand} S., {Acker} A., {Zijlstra} A., 1998, \aaps, 132, 13

\bibitem[{{Efron}(1979)}]{1979}
{Efron} B., 1979, The Annals of Statistics, 7, 1

\bibitem[{{Fisher}(1995)}]{FisherCircStat}
{Fisher} N.~I., 1995, Statistical Analysis of Circular Data. Cambridge
  University Press, Cambridge

\bibitem[{{Gaensler}(1998)}]{1998ApJ...493..781G}
{Gaensler} B.~M., 1998, \apj, 493, 781

\bibitem[{{Gesicki} \& {Zijlstra}(2007)}]{2007A&A...467L..29G}
{Gesicki} K., {Zijlstra} A.~A., 2007, \aap, 467, L29

\bibitem[{{Gesicki} {et~al}\mbox{.}(2006){Gesicki}, {Zijlstra}, {Acker},
  {G{\'o}rny}, {Gozdziewski}, \& {Walsh}}]{2006A&A...451..925G}
{Gesicki} K., {Zijlstra} A.~A., {Acker} A., {G{\'o}rny} S.~K., {Gozdziewski}
  K., {Walsh} J.~R., 2006, \aap, 451, 925

\bibitem[{{G{\'o}mez} {et~al}\mbox{.}(2009){G{\'o}mez}, {Tafoya}, {Anglada},
  {Miranda}, {Torrelles}, {Patel}, \& {Hern{\'a}ndez}}]{2009ApJ...695..930G}
{G{\'o}mez} Y., {Tafoya} D., {Anglada} G., {Miranda} L.~F., {Torrelles} J.~M.,
  {Patel} N.~A., {Hern{\'a}ndez} R.~F., 2009, \apj, 695, 930

\bibitem[{{Gonzalez}, {Brilliant} \& {Pompei}(2006){Gonzalez}, {Brilliant}, \&
  {Pompei}}]{EMMI}
{Gonzalez} J.-F., {Brilliant} S., {Pompei} E., 2006, EMMI: The ESO Multi-Mode
  Instrument, User's Manual, 5th edn. European Southern Observatory, doc. No.
  LSO-MAN-ESO-40100-0001/5.3

\bibitem[{{Herwig}(2005)}]{2005ARA&A..43..435H}
{Herwig} F., 2005, \araa, 43, 435

\bibitem[{Hoenen \& Gnaspini(1999)}]{Arach}
Hoenen S., Gnaspini P., 1999, The Journal of Arachnology, 27, 159

\bibitem[{{Howard} {et~al}\mbox{.}(2009){Howard}, {Rich}, {Clarkson},
  {Mallery}, {Kormendy}, {DePropris}, {Robin}, {Fux}, {Reitzel}, {Zhao},
  {Kuijken}, \& {Koch}}]{2009ApJ...702L.153H}
{Howard} C.~D. {et~al.}, 2009, \apjl, 702, L153

\bibitem[{{Jackson} \& {Jeffries}(2010)}]{2010MNRAS.402.1380J}
{Jackson} R.~J., {Jeffries} R.~D., 2010, \mnras, 402, 1380

\bibitem[{{Jammalamadaka} \& {SenGupta}(2001)}]{TopCircStat}
{Jammalamadaka} S.~R., {SenGupta} A., 2001, Topics In Circular Statistics.
  World Scientific, Singapore

\bibitem[{{Maercker} {et~al}\mbox{.}(2012){Maercker}, {Mohamed}, {Vlemmings},
  {Ramstedt}, {Groenewegen}, {Humphreys}, {Kerschbaum}, {Lindqvist},
  {Olofsson}, {Paladini}, {Wittkowski}, {de Gregorio-Monsalvo}, \&
  {Nyman}}]{2012Natur.490..232M}
{Maercker} M. {et~al.}, 2012, \nat, 490, 232

\bibitem[{{Malaroda}, {Levato} \& {Galliani}(2006){Malaroda}, {Levato}, \&
  {Galliani}}]{Malaroda2006}
{Malaroda} S., {Levato} H., {Galliani} S., 2006, Stellar radial velocities
  bibliographic catalog; from vizier service, cds, strasbourg, france.
  \url{http://cdsarc.u-strasbg.fr/viz-bin/VizieR?-meta.foot\&-source=III/249}

\bibitem[{{Matsumoto}, {Nakazato} \& {Tomisaka}(2006){Matsumoto}, {Nakazato},
  \& {Tomisaka}}]{2006ApJ...637L.105M}
{Matsumoto} T., {Nakazato} T., {Tomisaka} K., 2006, \apjl, 637, L105

\bibitem[{{Melnick} \& {Harwit}(1975)}]{1975MNRAS.171..441M}
{Melnick} G., {Harwit} M., 1975, \mnras, 171, 441

\bibitem[{{M{\'e}nard} \& {Duch{\^e}ne}(2004)}]{2004A&A...425..973M}
{M{\'e}nard} F., {Duch{\^e}ne} G., 2004, \aap, 425, 973

\bibitem[{{Minniti} \& {Zoccali}(2008)}]{2008IAUS..245..323M}
{Minniti} D., {Zoccali} M., 2008, in {Bureau} M., {Athanassoula} E., {Barbuy} B.,
  eds., IAU Symposium, Vol. 245, Formation and
  Evolution of Galaxy Bulges,  p. 323

\bibitem[{{Morris}(1998)}]{1998IAUS..184..331M}
{Morris} M., 1998, in {Y.~Sofue}, ed., IAU Symposium, Vol. 184, The Central Regions of the
  Galaxy and Galaxies,  p. 331

\bibitem[{{Nakamura} \& {Li}(2008)}]{2008ApJ...687..354N}
{Nakamura} F., {Li} Z.-Y., 2008, \apj, 687, 354

\bibitem[{{Parker} {et~al}\mbox{.}(2006){Parker}, {Acker}, {Frew}, {Hartley},
  {Peyaud}, {Ochsenbein}, {Phillipps}, {Russeil}, {Beaulieu}, {Cohen},
  {K{\"o}ppen}, {Miszalski}, {Morgan}, {Morris}, {Pierce}, \&
  {Vaughan}}]{2006MNRAS.373...79P}
{Parker} Q.~A. {et~al.}, 2006, \mnras, 373, 79

\bibitem[{{Protheroe}(1985)}]{1985ICRC....3..485P}
{Protheroe} R.~J., 1985, in {F.~C.~Jones}, ed., International Cosmic Ray Conference, Vol.~3,
  International Cosmic Ray Conference,  p. 485

\bibitem[{{Rich}(2001)}]{2001ASPC..245..216R}
{Rich} R.~M., 2001, in {von Hippel} T., {Simpson} C.,
  {Manset} N., eds., Astronomical Society of the Pacific Conference Series,
  Vol. 245, Astrophysical Ages and Times Scales,  p. 216

\bibitem[{{Sabin}, {Zijlstra} \& {Greaves}(2007){Sabin}, {Zijlstra}, \&
  {Greaves}}]{2007MNRAS.376..378S}
{Sabin} L., {Zijlstra} A.~A., {Greaves} J.~S., 2007, \mnras, 376, 378

\bibitem[{{Schwarz}, {Monteiro} \& {Peterson}(2008){Schwarz}, {Monteiro}, \&
  {Peterson}}]{2008ApJ...675..380S}
{Schwarz} H.~E., {Monteiro} H., {Peterson} R., 2008, \apj, 675, 380

\bibitem[{{Shappee} \& {Thompson}(2013)}]{2013ApJ...766...64S}
{Shappee} B.~J., {Thompson} T.~A., 2013, \apj, 766, 64

\bibitem[{{Si{\'o}dmiak} \& {Tylenda}(2001)}]{S&T2001}
{Si{\'o}dmiak} N., {Tylenda} R., 2001, \aap, 373, 1032

\bibitem[{{Soker}(2006)}]{2006PASP..118..260S}
{Soker} N., 2006, \pasp, 118, 260

\bibitem[{{Tang} {et~al}\mbox{.}(2009){Tang}, {Ho}, {Koch}, {Girart}, {Lai}, \&
  {Rao}}]{2009ApJ...700..251T}
{Tang} Y.-W., {Ho} P.~T.~P., {Koch} P.~M., {Girart} J.~M., {Lai} S.-P., {Rao}
  R., 2009, \apj, 700, 251

\bibitem[{{Testori}, {Reich} \& {Reich}(2008){Testori}, {Reich}, \&
  {Reich}}]{2008A&A...484..733T}
{Testori} J.~C., {Reich} P., {Reich} W., 2008, \aap, 484, 733

\bibitem[{{van Winckel}(2003)}]{2003ARA&A..41..391V}
{van Winckel} H., 2003, \araa, 41, 391

\bibitem[{{Vink} {et~al}\mbox{.}(2005){Vink}, {Drew}, {Harries}, {Oudmaijer},
  \& {Unruh}}]{2005MNRAS.359.1049V}
{Vink} J.~S., {Drew} J.~E., {Harries} T.~J., {Oudmaijer} R.~D., {Unruh} Y.,
  2005, \mnras, 359, 1049

\bibitem[{{Vlemmings} \& {van Langevelde}(2008)}]{2008A&A...488..619V}
{Vlemmings} W.~H.~T., {van Langevelde} H.~J., 2008, \aap, 488, 619

\bibitem[{{Weidmann} \& {D{\'{\i}}az}(2008)}]{2008PASP..120..380W}
{Weidmann} W.~A., {D{\'{\i}}az} R.~J., 2008, \pasp, 120, 380

\bibitem[{{Zijlstra}(1990)}]{1990A&A...234..387Z}
{Zijlstra} A.~A., 1990, \aap, 234, 387

\bibitem[{{Zijlstra}(2007)}]{2007BaltA..16...79Z}
{Zijlstra} A.~A., 2007, Baltic Astronomy, 16, 79

\end{thebibliography}

\end{document}


\setcounter{figure}{8}
\setcounter{table}{3}

\begin{figure*}
\begin{flushleft}
\section{Contour Plots}
\label{contplotlobe}
\end{flushleft}
    \begin{center}

$
    \begin{array}{cc}
    \includegraphics[width=8.1cm]{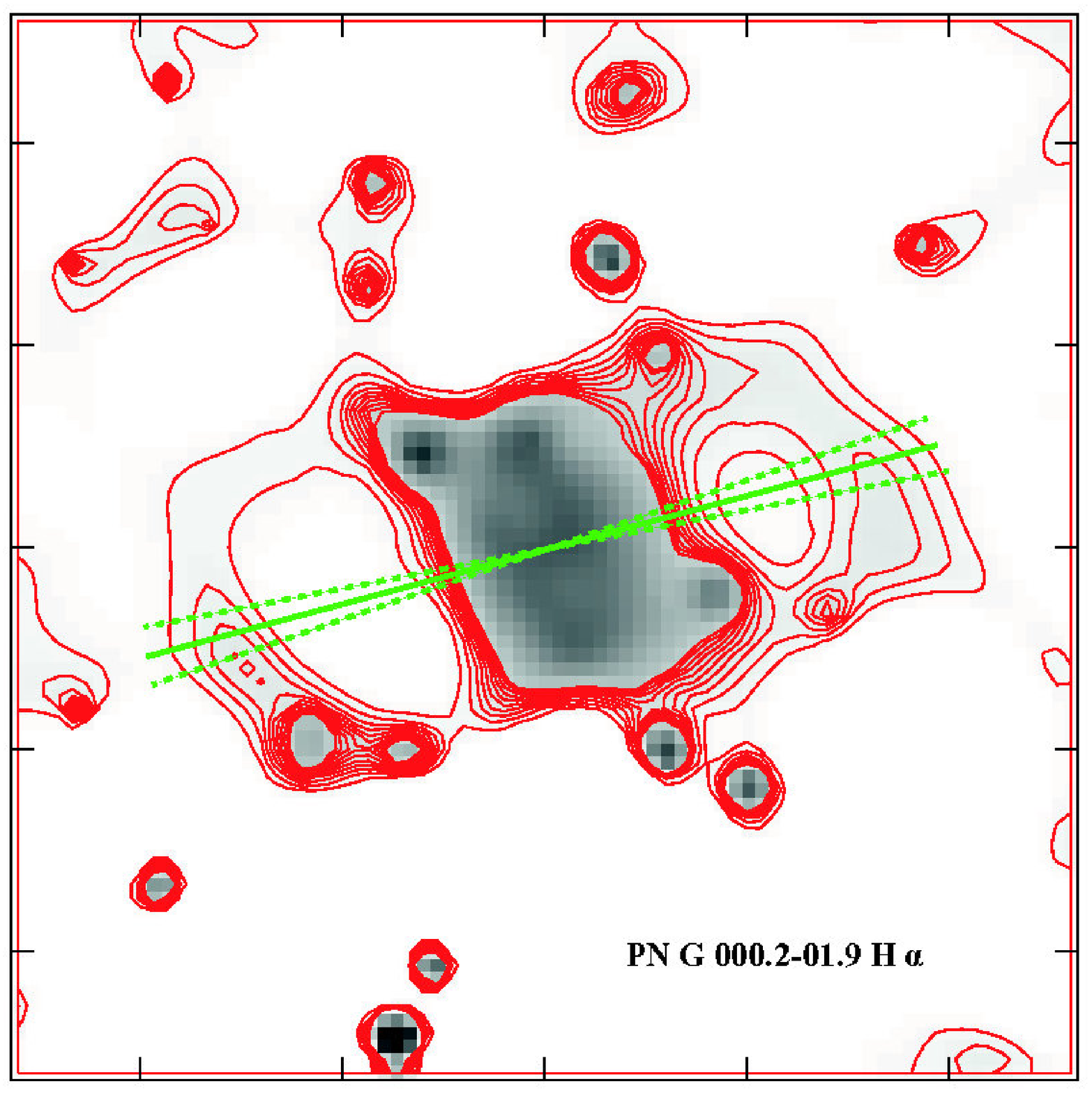} &
    \includegraphics[width=8.1cm]{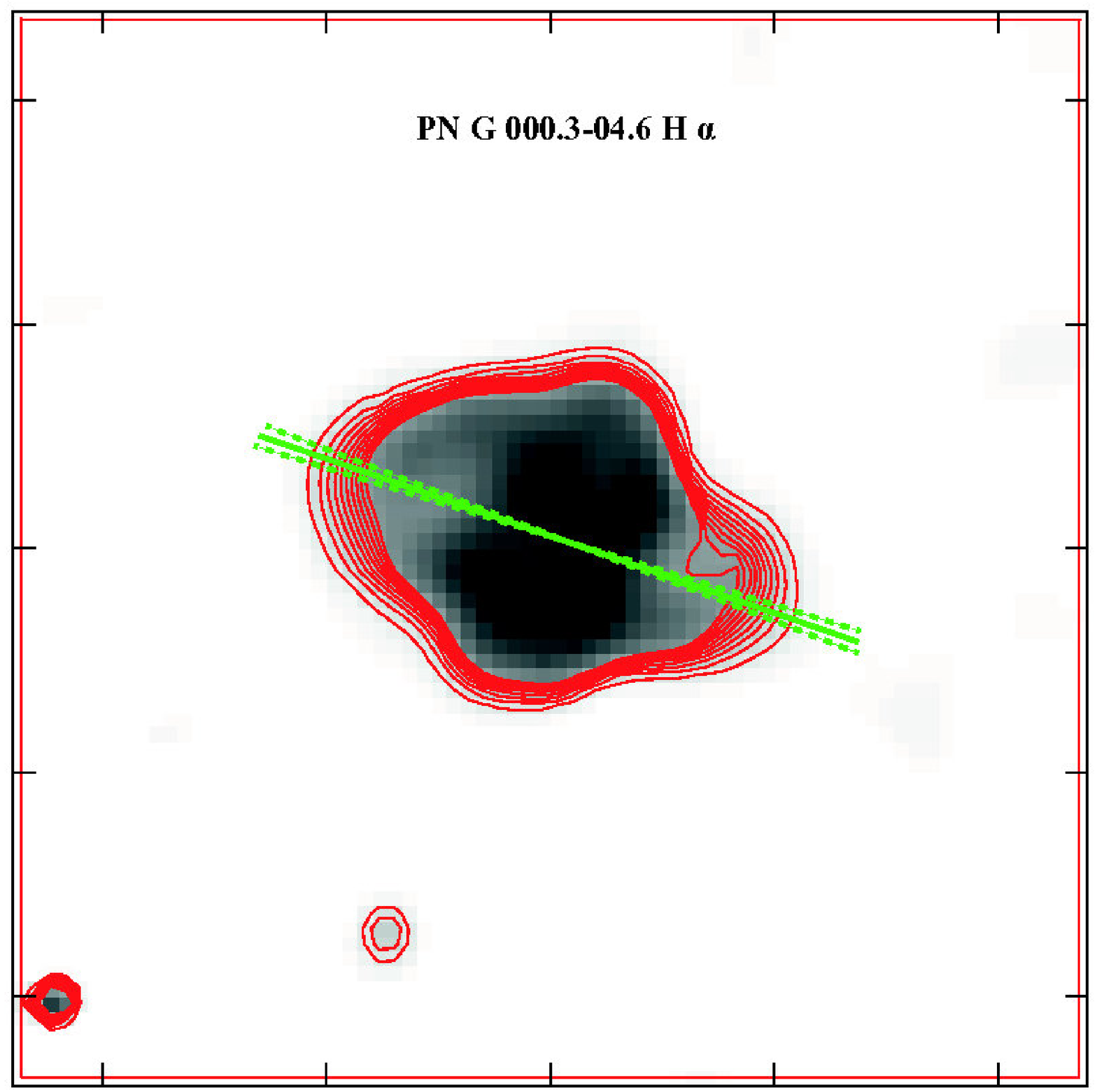}
\\
    \includegraphics[width=8.1cm]{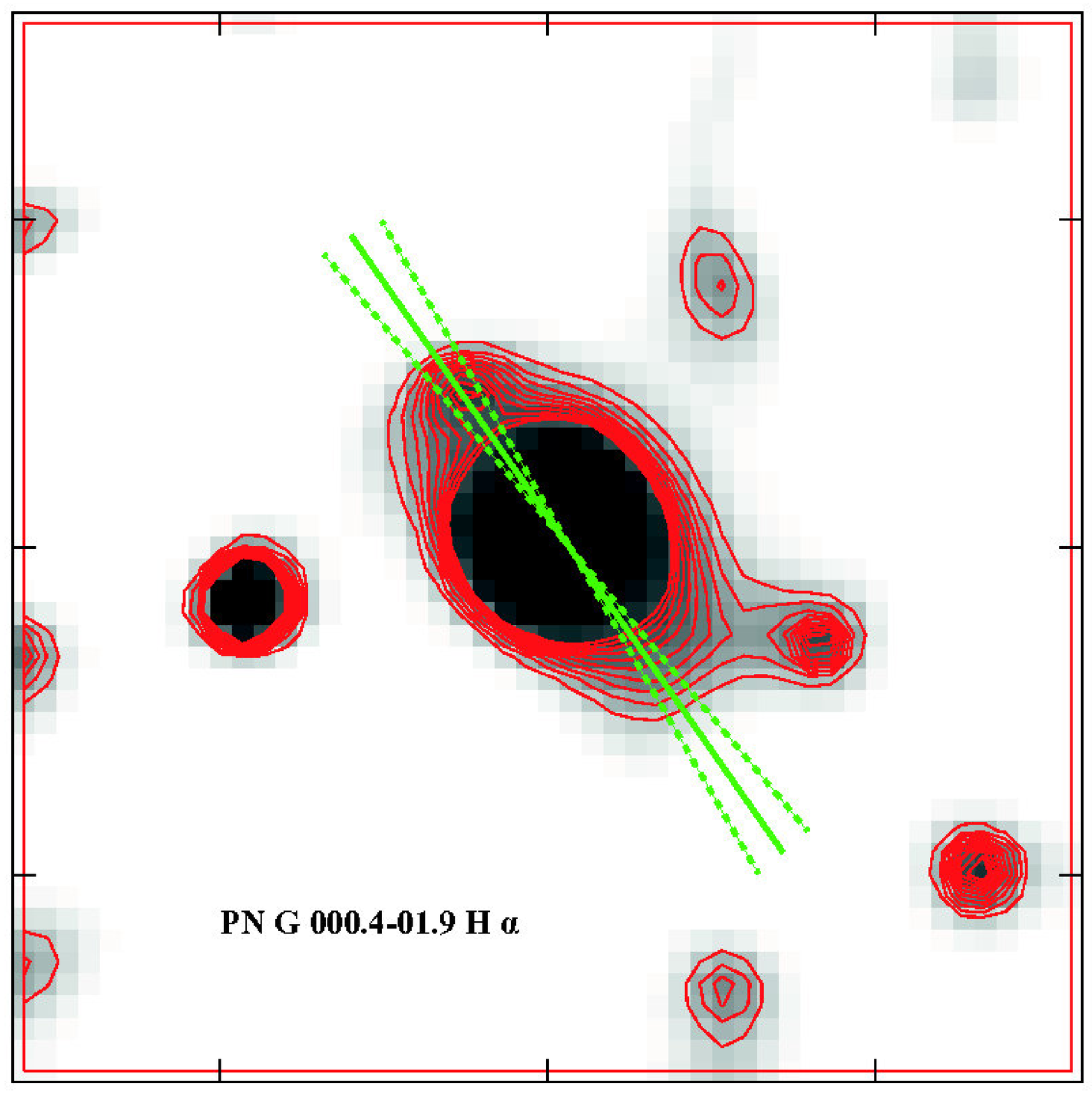} &
    \includegraphics[width=8.1cm]{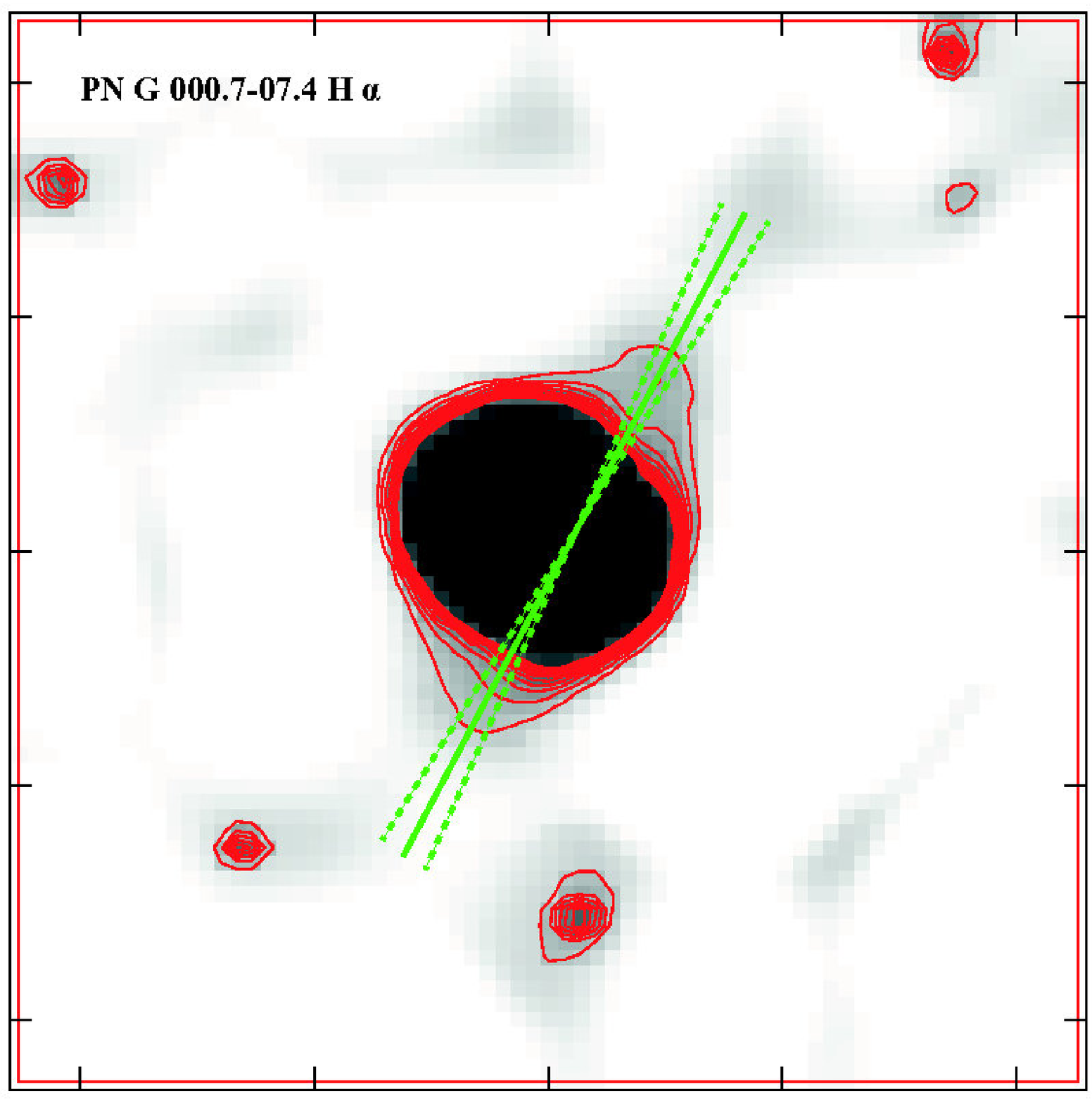}

    \end{array}
$

    \end{center}
    \caption{Greyscale images of the bipolar PNe, each overlaid with isophotes of its lobes together with lines indicating the measured PA and its uncertainty.  The images are  in negative form and those observed using the NTT were deconvolved. The maximum isophote is at 10 per cent of the peak PN intensity in the image and the isophotes are at 1 per cent intervals of that intensity. Each NTT plot has North at the top and East at the left whereas the \textit{HST} plots have their orientations marked on them. Tick marks on the plots are at 5 arcsec intervals.  Scale bars are shown on those plots without tick marks.}
\label{plots1}
\end{figure*}

\begin{figure*}
    \begin{center}

$
    \begin{array}{cc}
    \includegraphics[width=8.1cm]{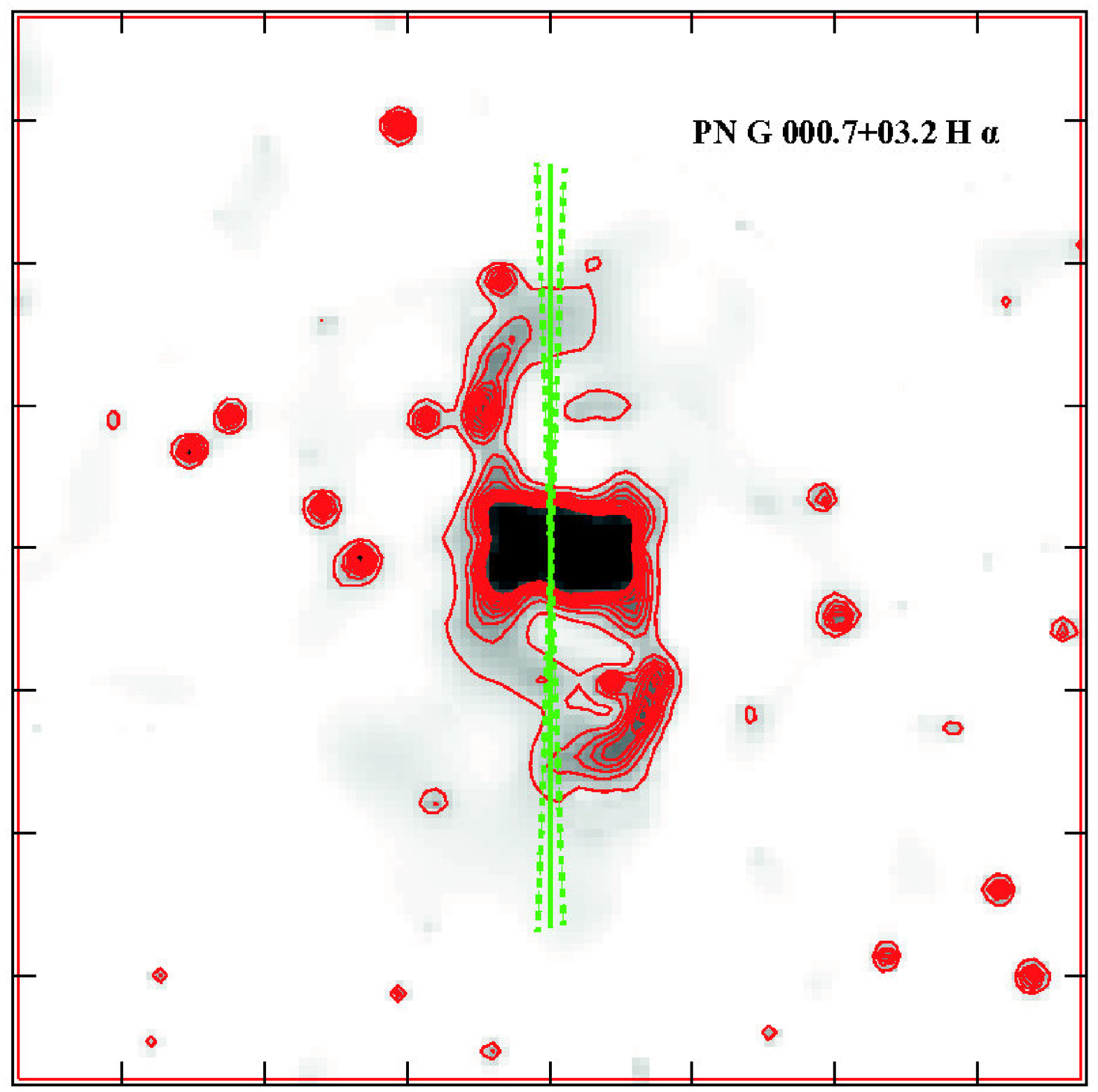} &
    \includegraphics[width=8.1cm]{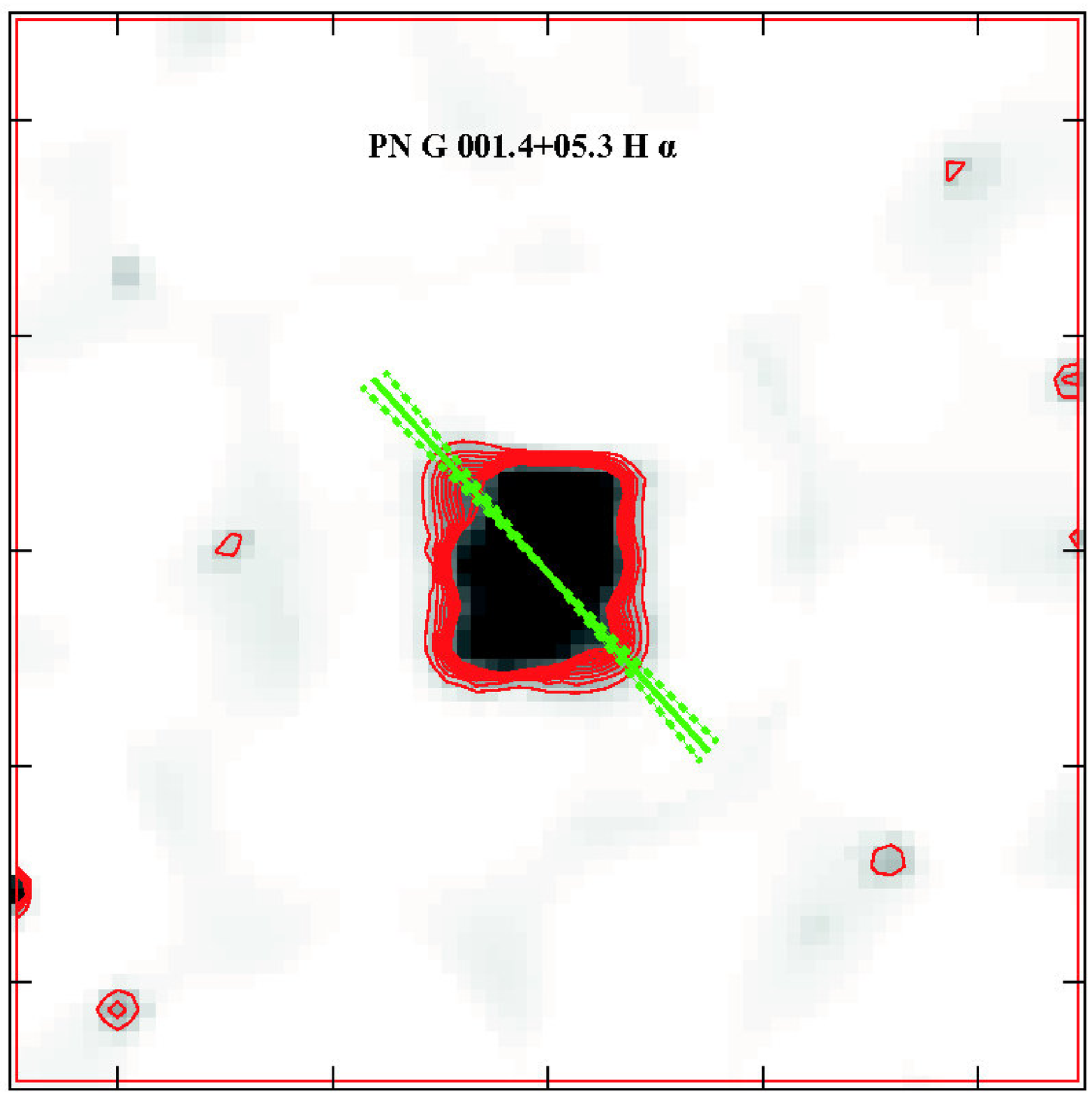}
\\
    \includegraphics[width=8.1cm]{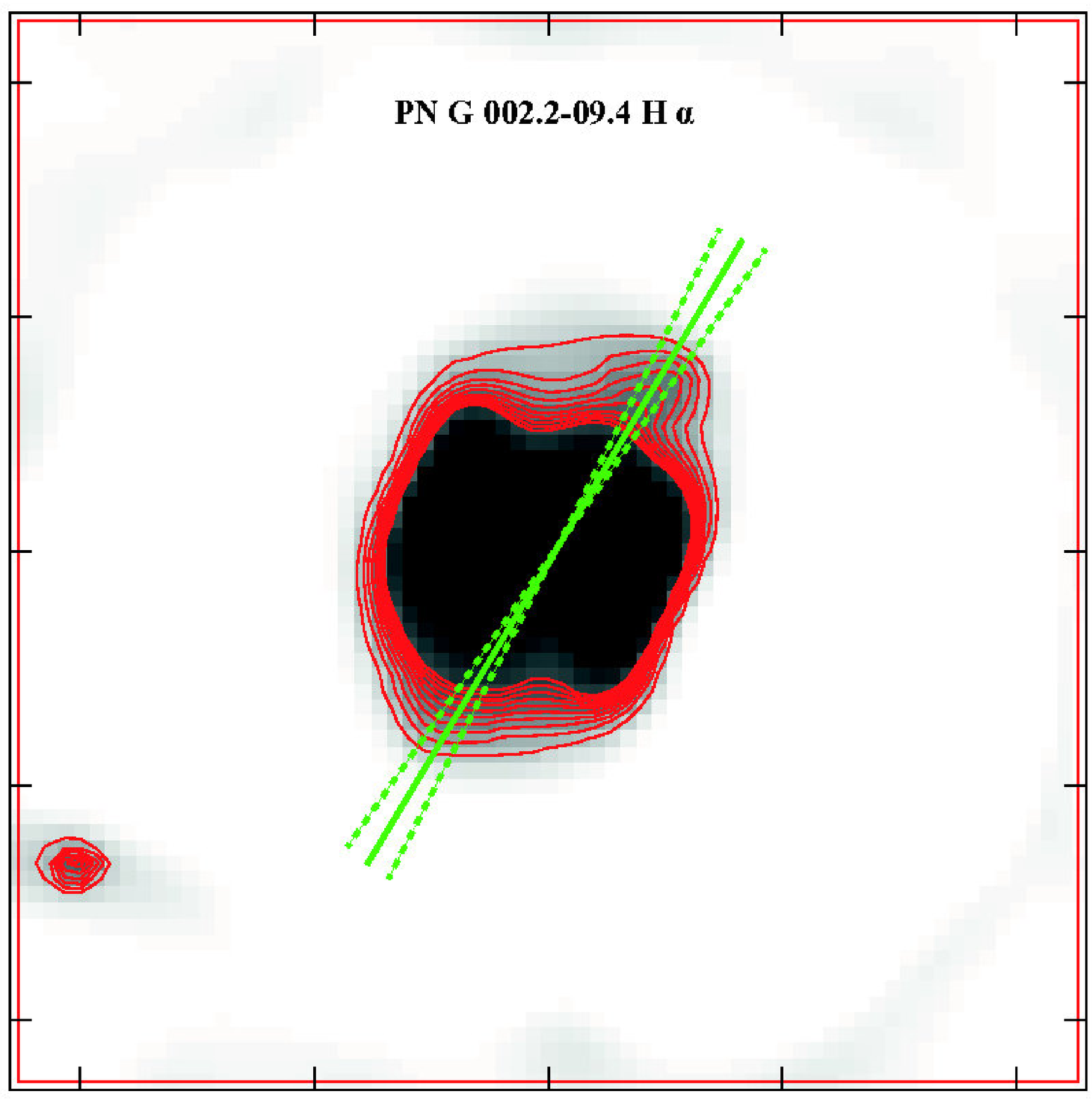} &
    \includegraphics[width=8.1cm]{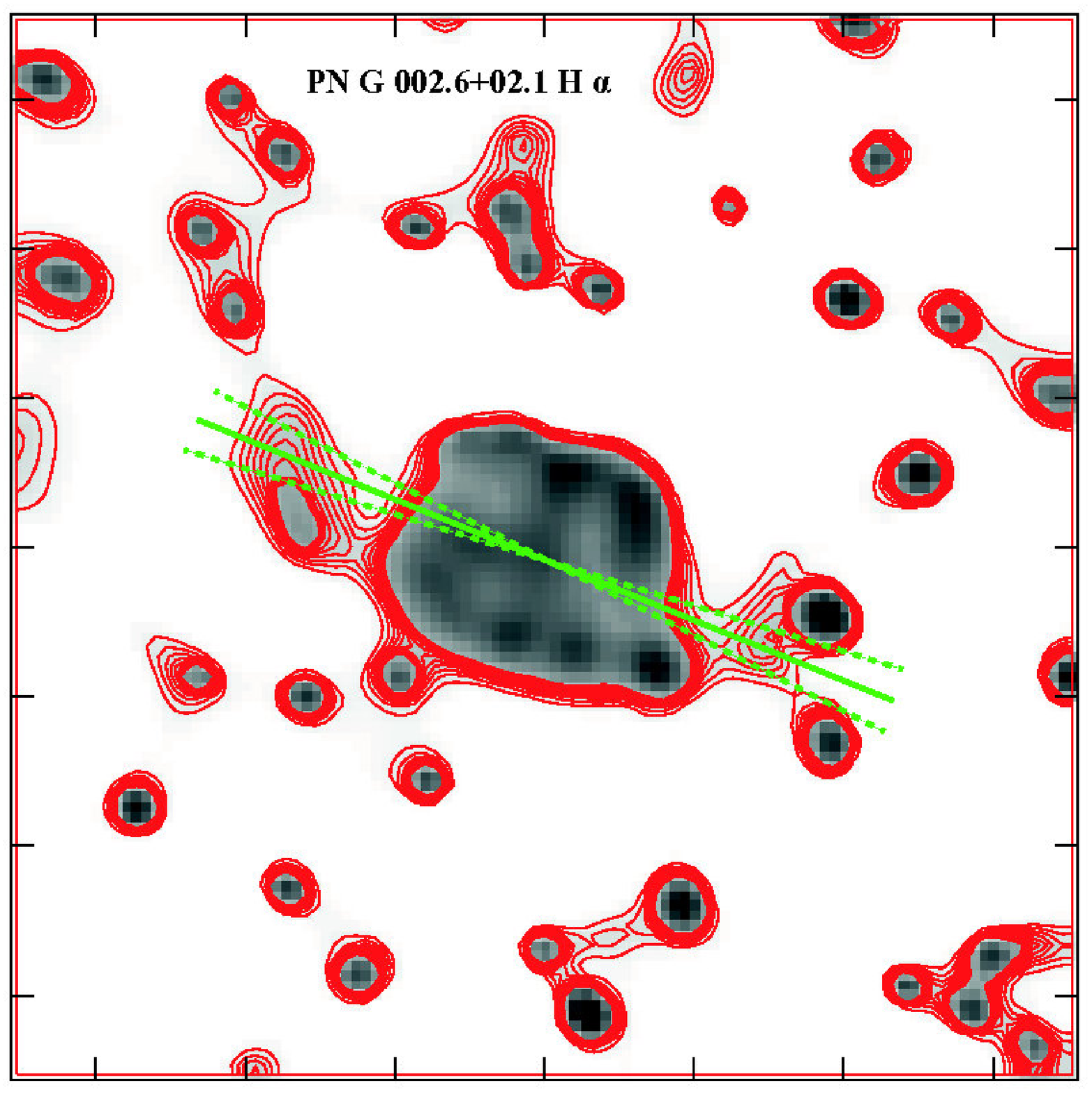}
\\
    \includegraphics[width=8.1cm]{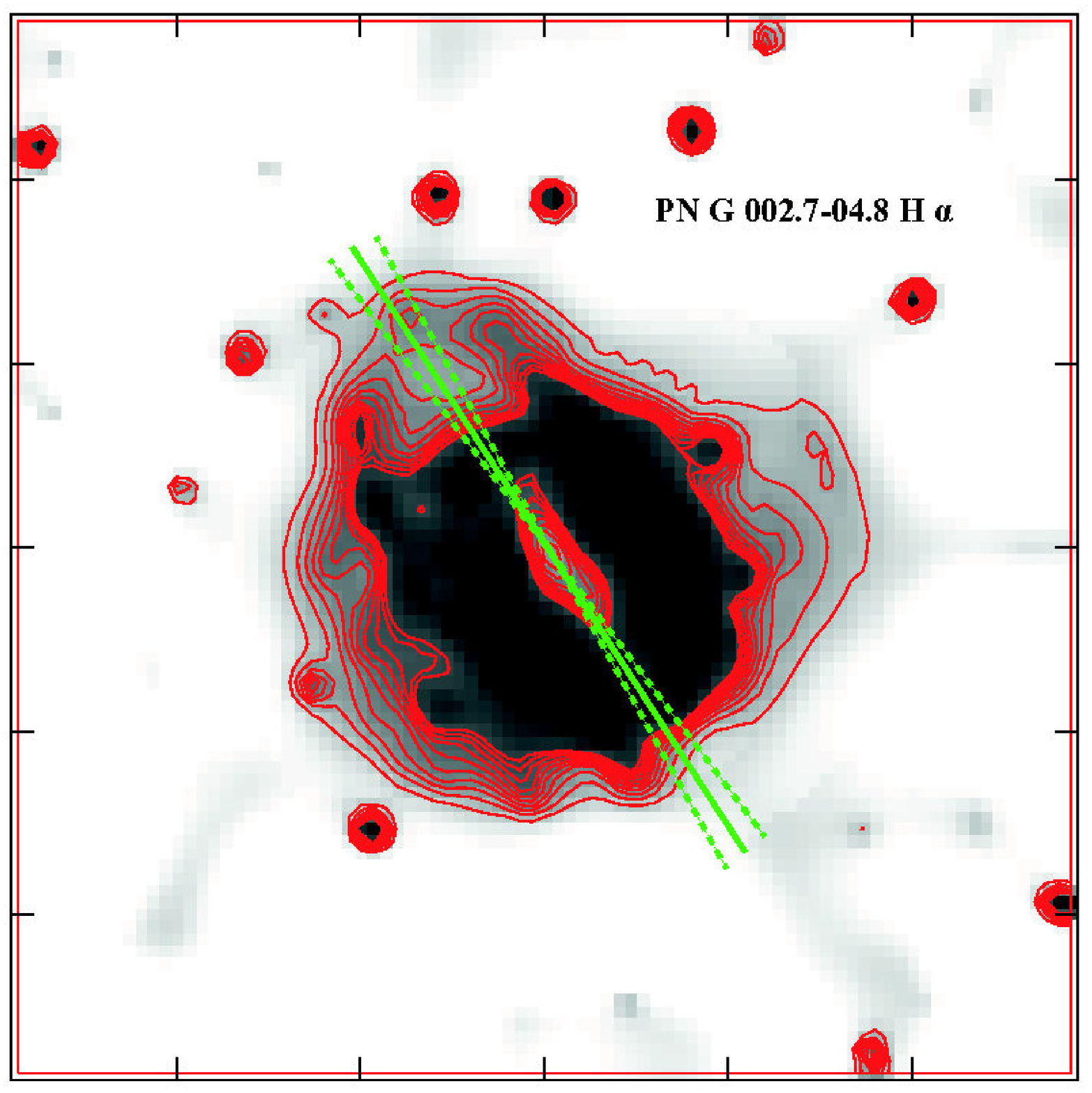} &
    \includegraphics[width=8.1cm]{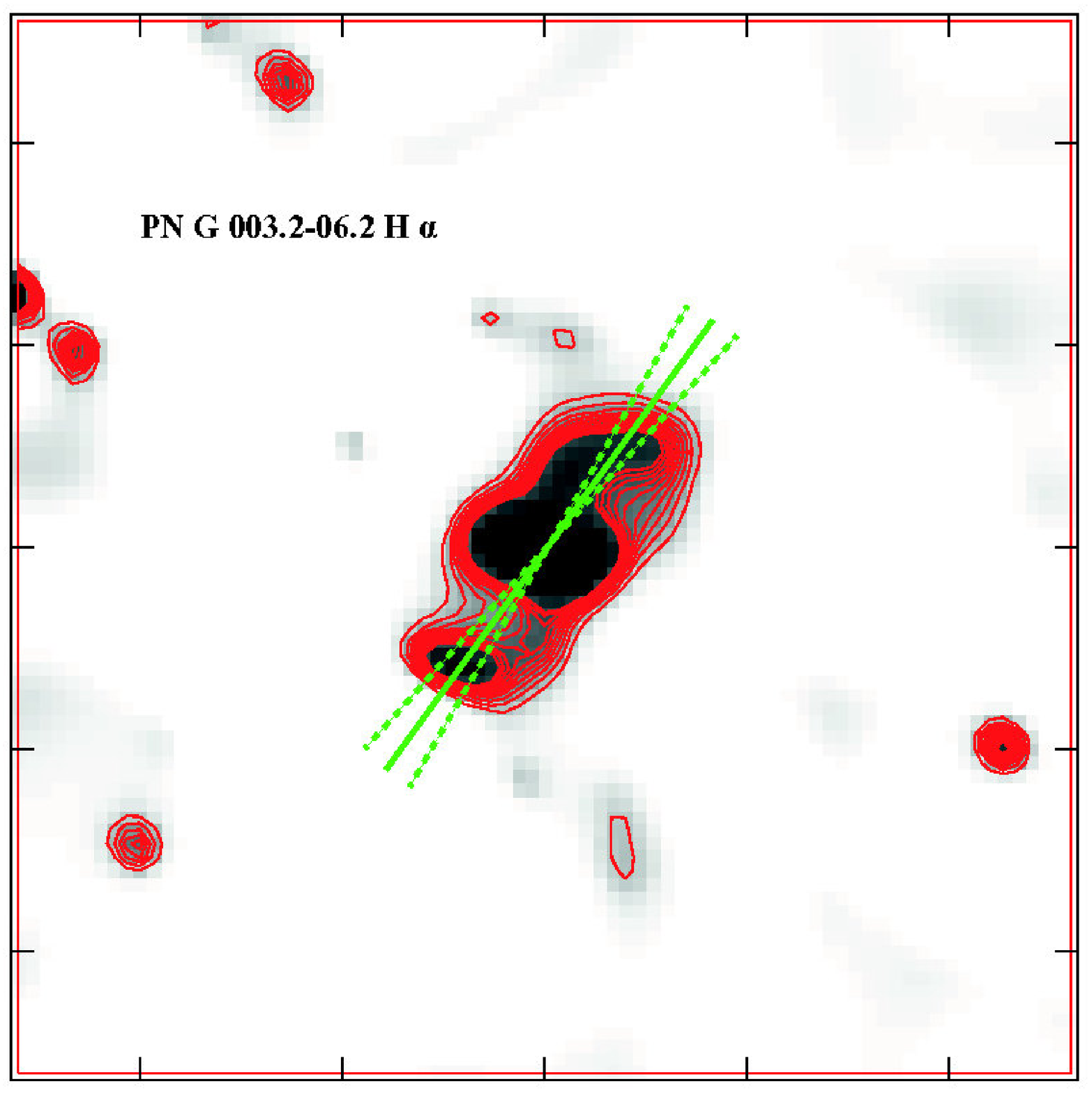}

    \end{array}
$

    \end{center}
    \contcaption{}
\label{plots1b}
\end{figure*}

\begin{figure*}
    \begin{center}

$
    \begin{array}{cc}
    \includegraphics[width=8.1cm]{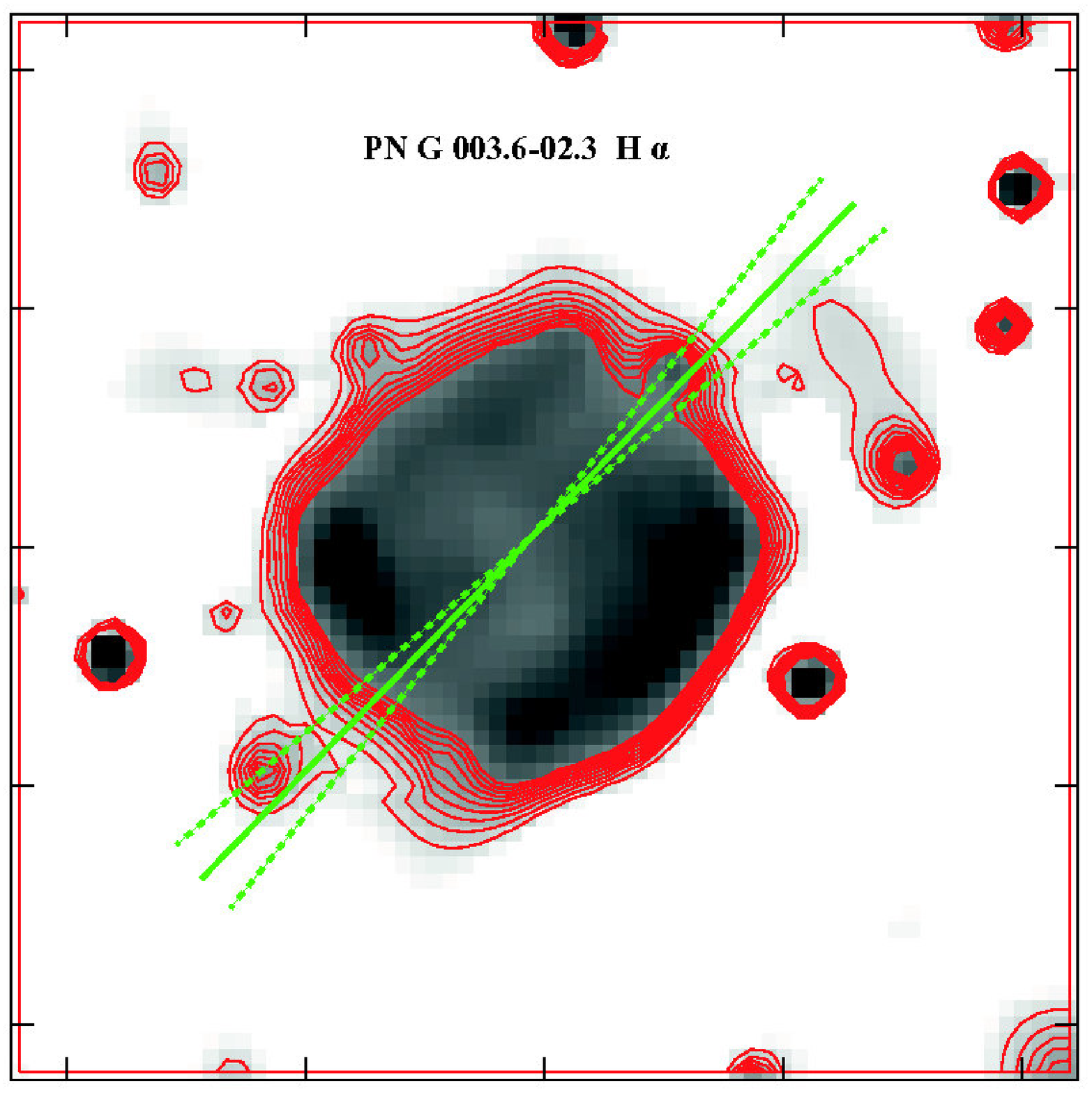} &
    \includegraphics[width=8.1cm]{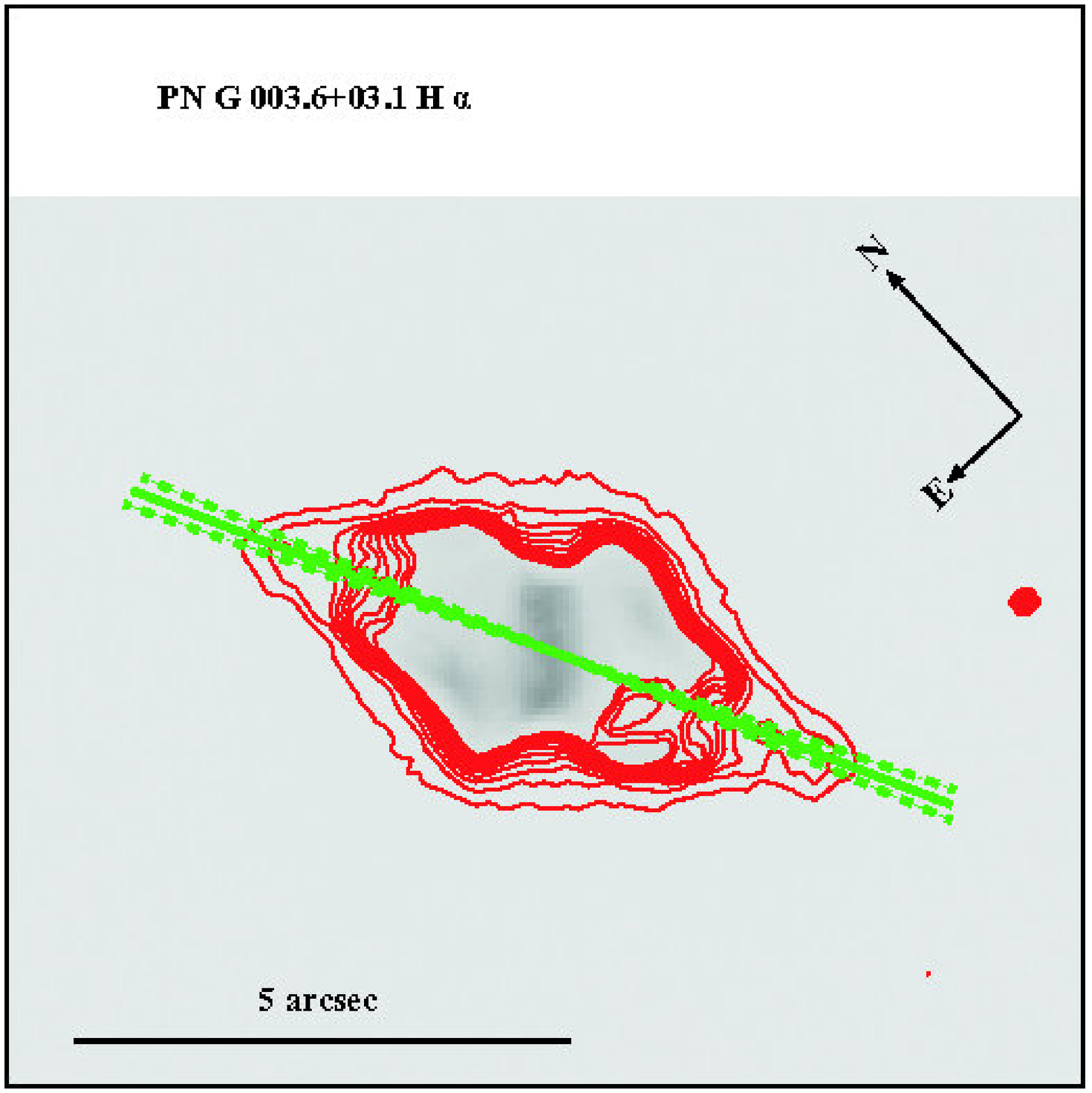}
\\
    \includegraphics[width=8.1cm]{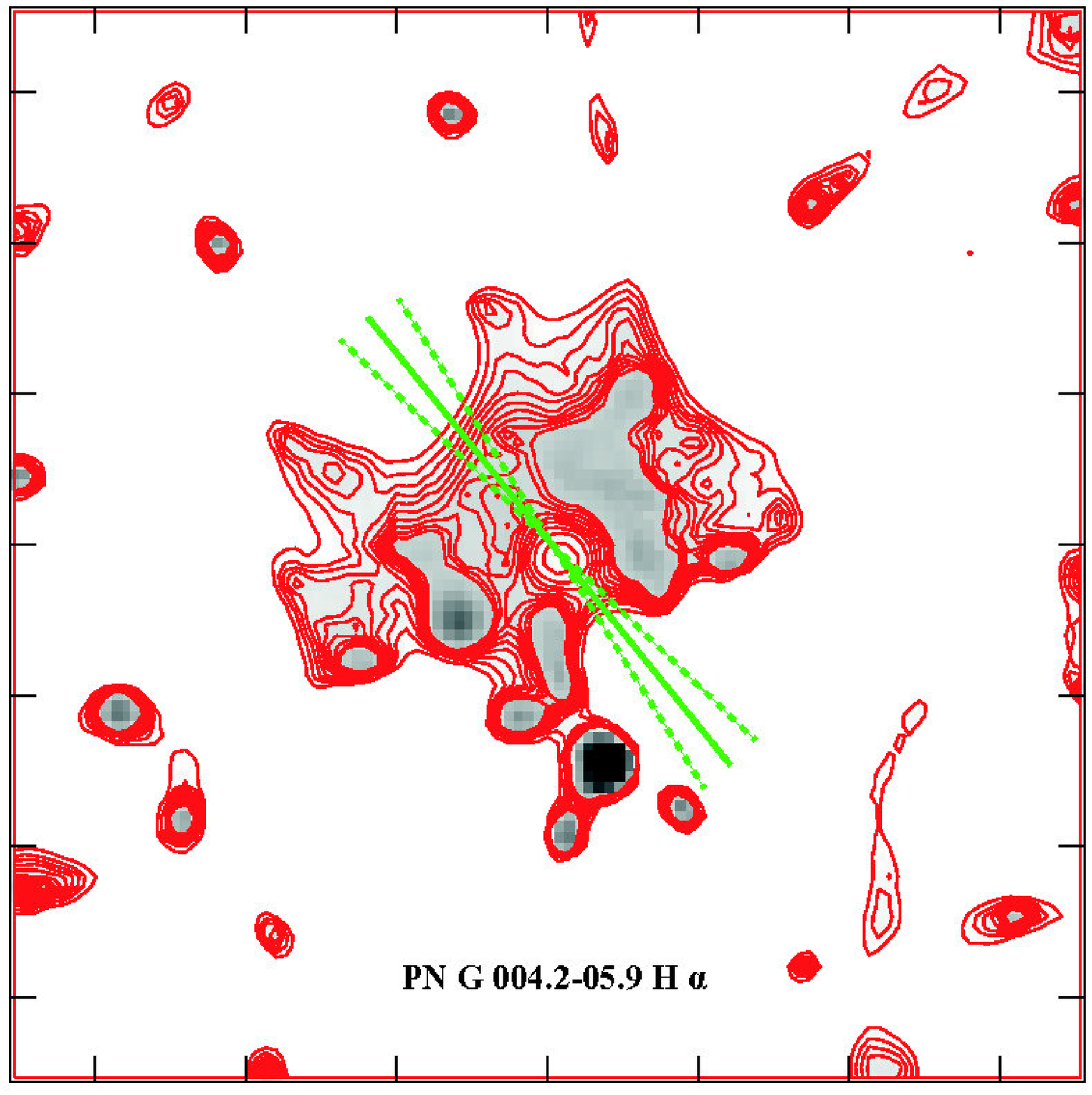} &
    \includegraphics[width=8.1cm]{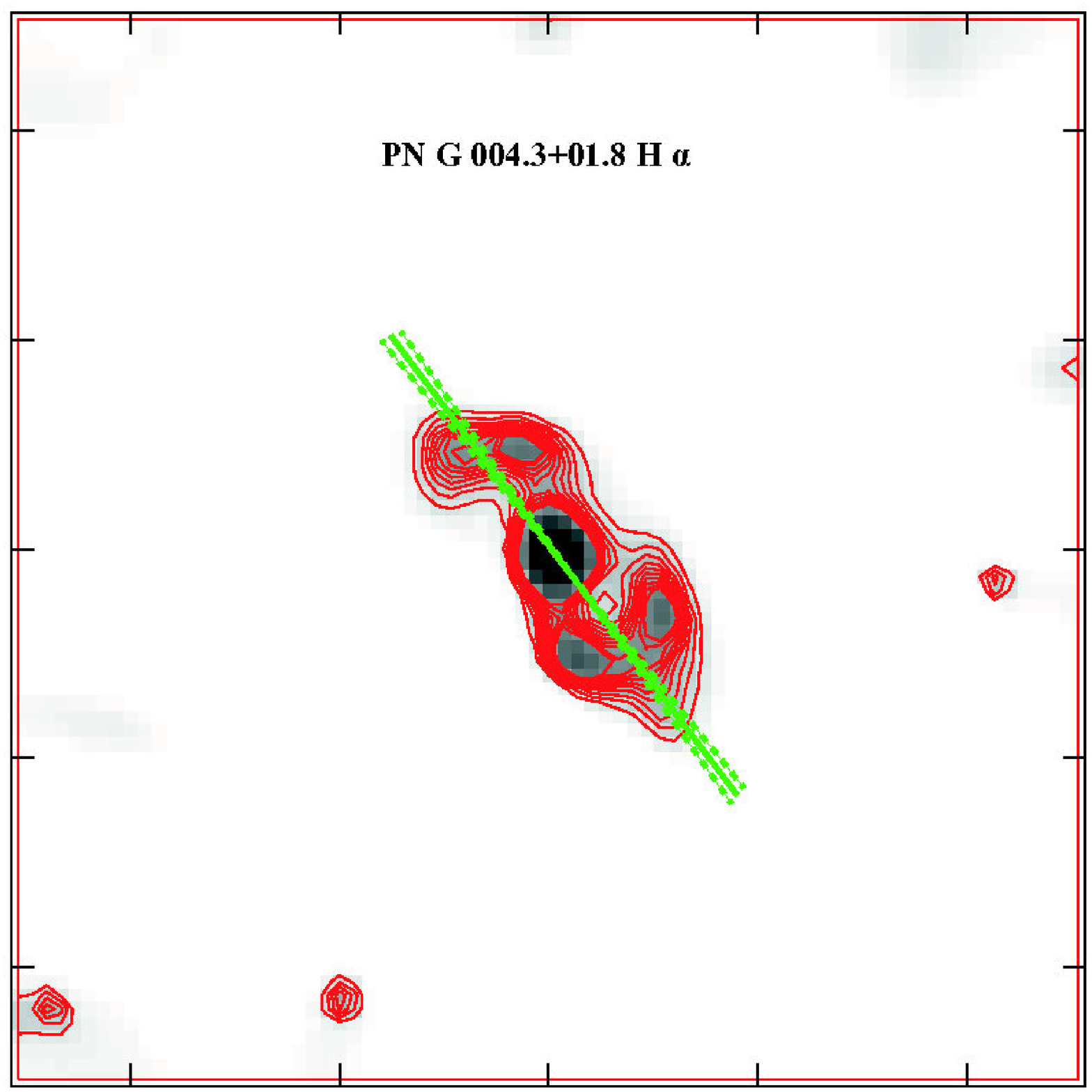}
\\
    \includegraphics[width=8.1cm]{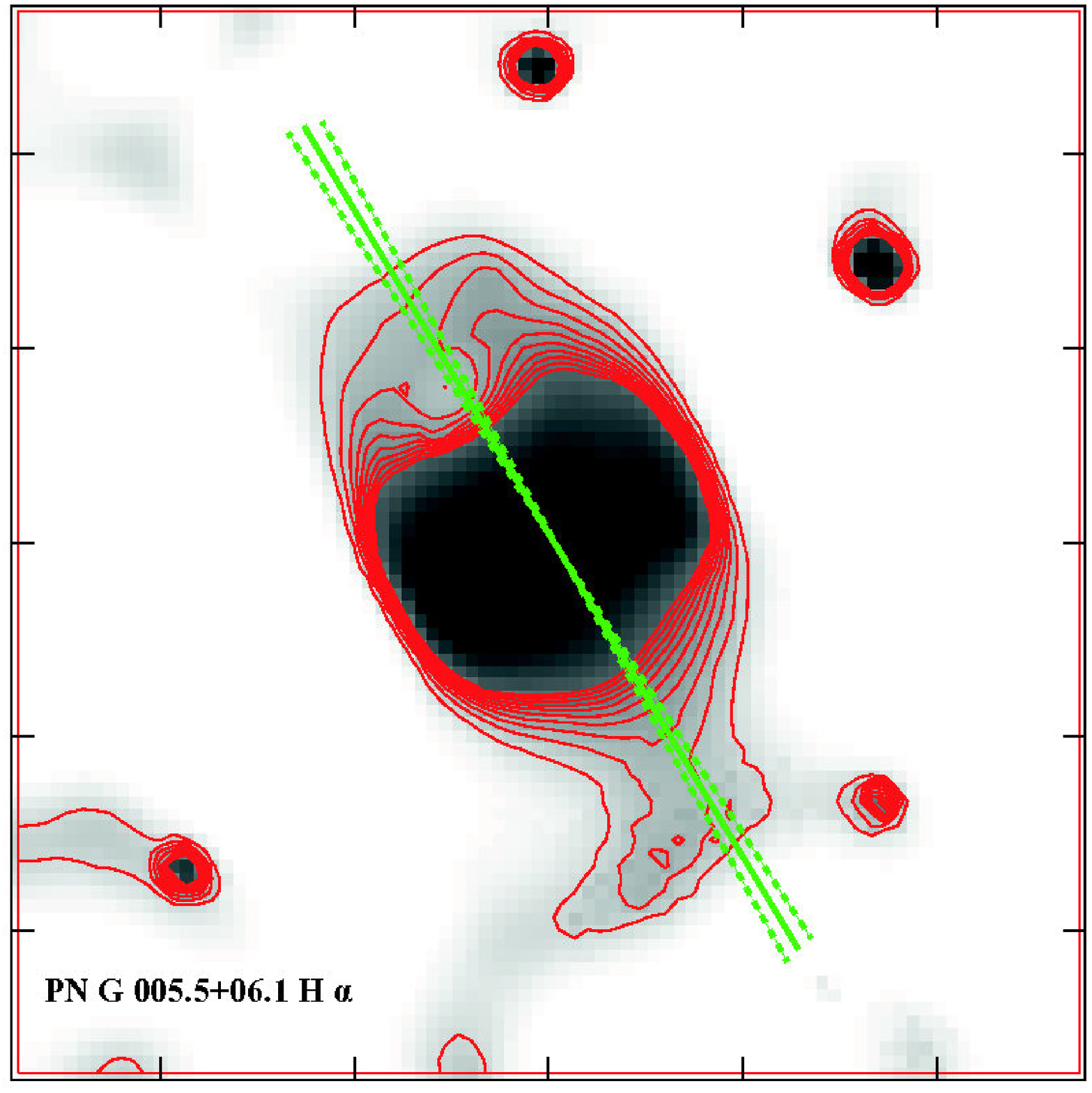} &
    \includegraphics[width=8.1cm]{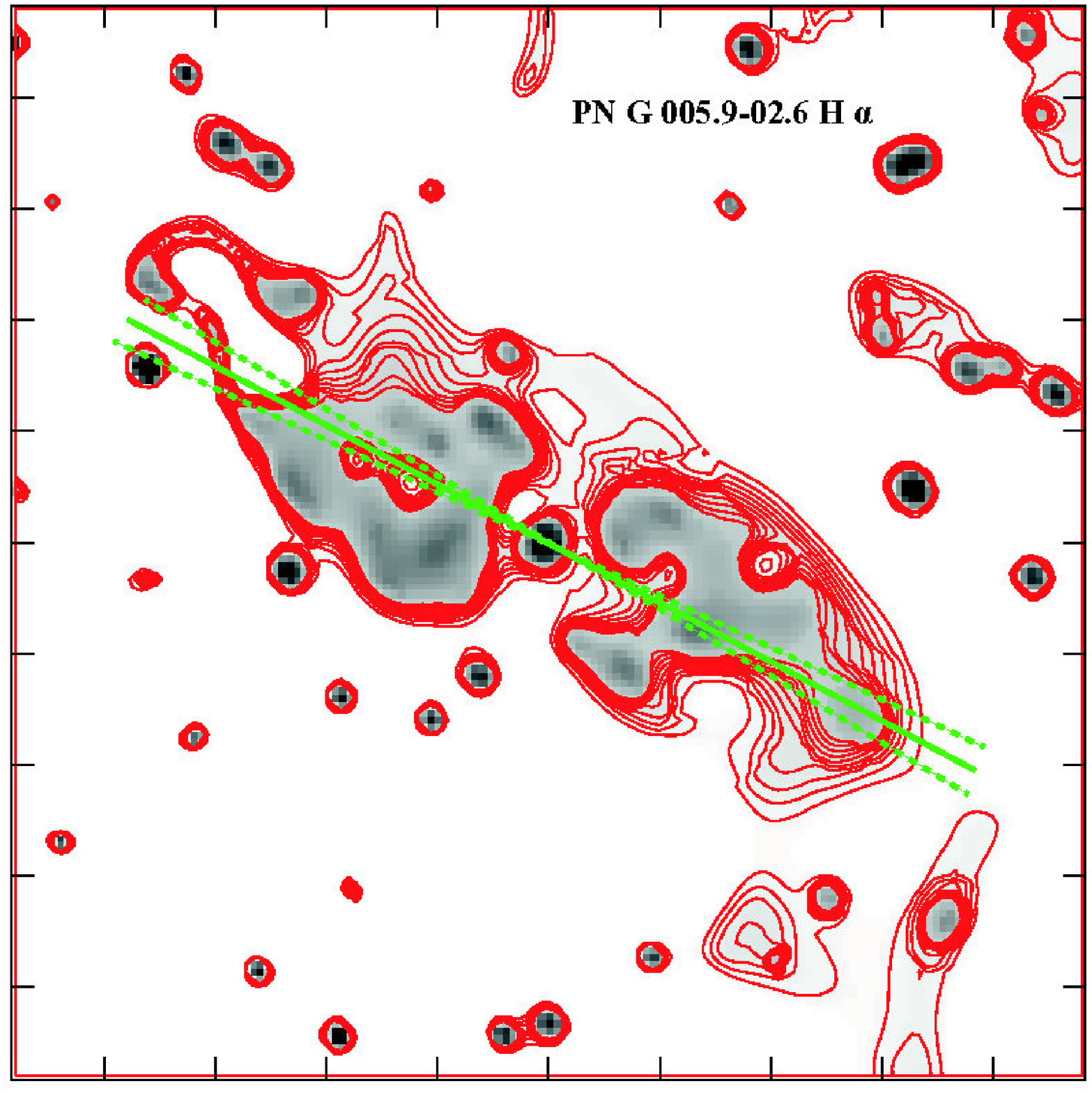}

    \end{array} 
$

    \end{center}
    \contcaption{}
\label{plots1c}
\end{figure*}

\begin{figure*}
    \begin{center}

$
    \begin{array}{cc}
    \includegraphics[width=8.1cm]{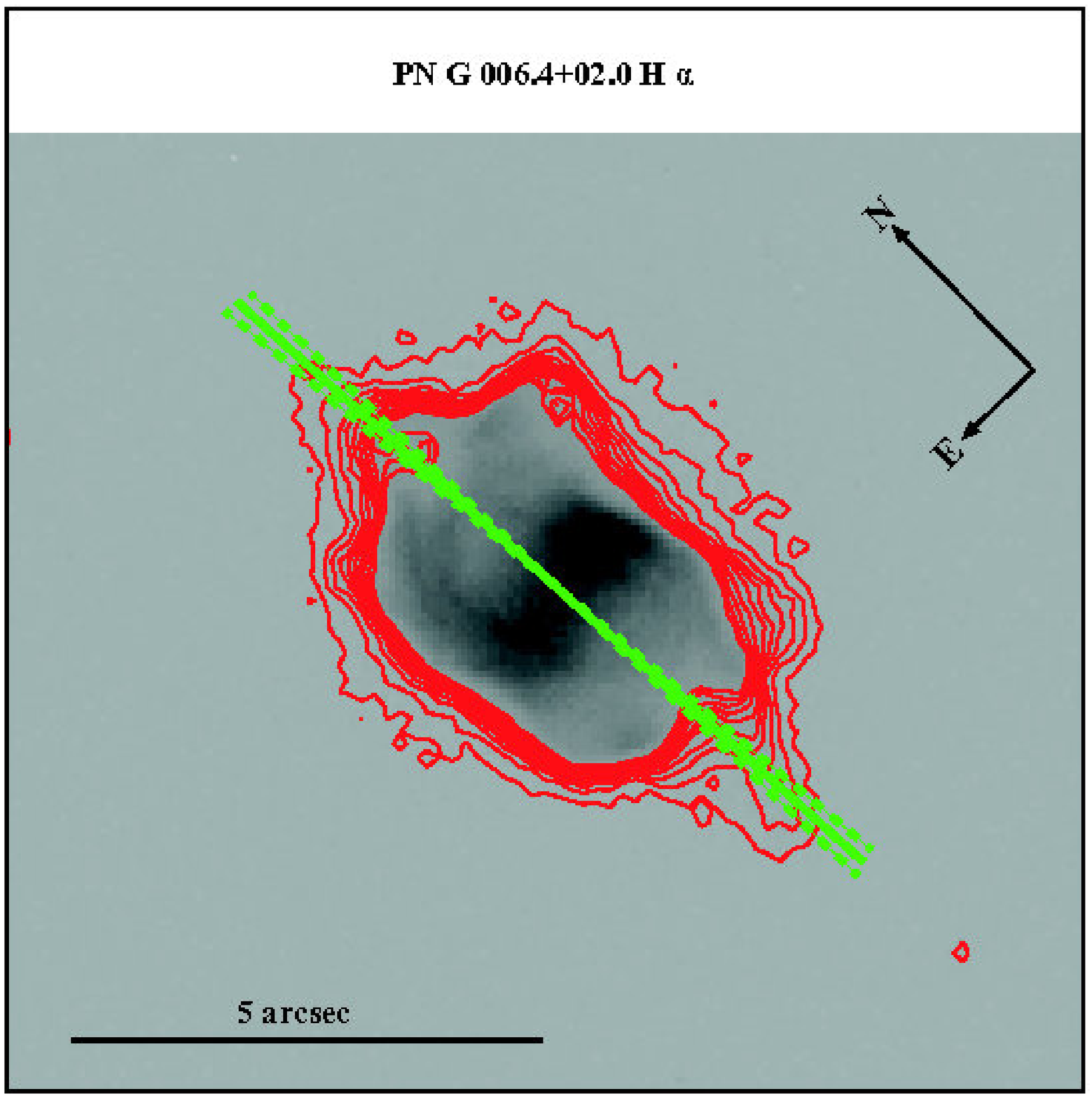} &
    \includegraphics[width=8.1cm]{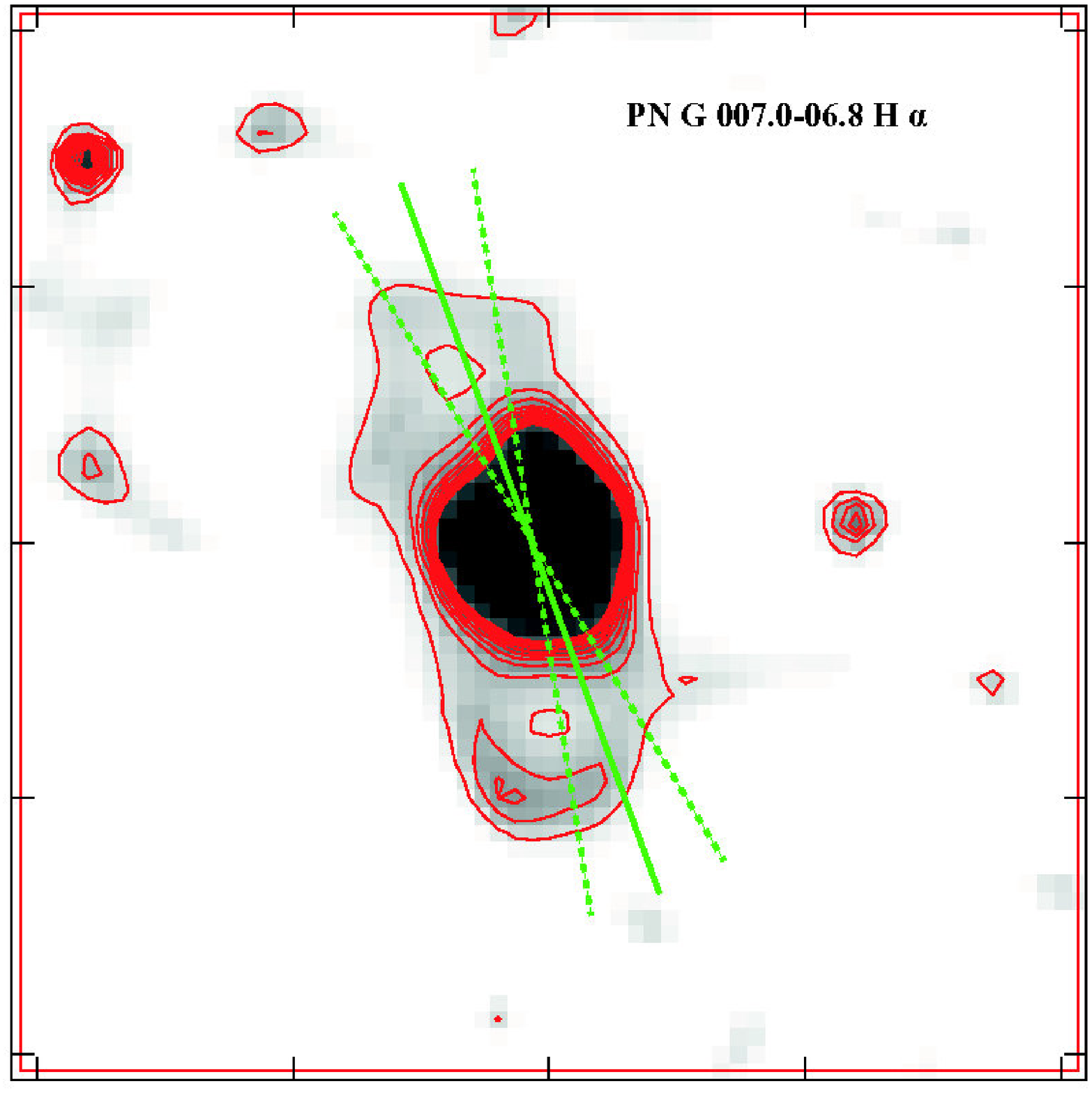}
\\
    \includegraphics[width=8.1cm]{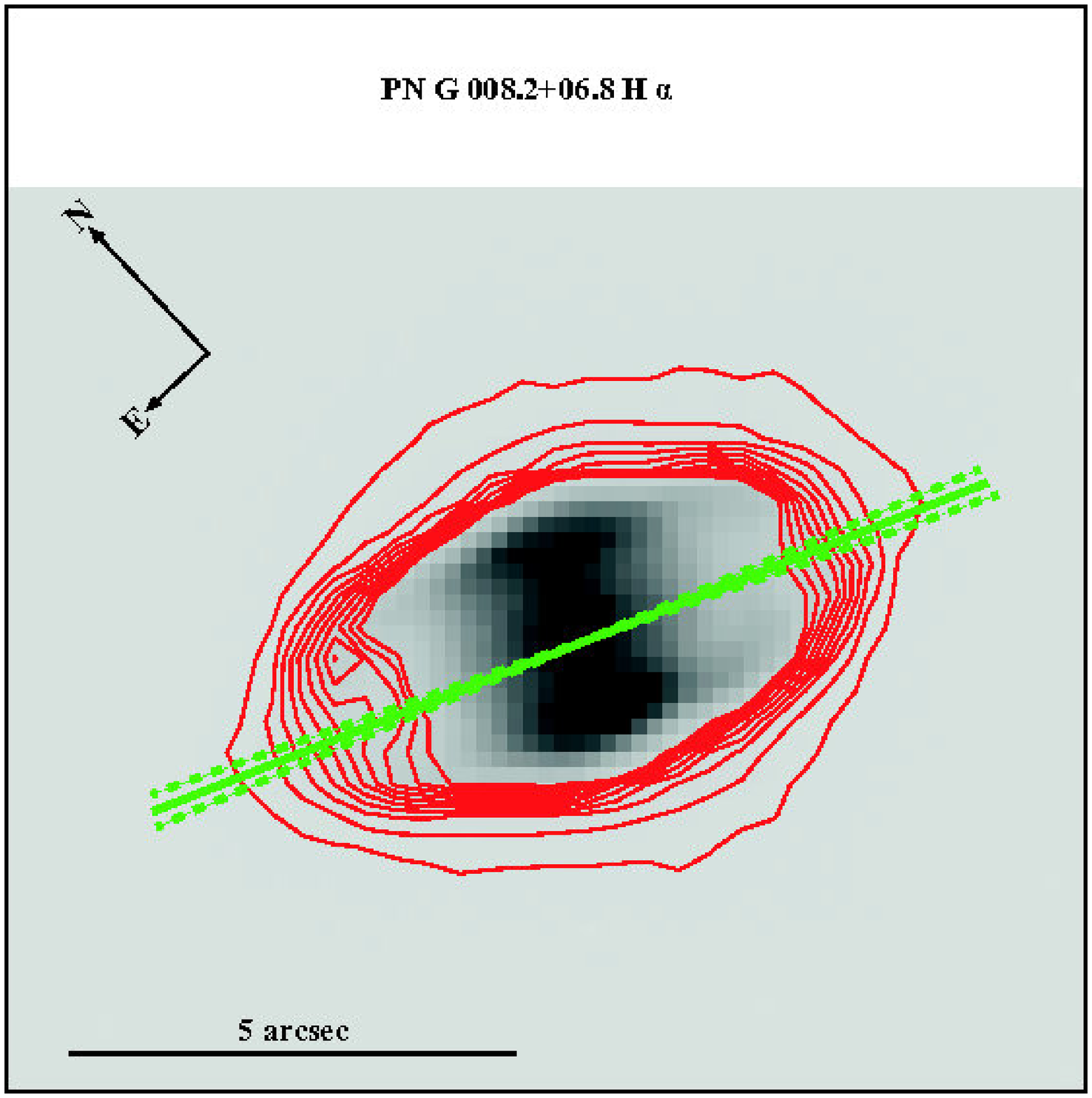} &
    \includegraphics[width=8.1cm]{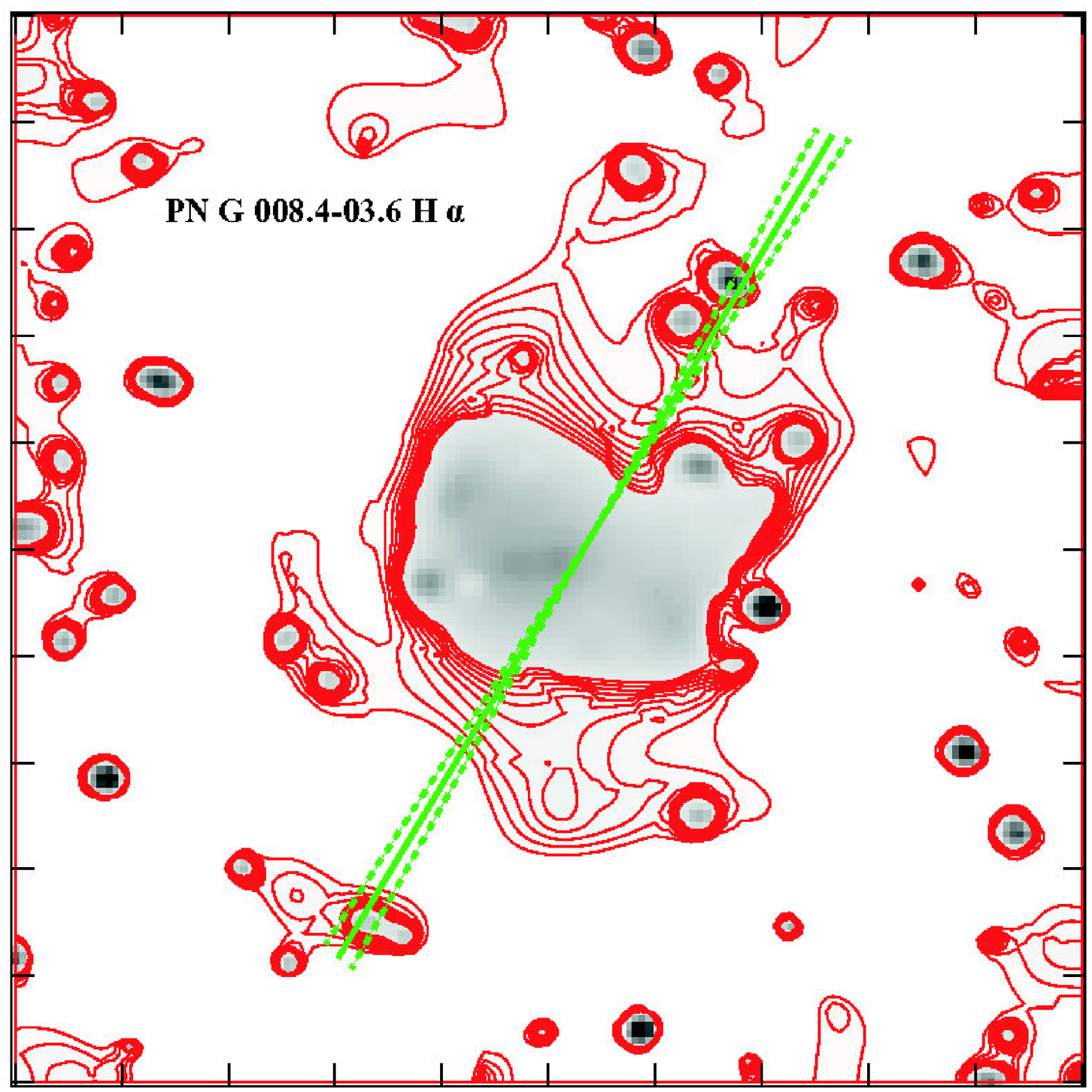}
\\
    \includegraphics[width=8.1cm]{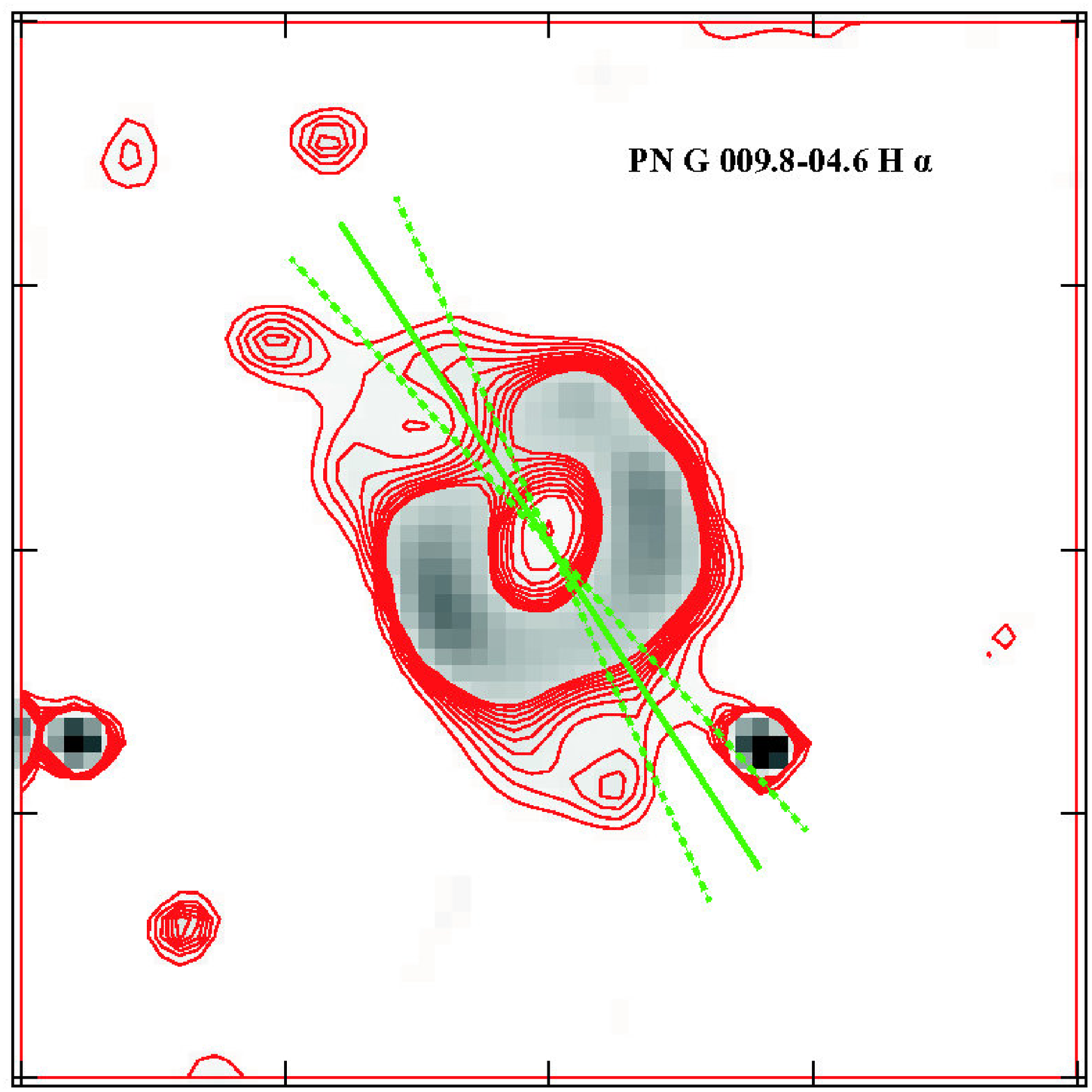} &
    \includegraphics[width=8.1cm]{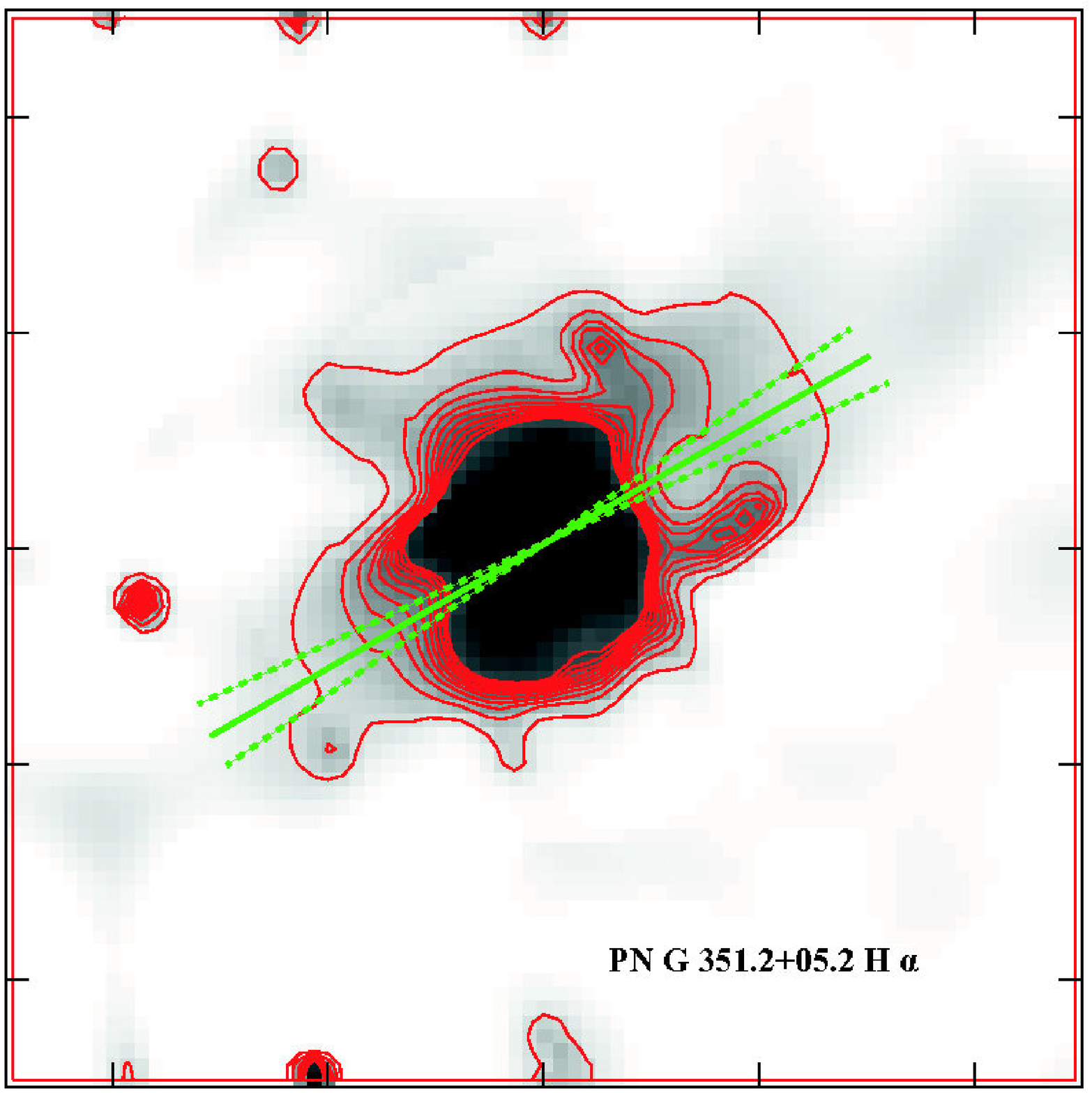}

    \end{array}
$

    \end{center}
    \contcaption{}
\label{plots1d}
\end{figure*}

\begin{figure*}
    \begin{center}

$
    \begin{array}{cc}
    \includegraphics[width=8.1cm]{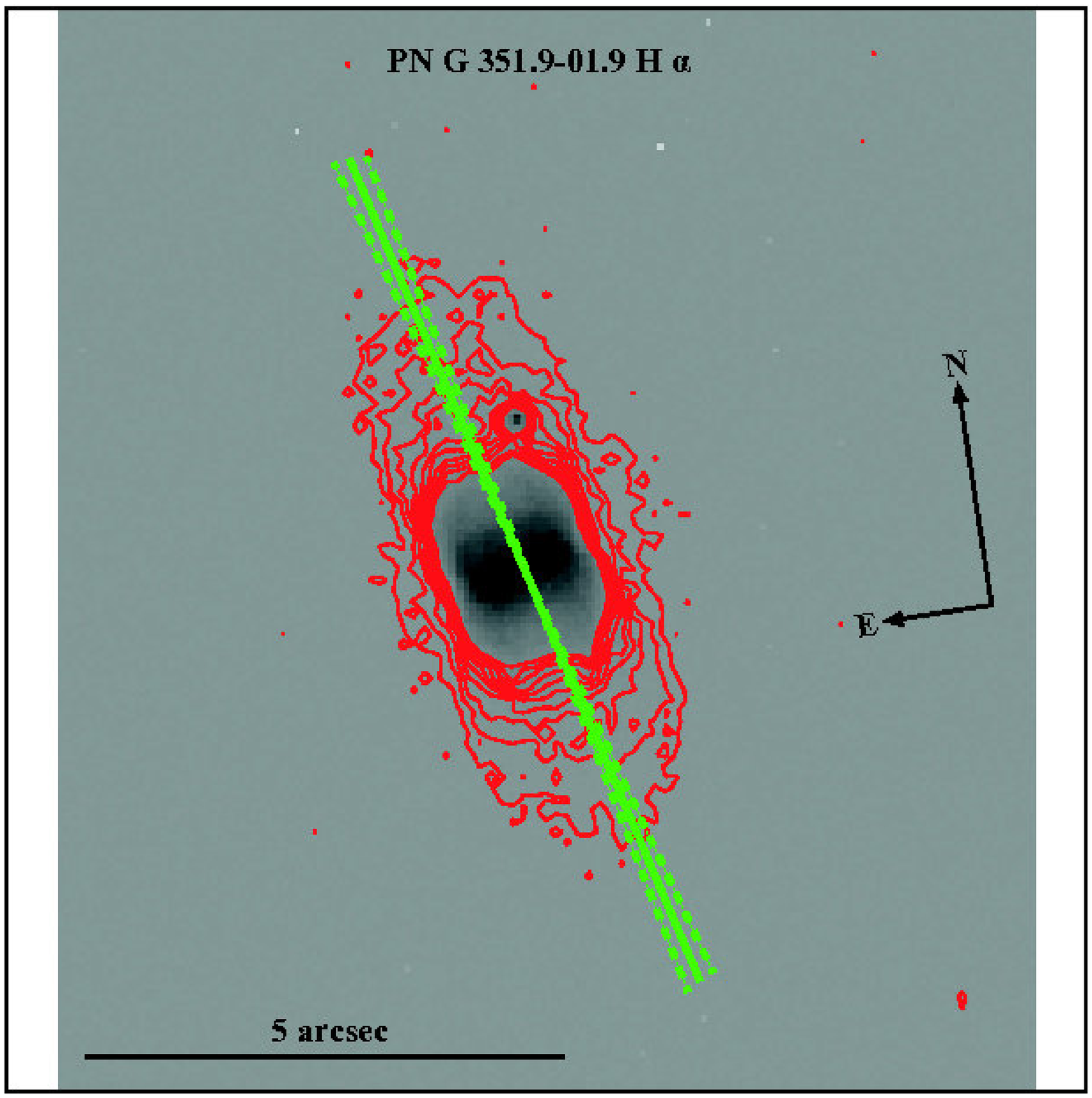} &
    \includegraphics[width=8.1cm]{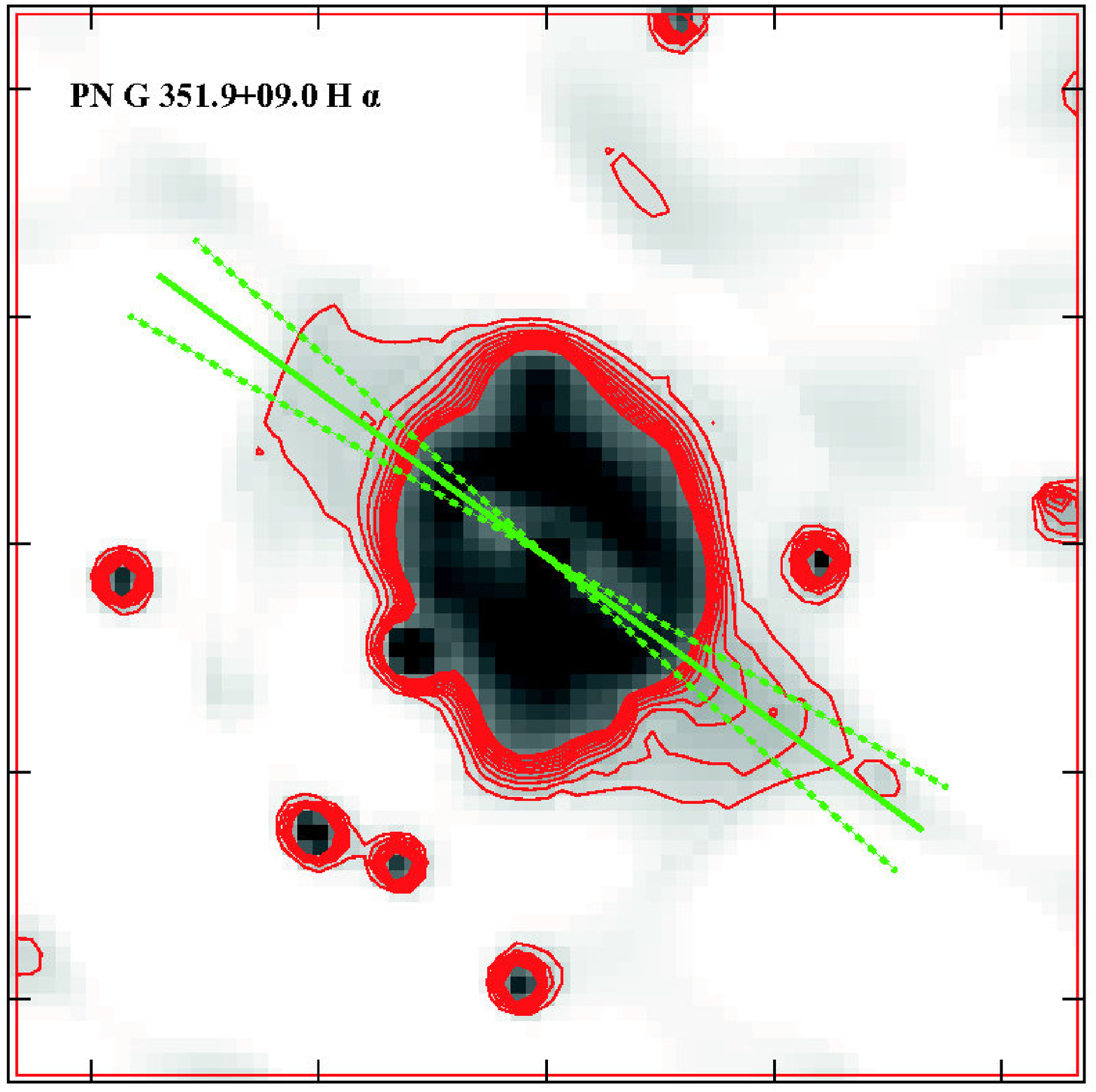}
\\
    \includegraphics[width=8.1cm]{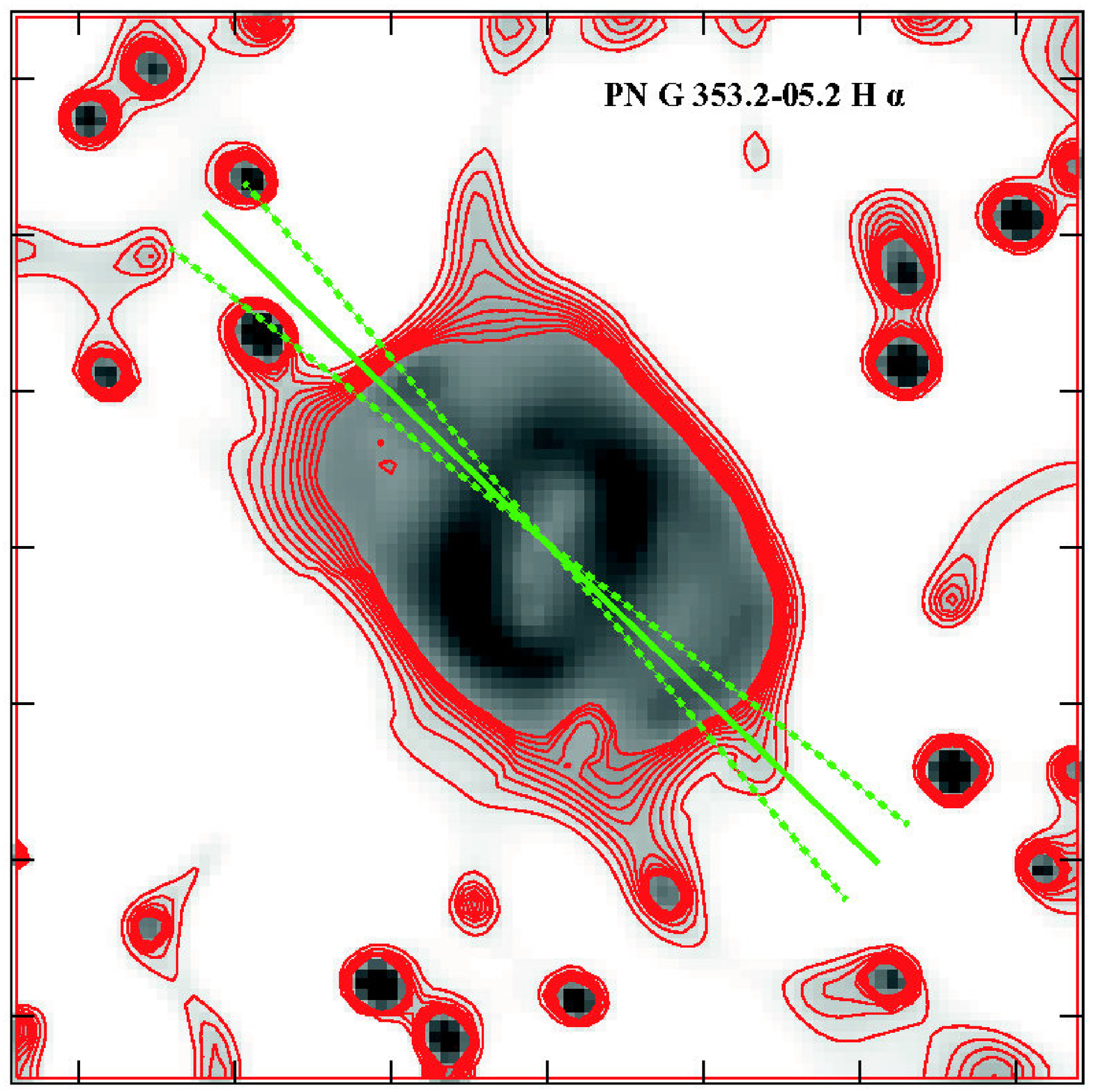} &
    \includegraphics[width=8.1cm]{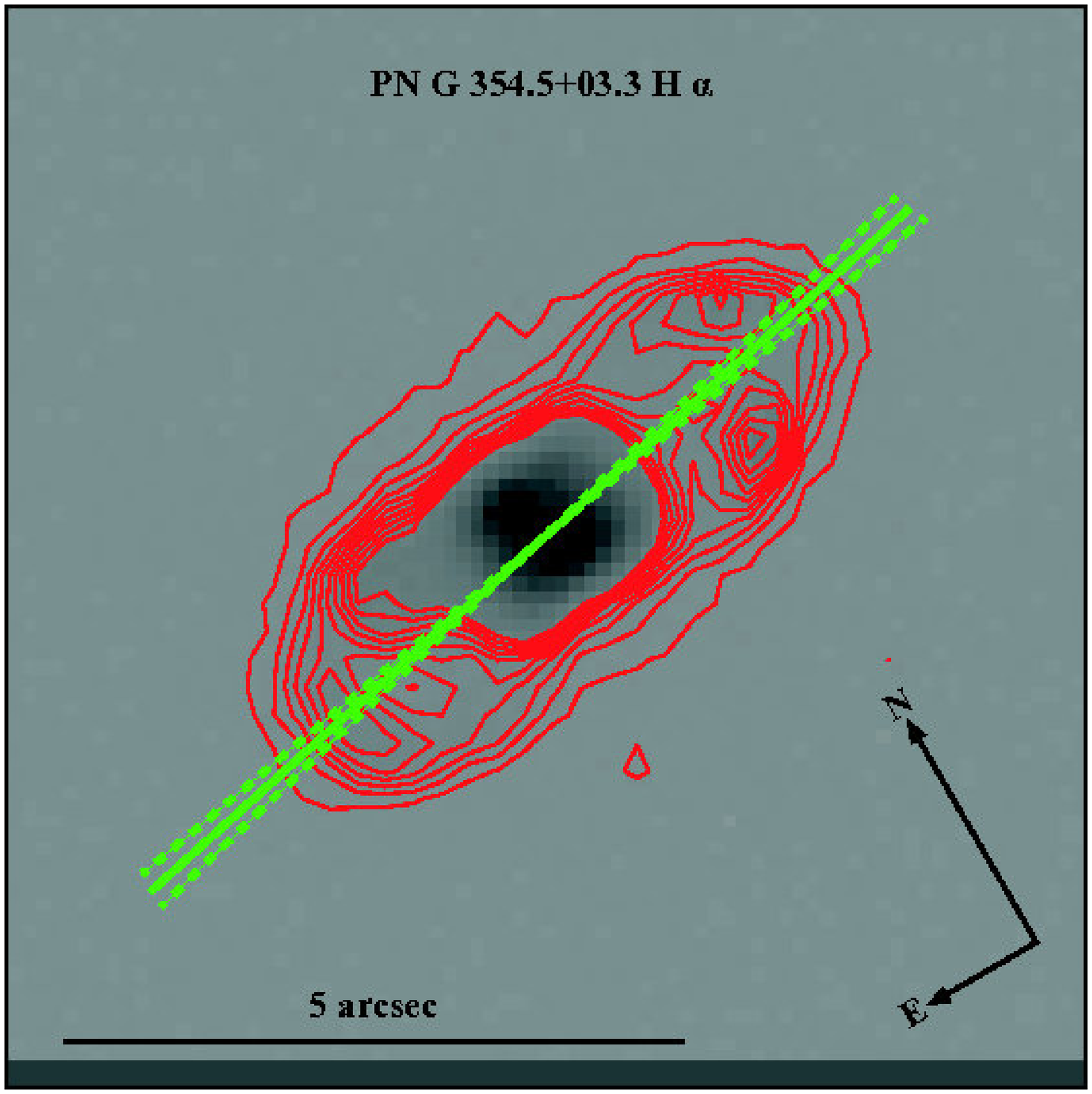}
\\
    \includegraphics[width=8.1cm]{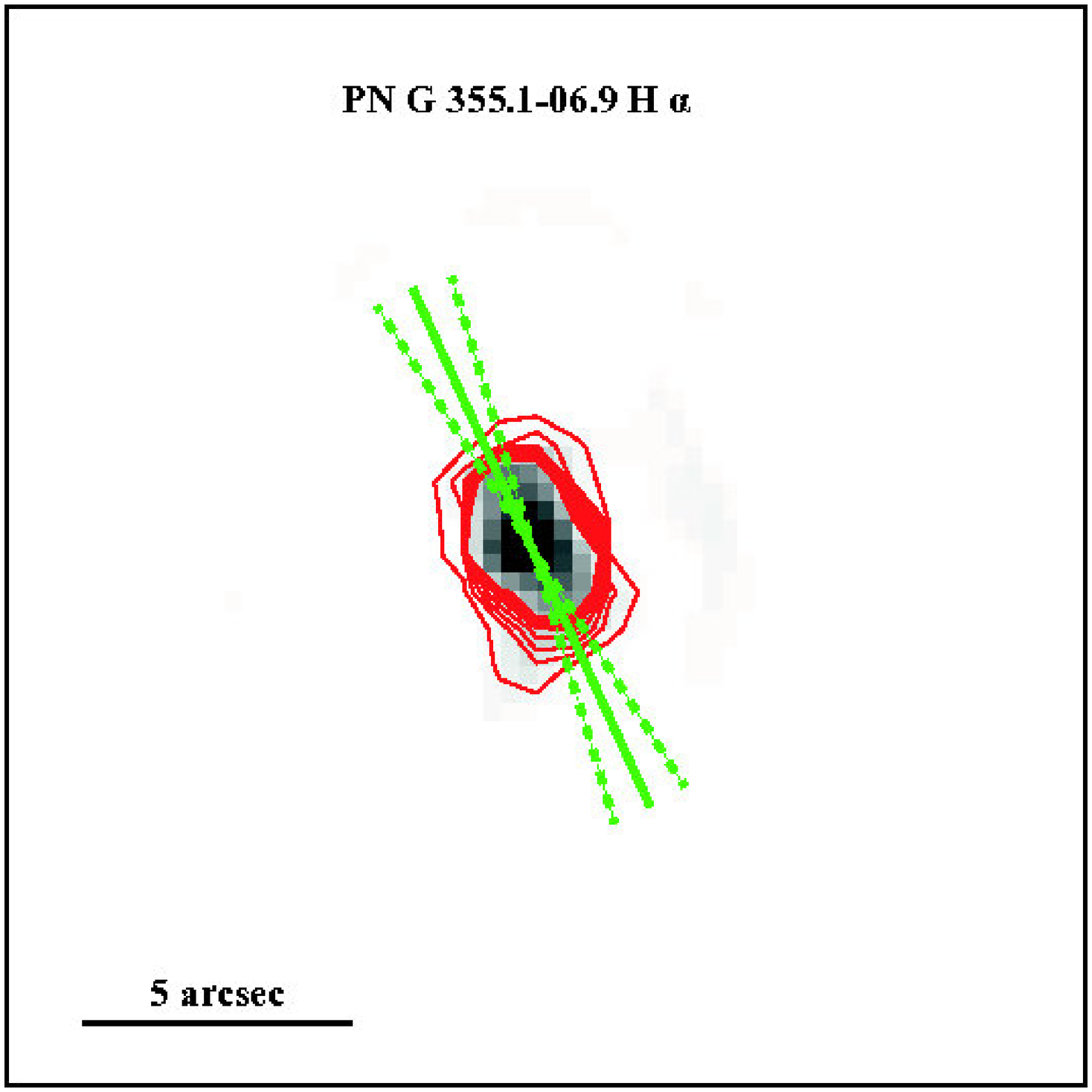} &
    \includegraphics[width=8.1cm]{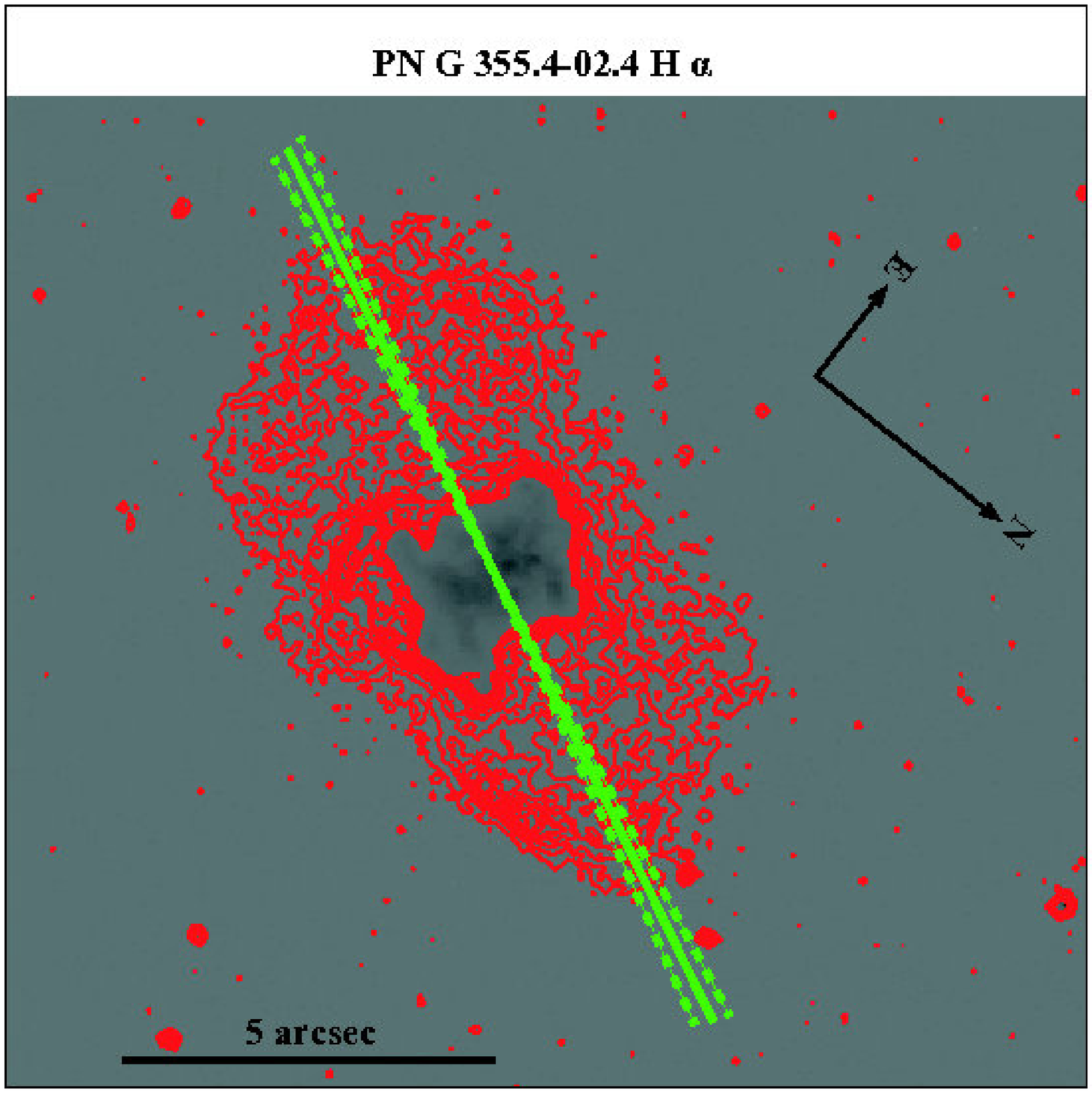}

    \end{array}
$

    \end{center}
    \contcaption{}
\label{plots1e}
\end{figure*}

\begin{figure*}
    \begin{center}

$
    \begin{array}{cc}
    \includegraphics[width=8.1cm]{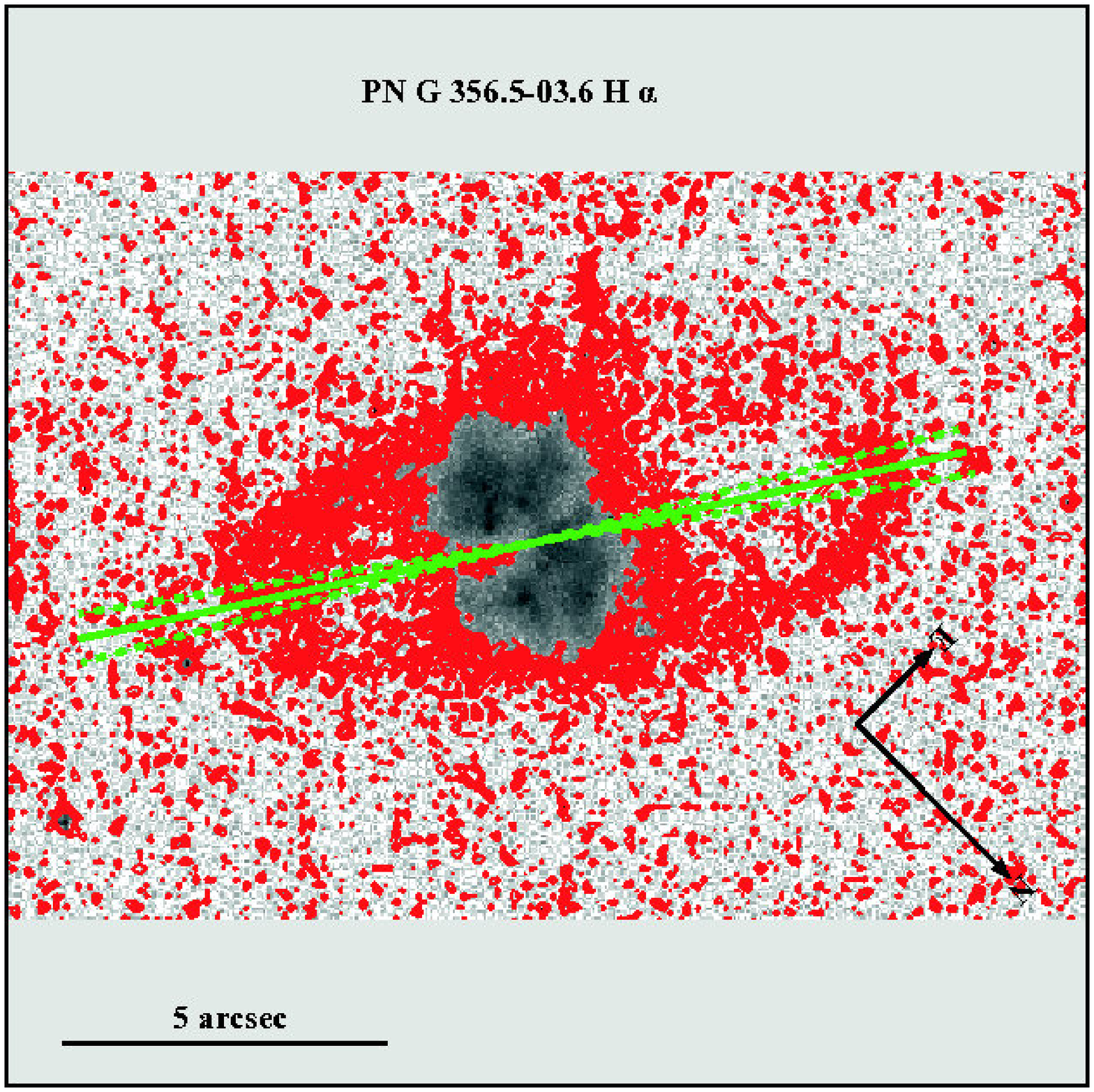} &
    \includegraphics[width=8.1cm]{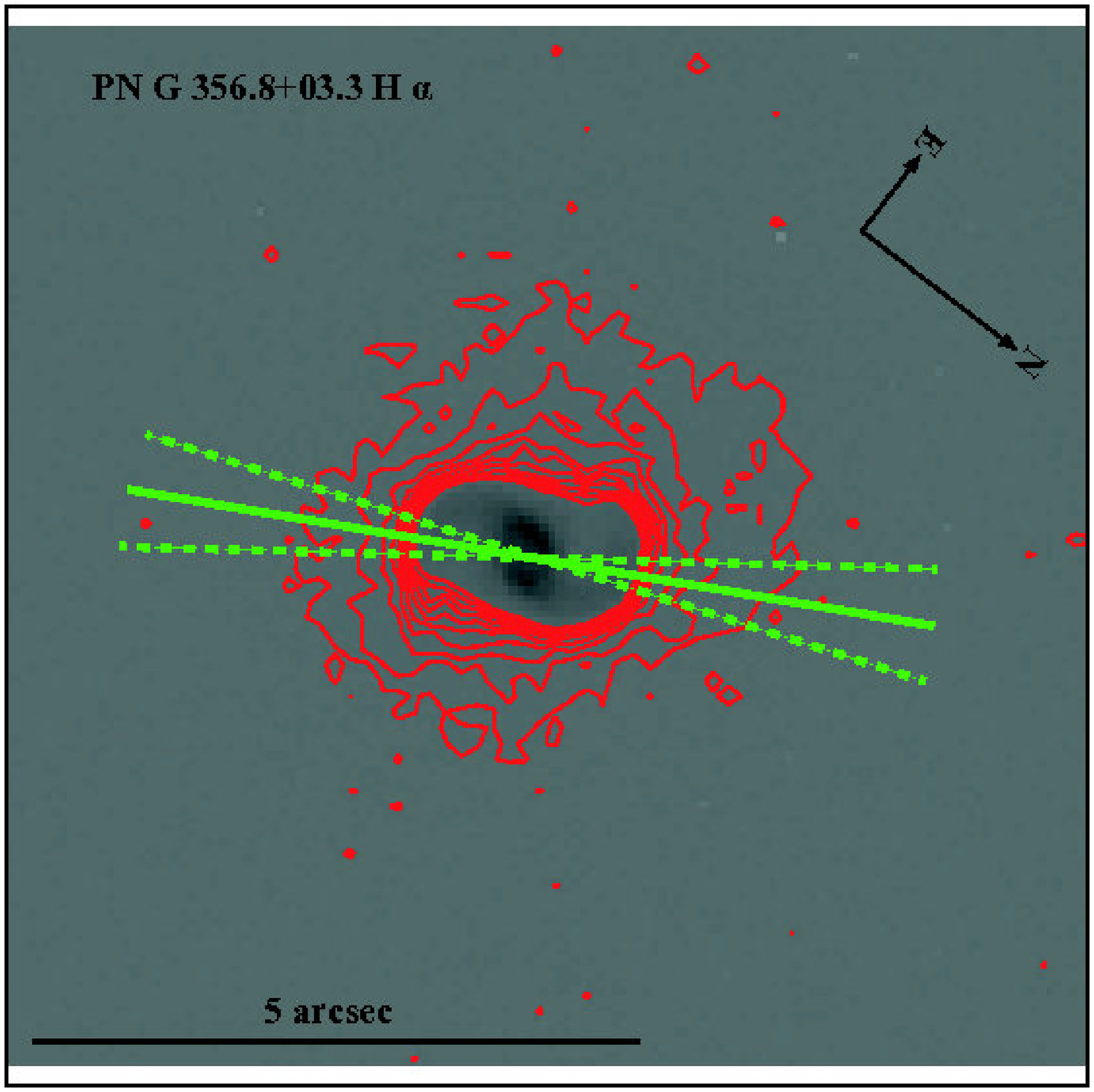}
\\
    \includegraphics[width=8.1cm]{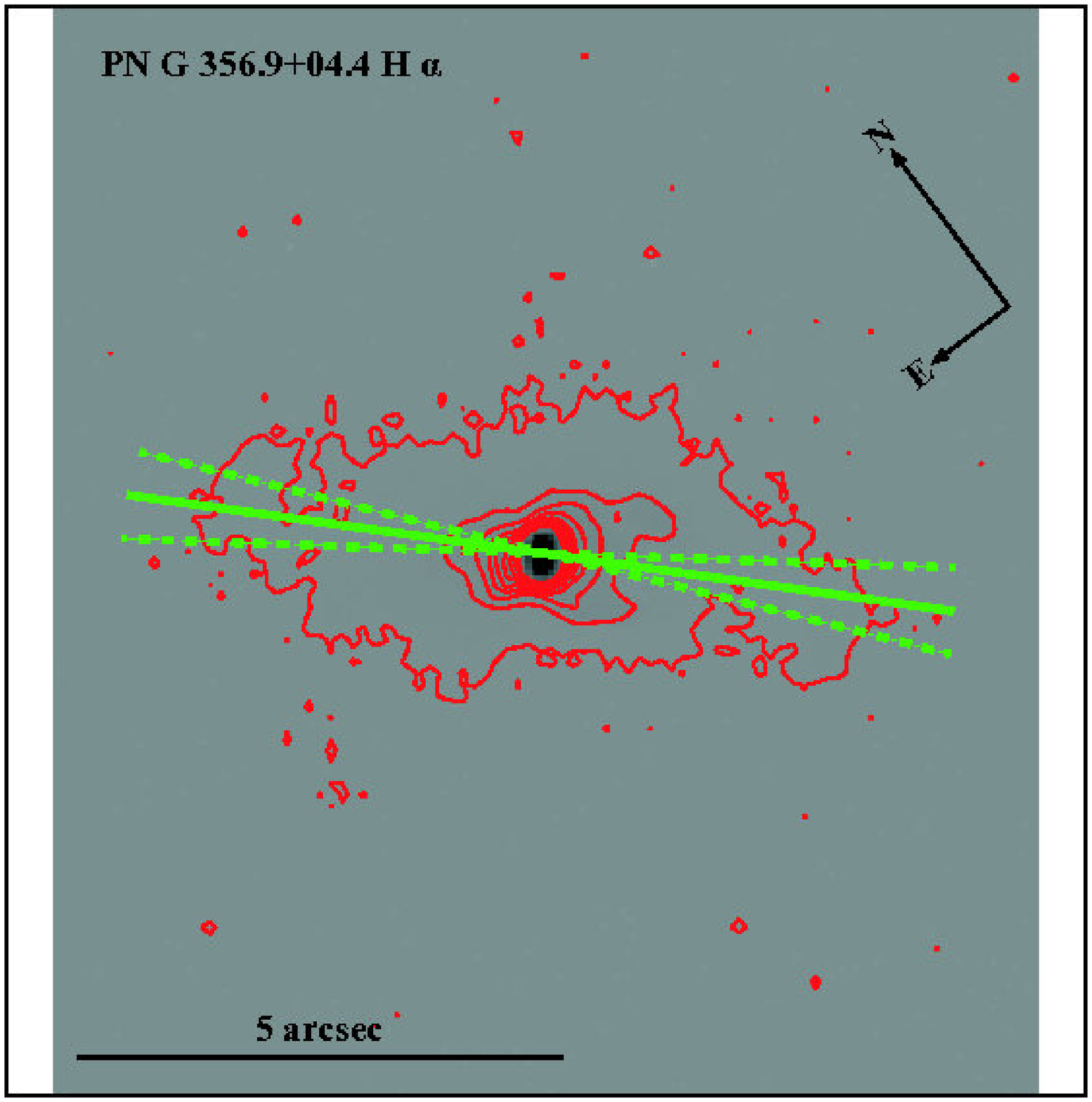} &
    \includegraphics[width=8.1cm]{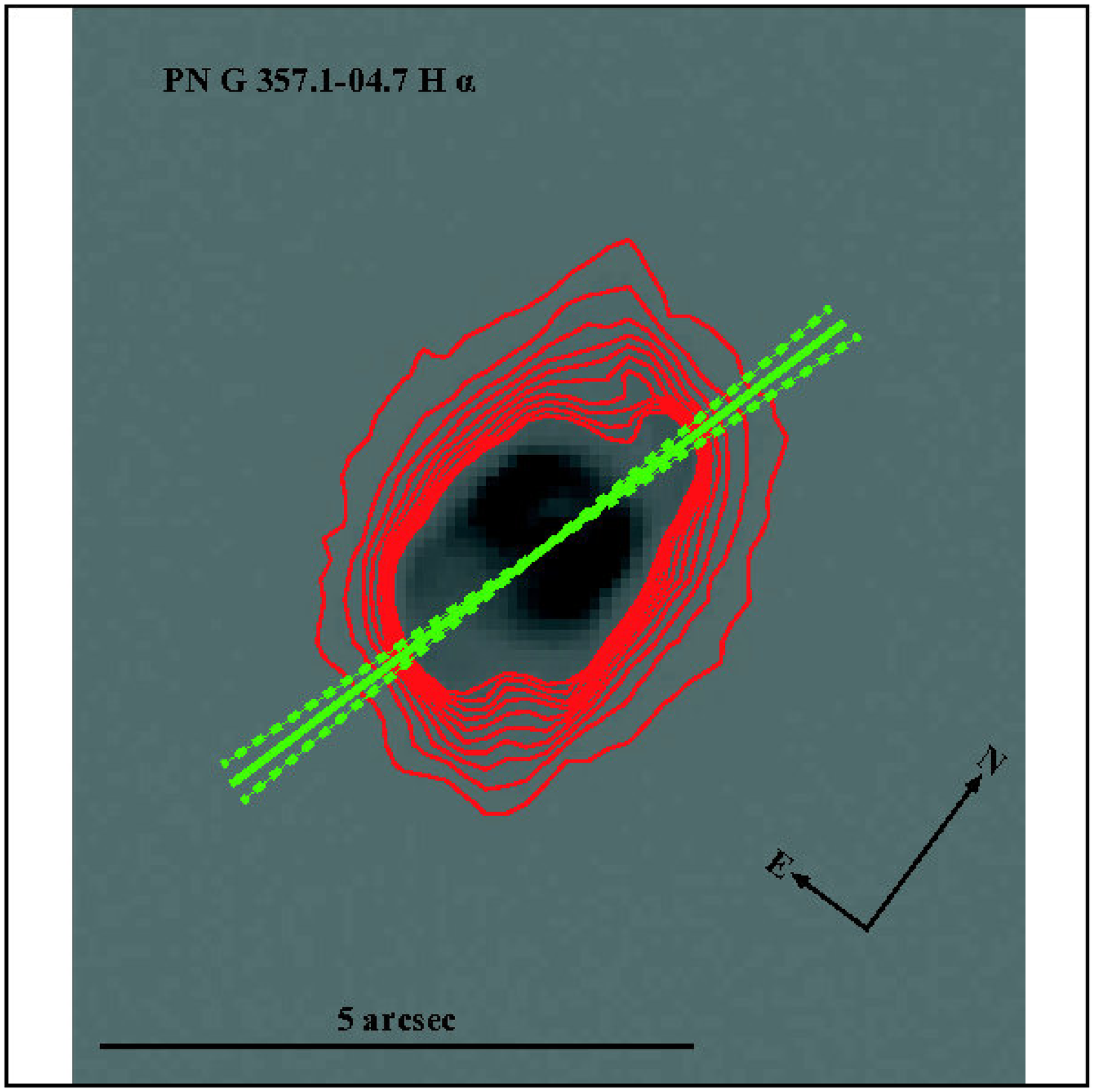}
\\
    \includegraphics[width=8.1cm]{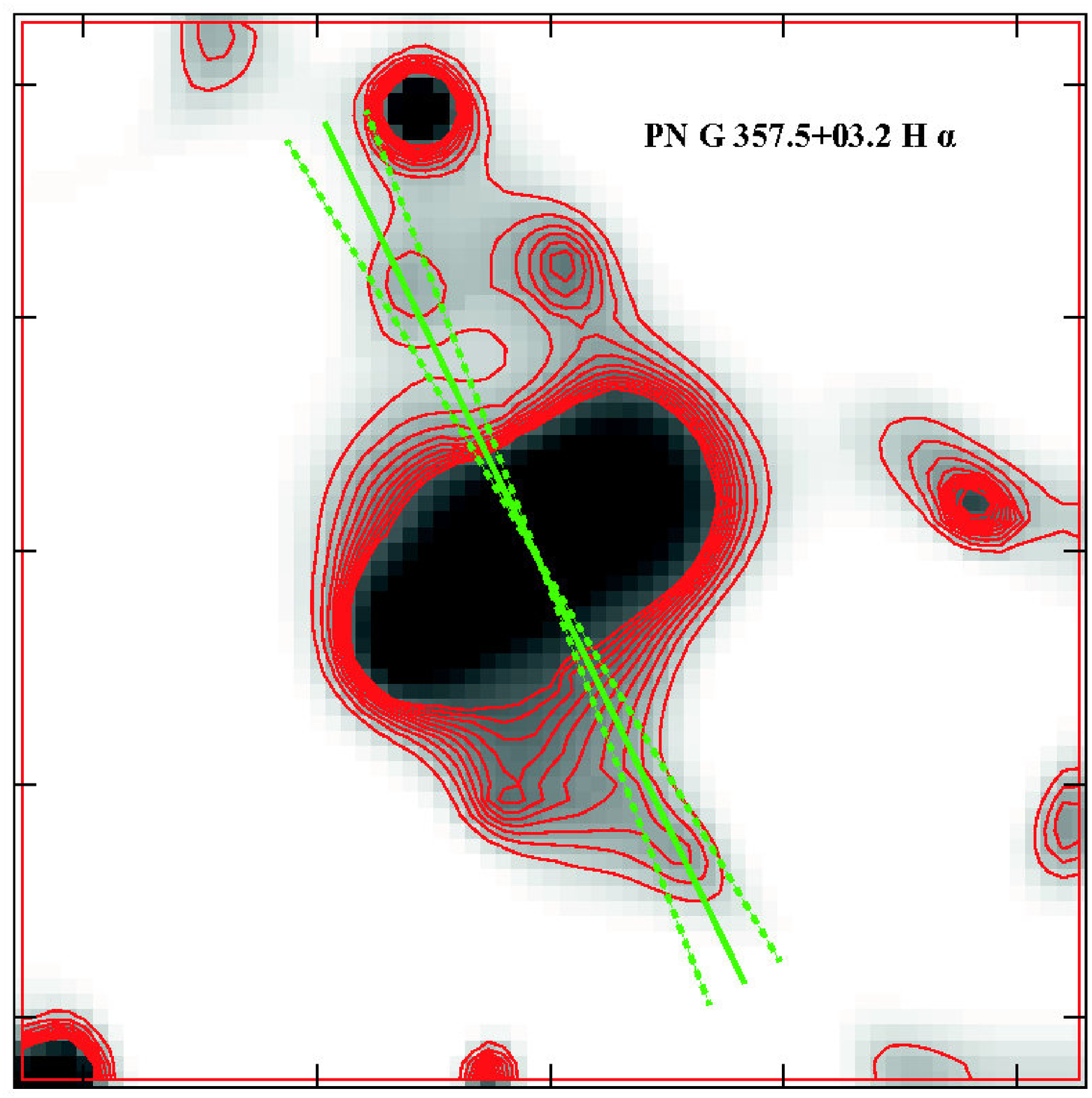} &
    \includegraphics[width=8.1cm]{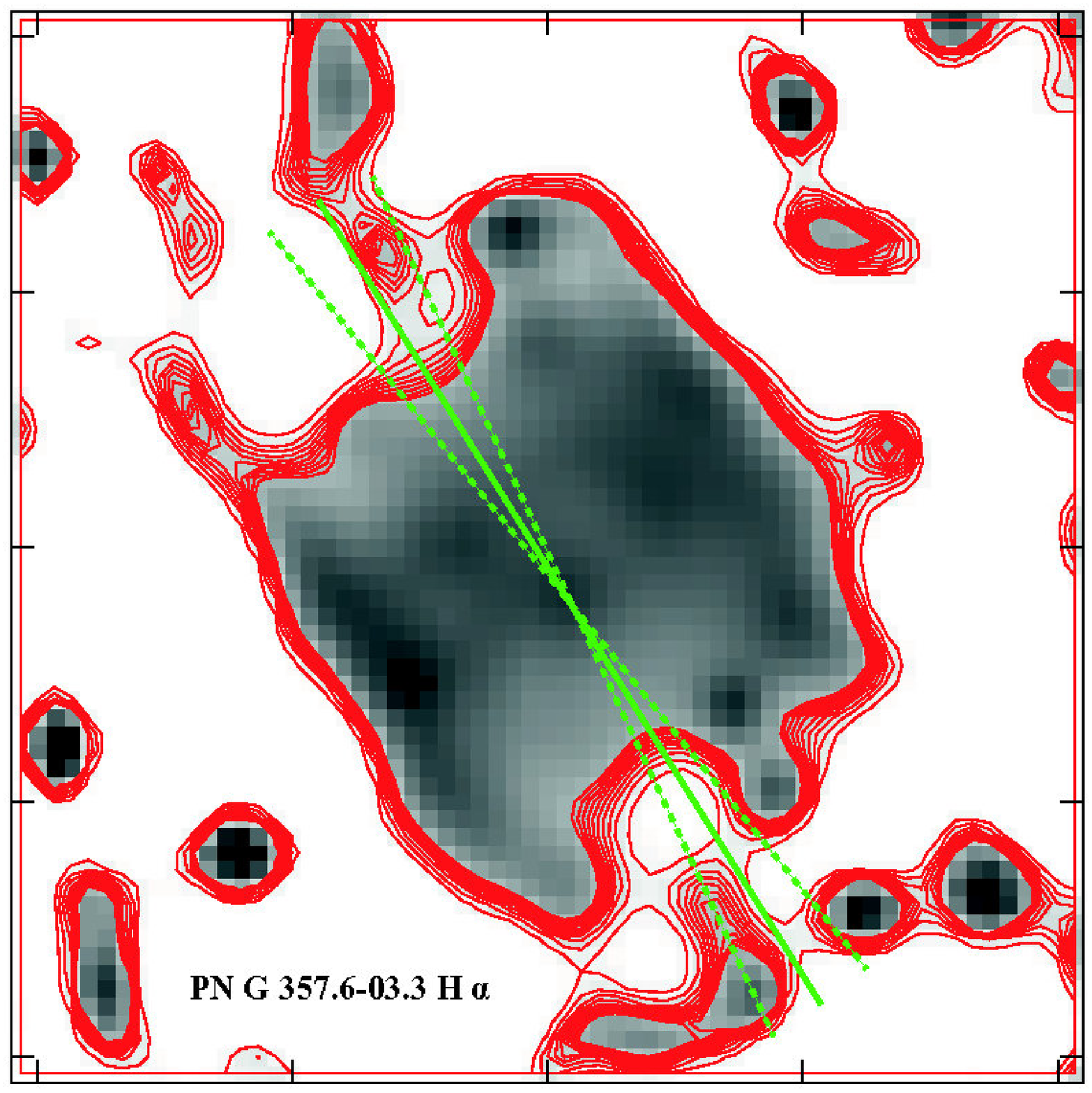}

    \end{array}
$

    \end{center}
    \contcaption{}
\label{plots1f}
\end{figure*}

\begin{figure*}
    \begin{center}

$
    \begin{array}{cc}
    \includegraphics[width=8.1cm]{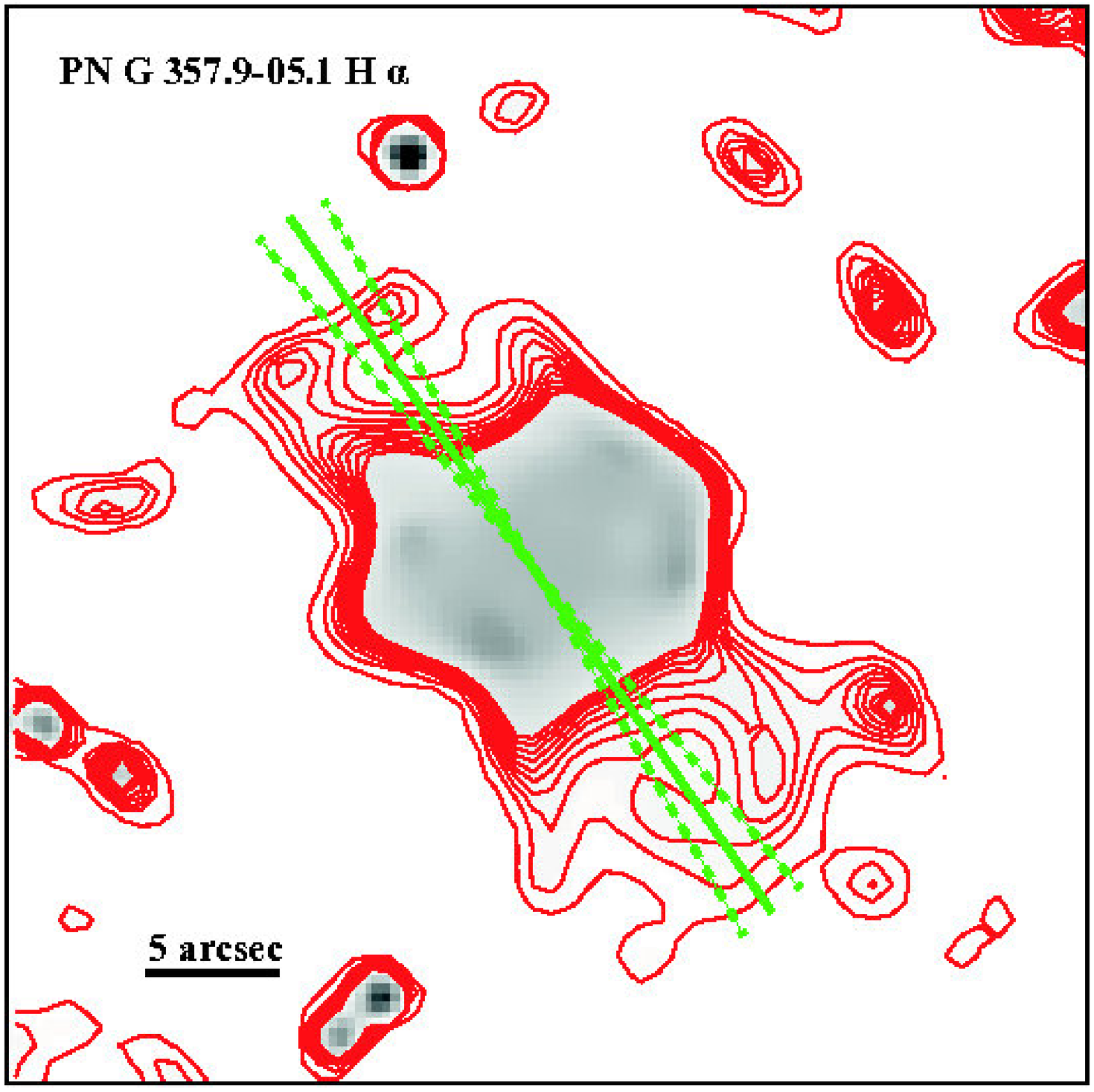} &
    \includegraphics[width=8.1cm]{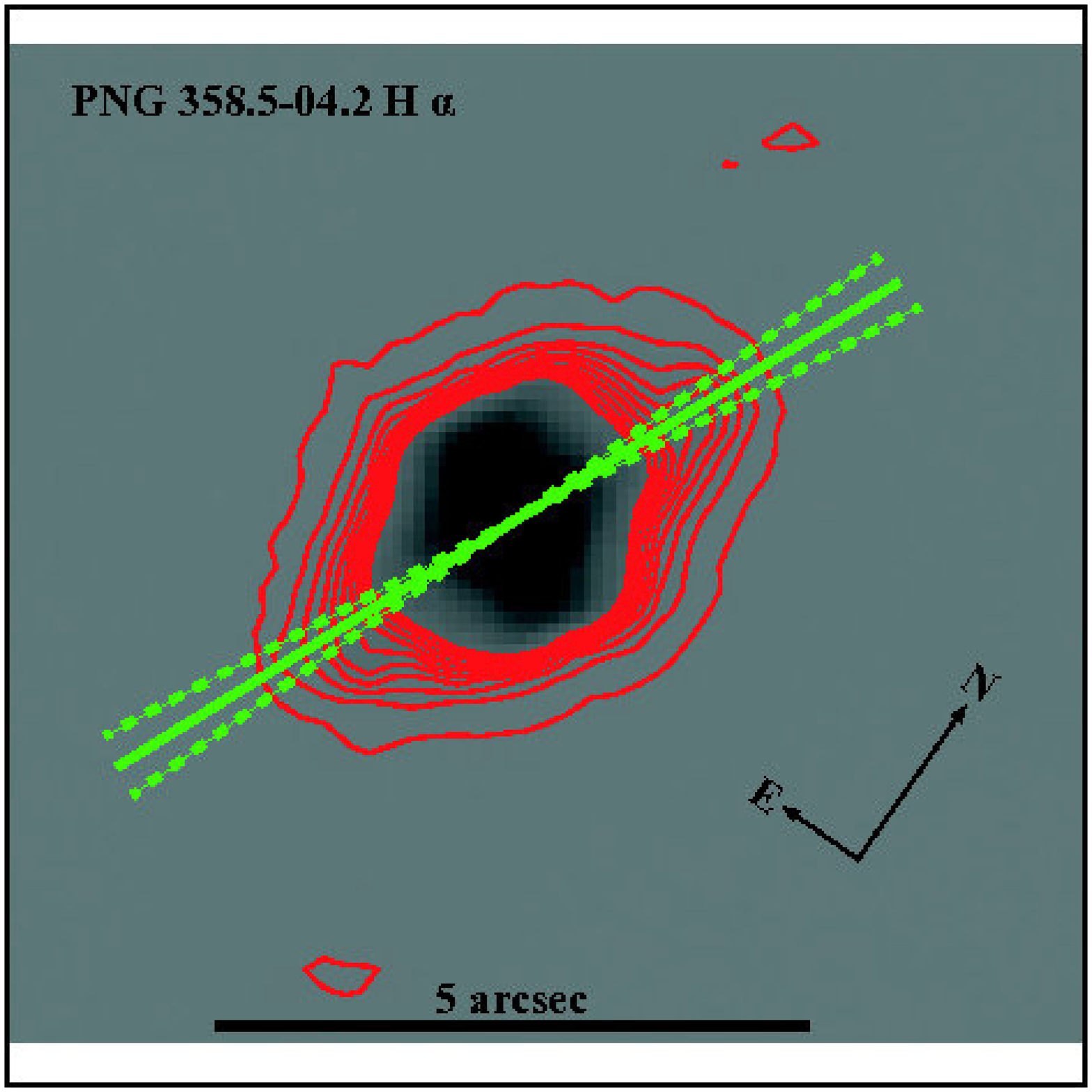}    
\\
    \includegraphics[width=8.1cm]{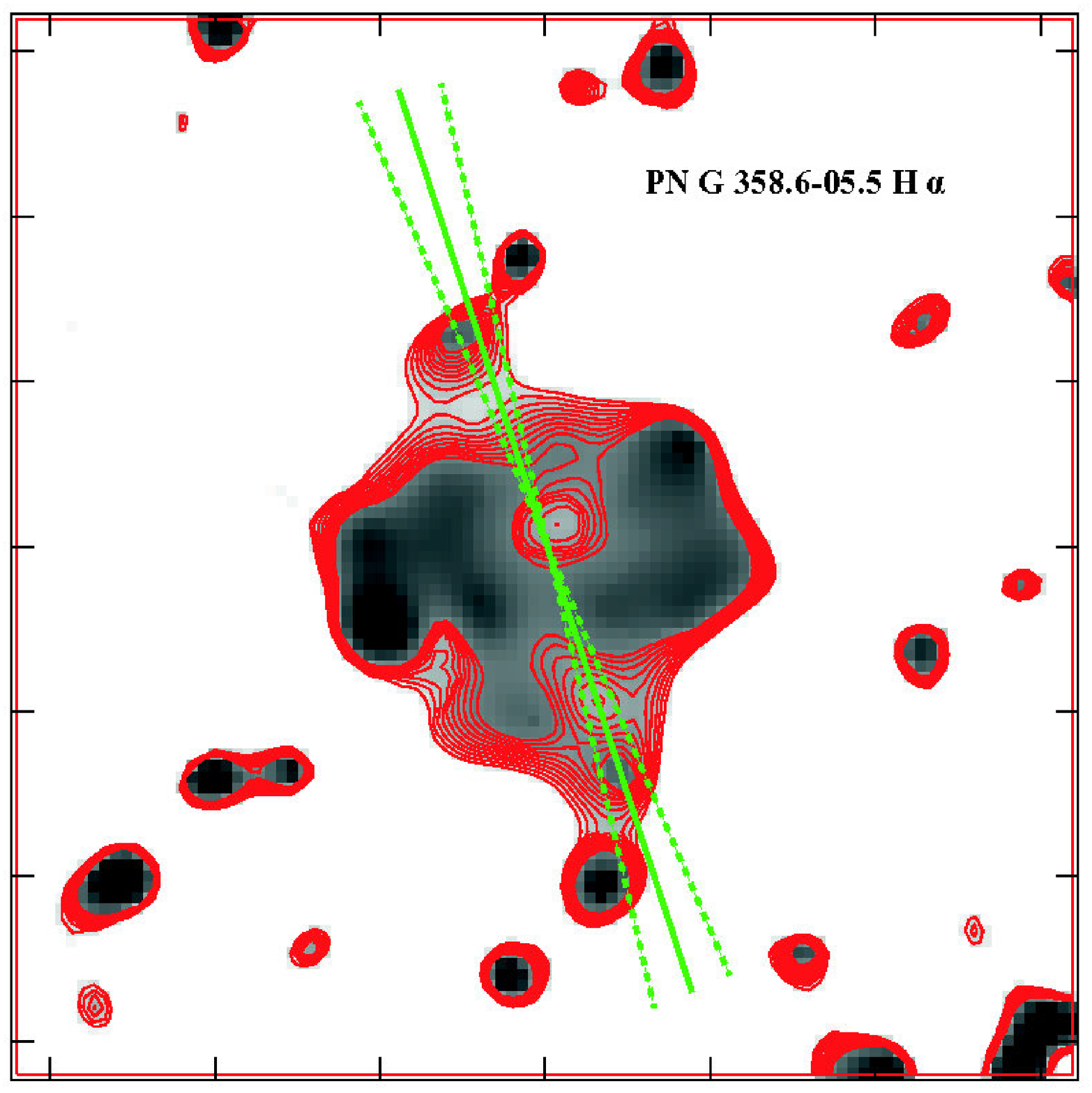} &
    \includegraphics[width=8.1cm]{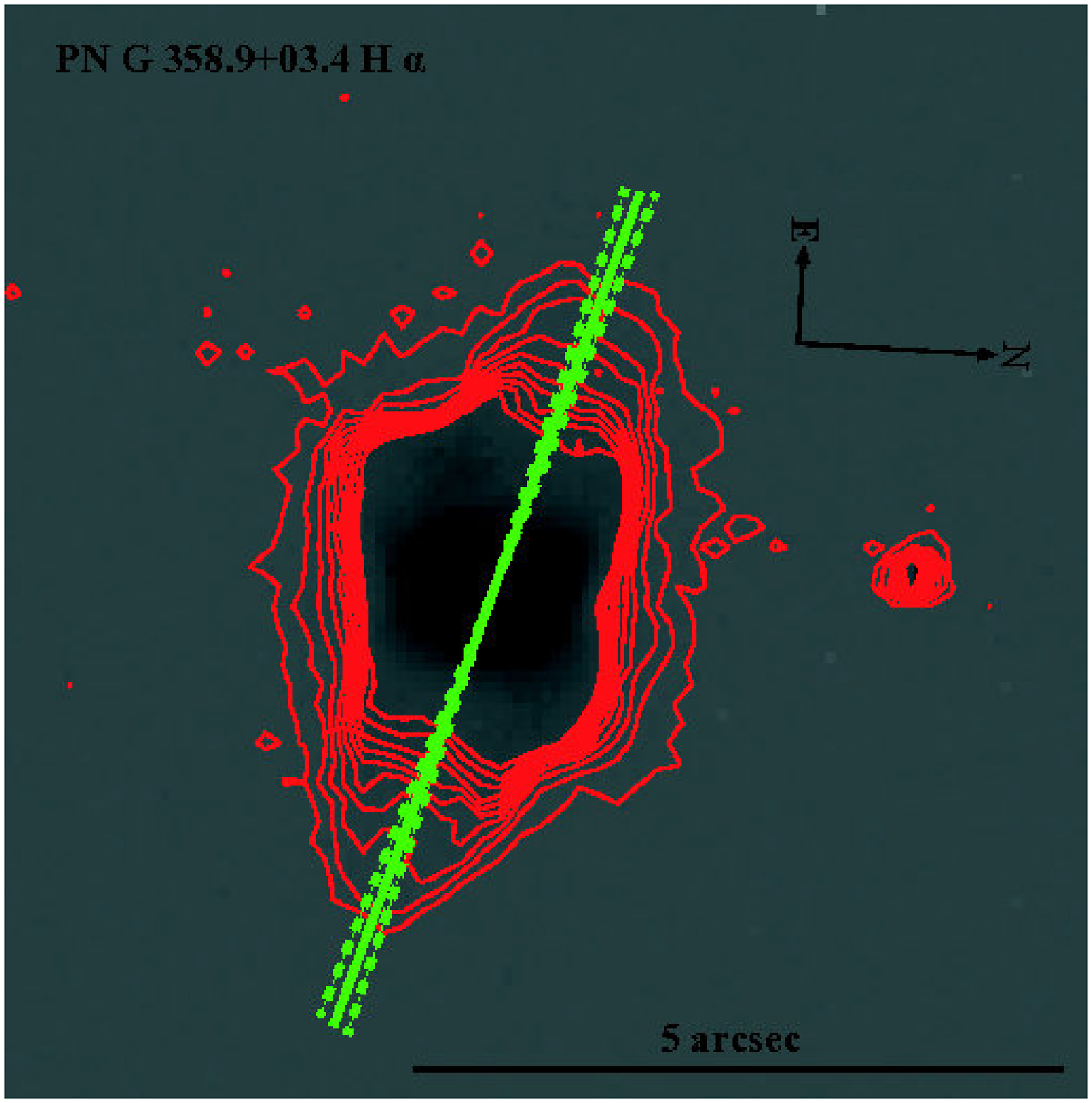}
\\
    \includegraphics[width=8.1cm]{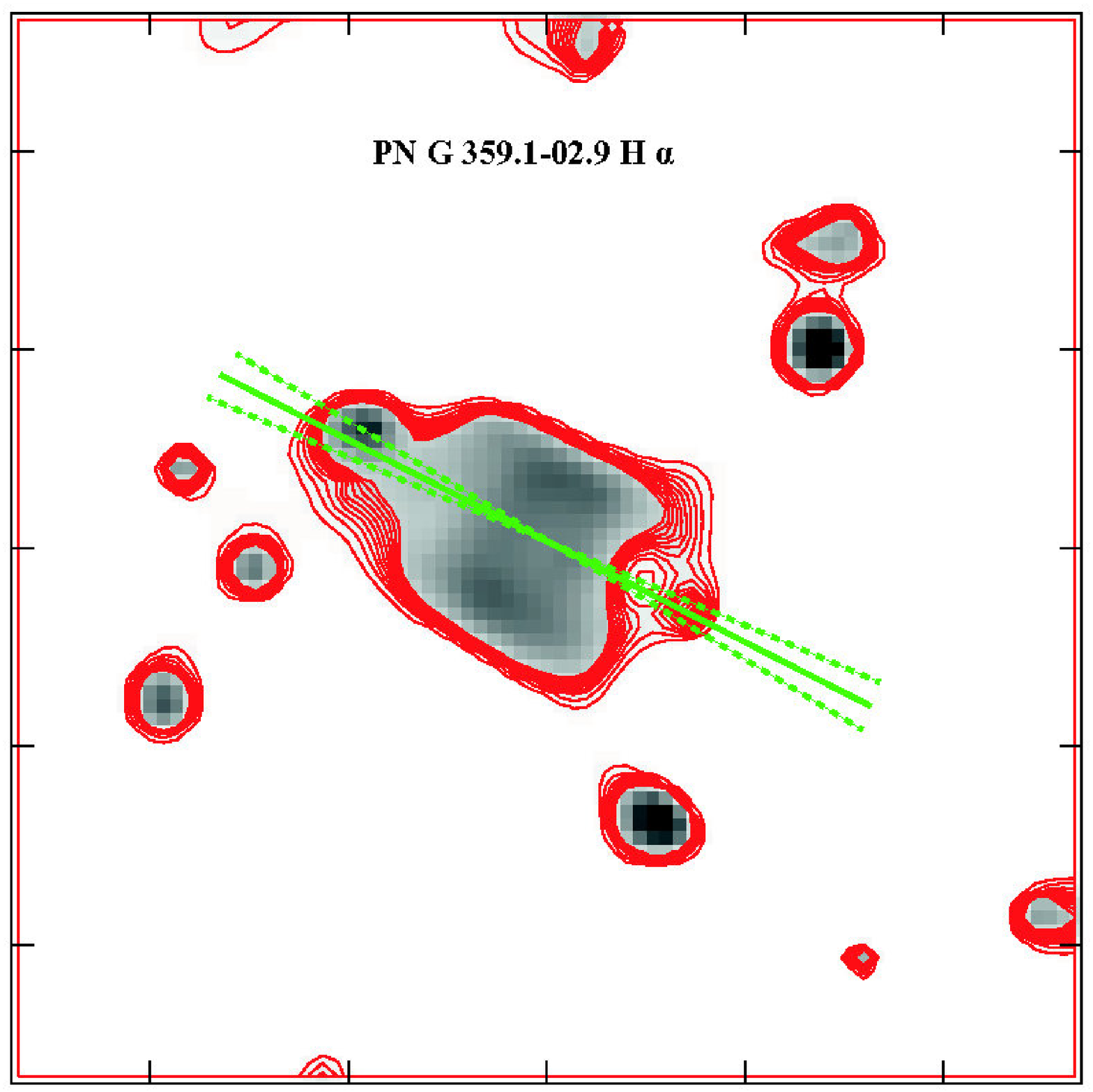} &
    \includegraphics[width=8.1cm]{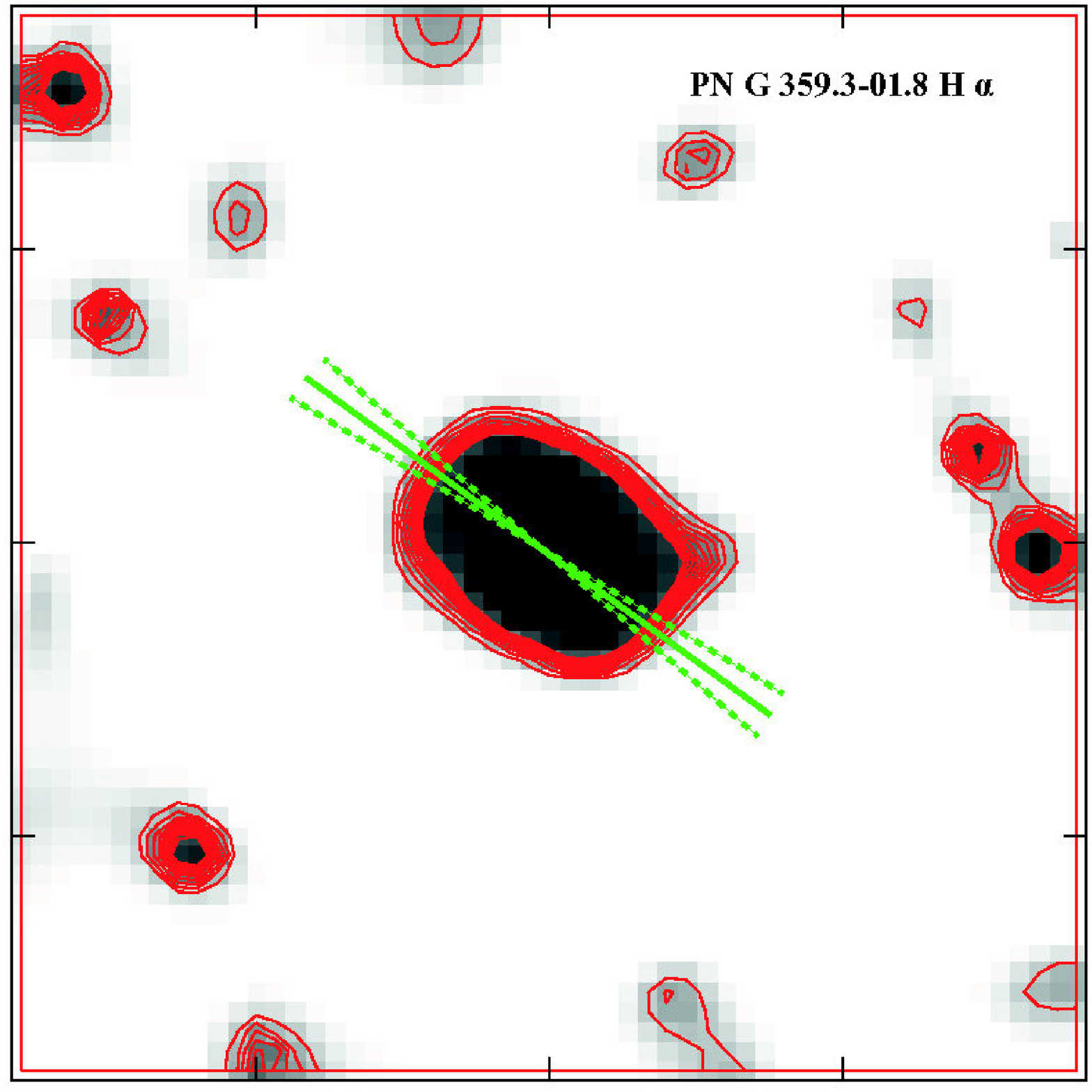}

    \end{array}
$

    \end{center}
    \contcaption{}
\label{plots1g}
\end{figure*}

\begin{figure*}
    \begin{center}

$
    \begin{array}{cc}
    \includegraphics[width=8.1cm]{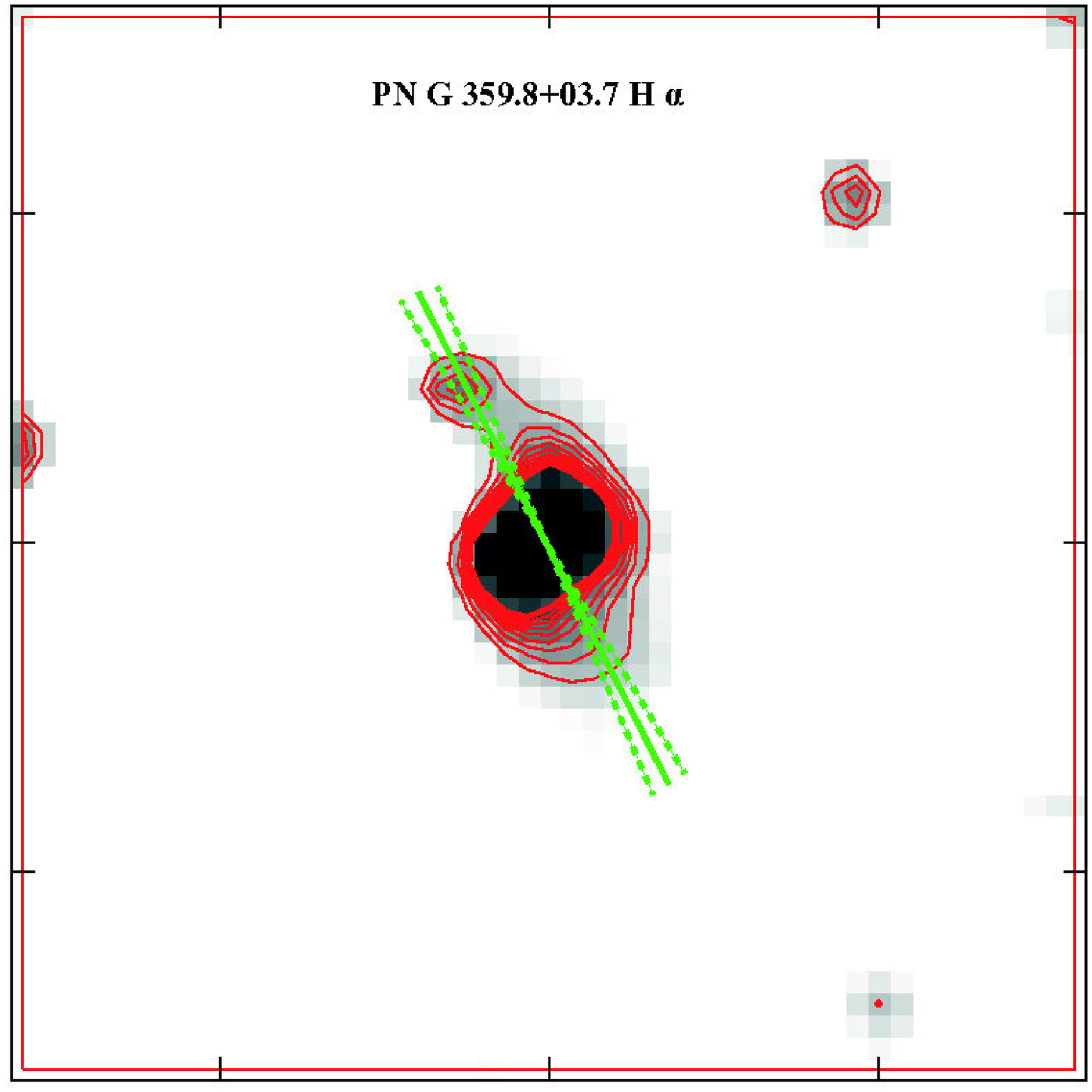} &
    \includegraphics[width=8.1cm]{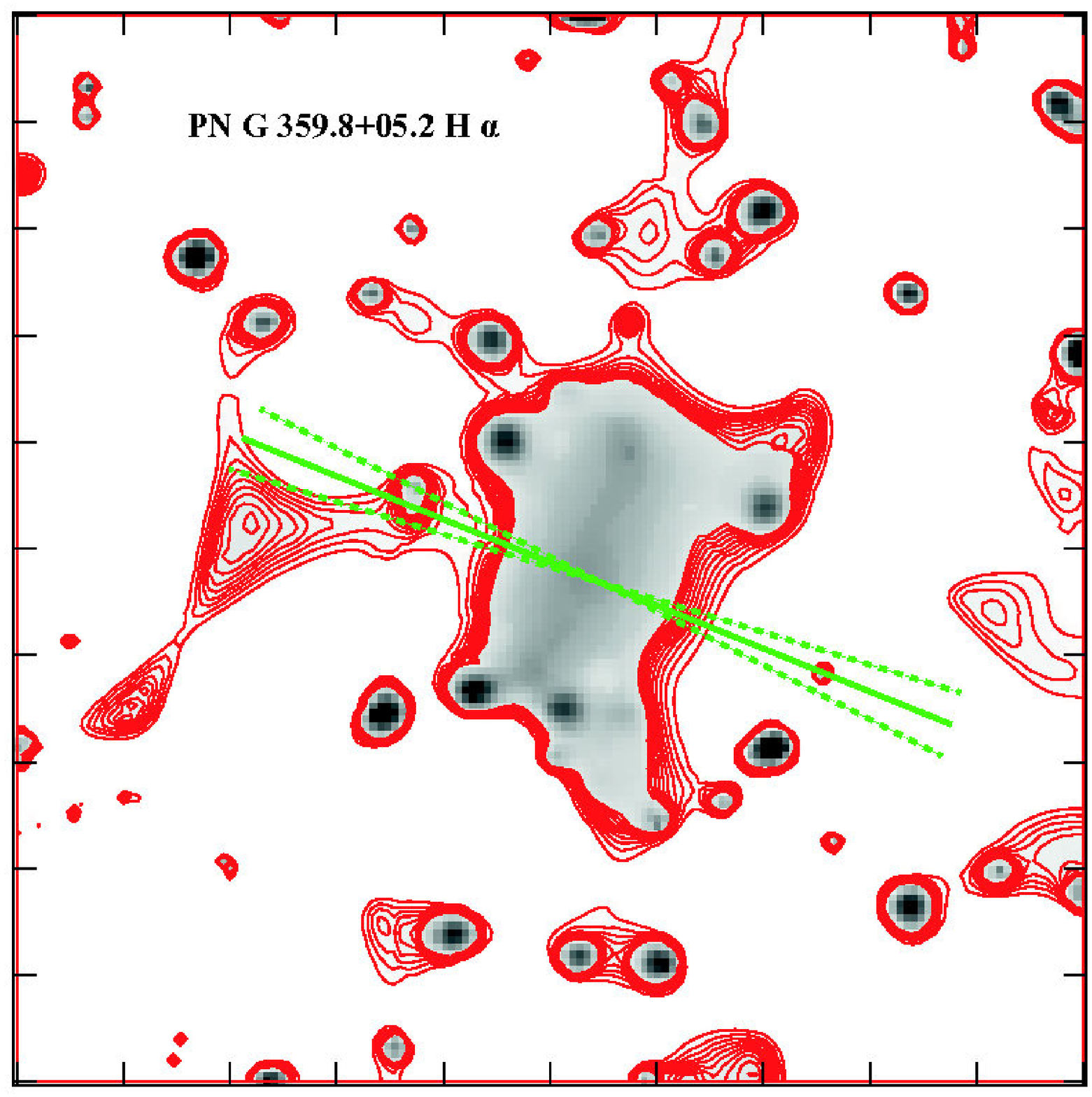}
\\
    \includegraphics[width=8.1cm]{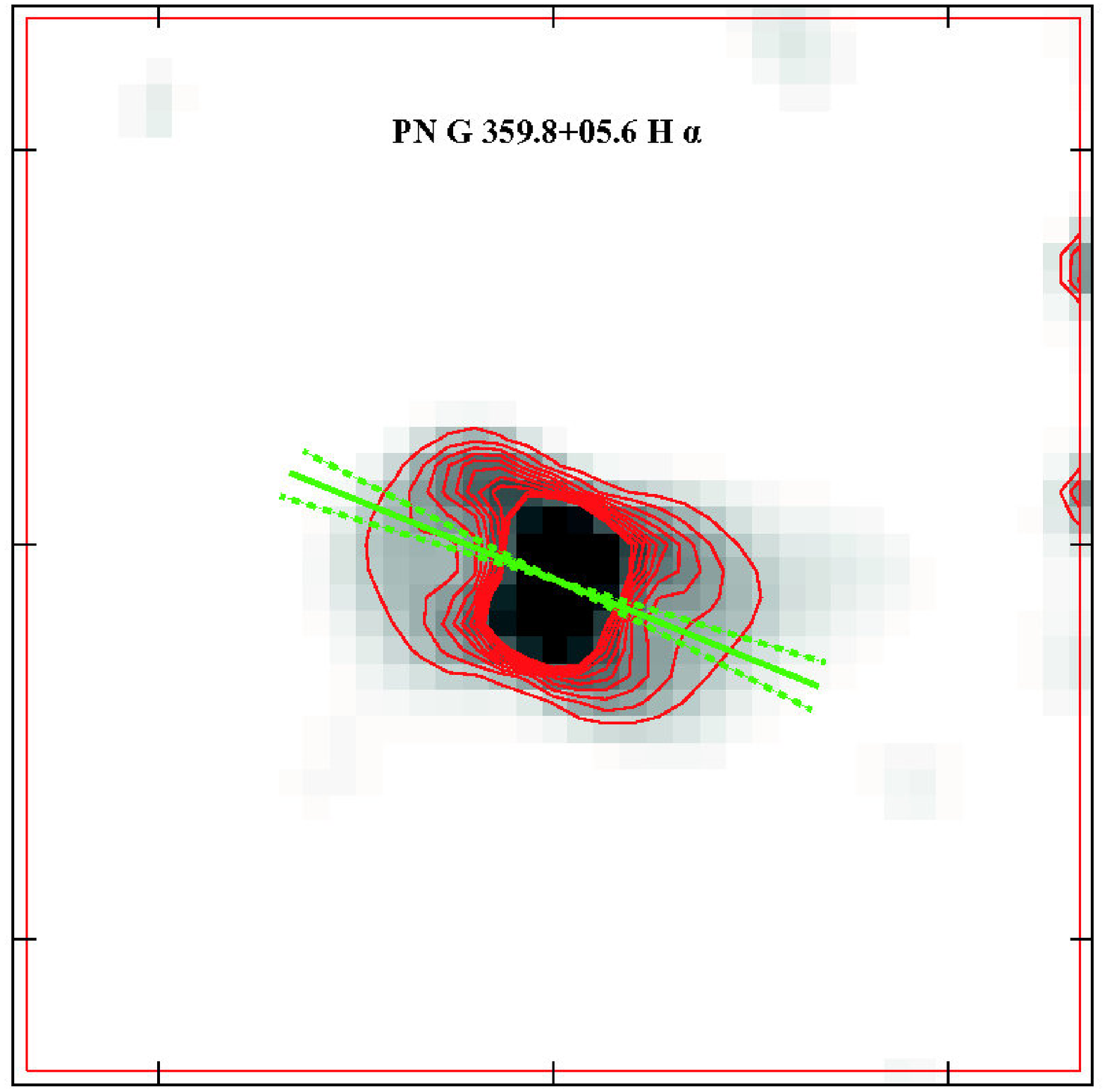} &
    \includegraphics[width=8.1cm]{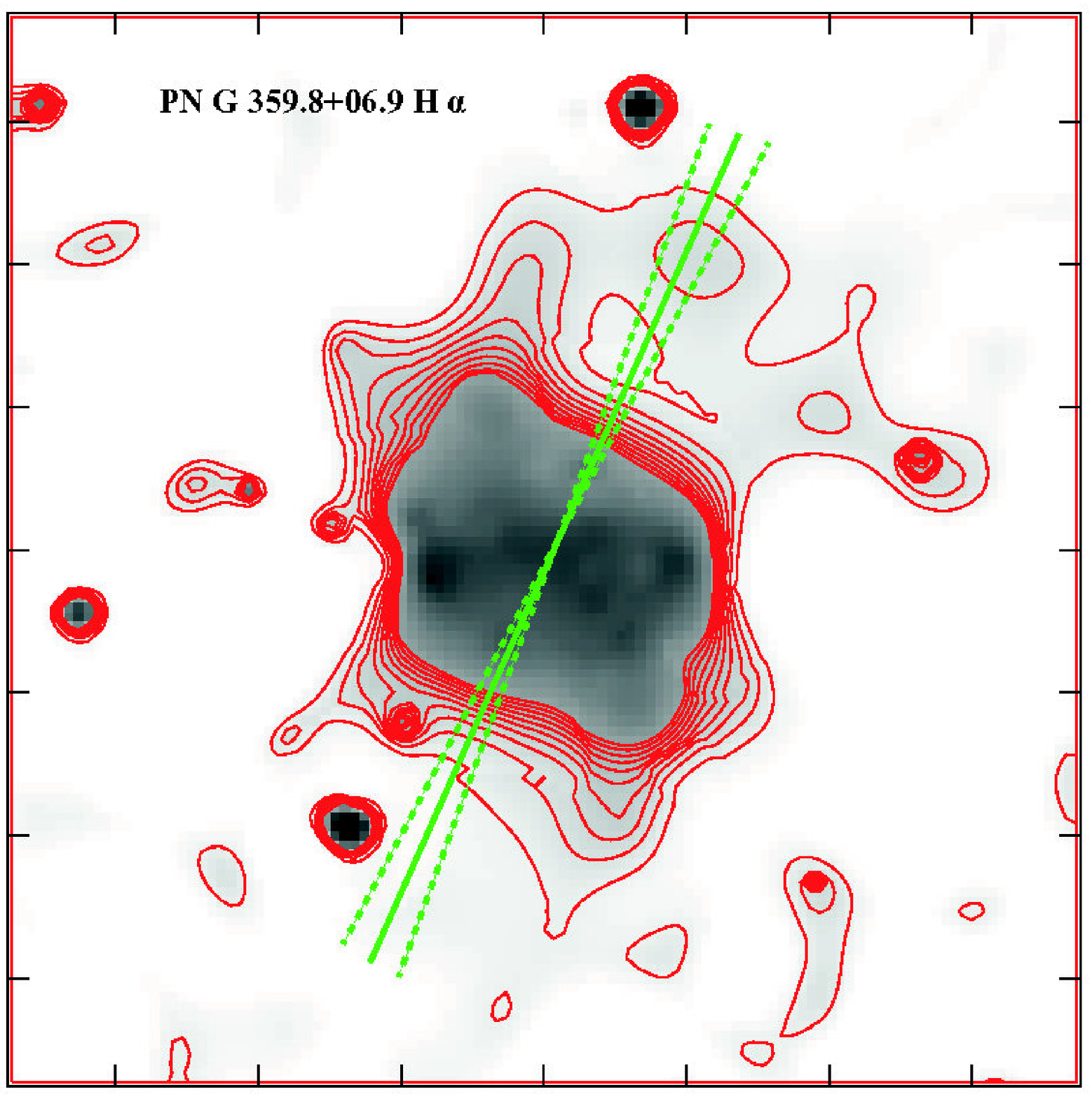}
    \end{array}
$
    \end{center}
    \contcaption{}
\label{plots1h}
\end{figure*}

\begin{figure*}
    \begin{center}

$
    \begin{array}{cc}    
    \includegraphics{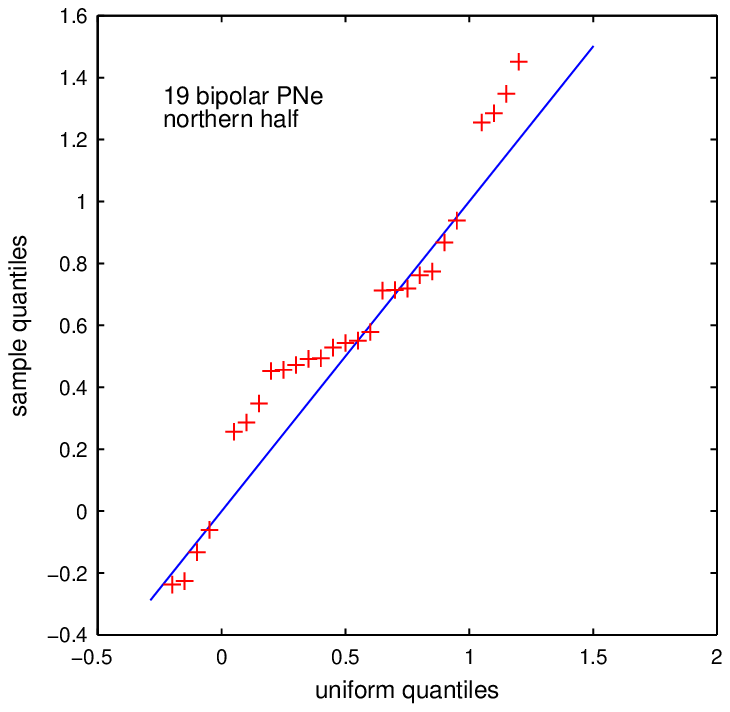} &
    \includegraphics{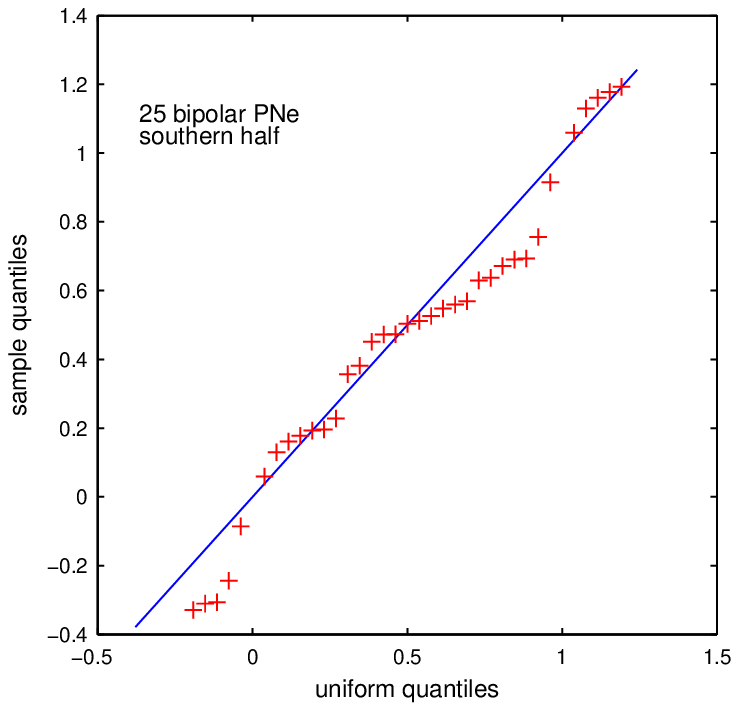}
\\
    \includegraphics{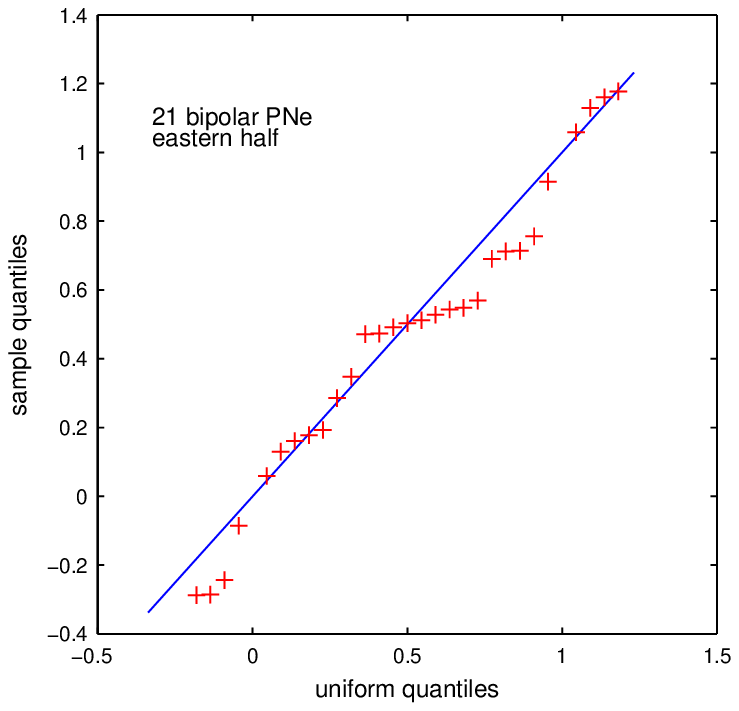} &
    \includegraphics{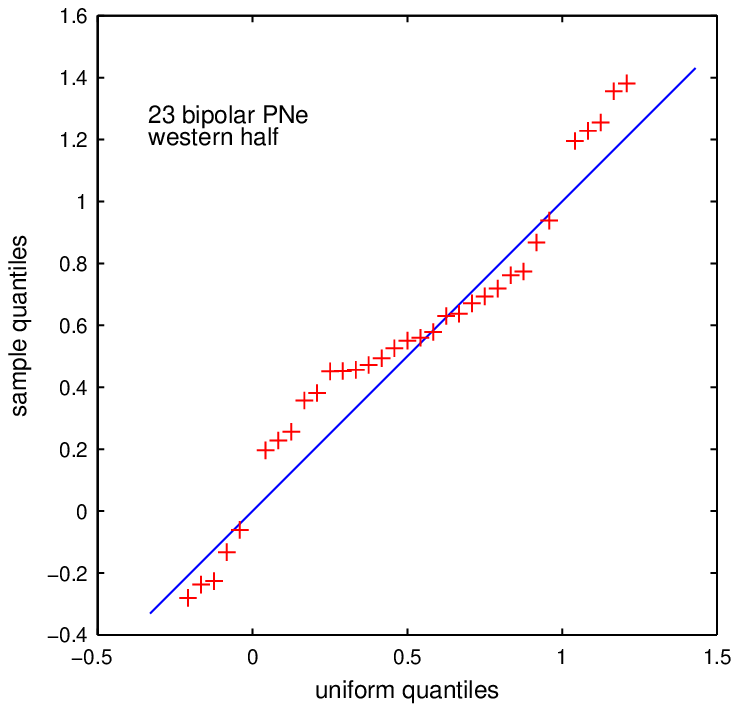}

    \end{array}
$
    \end{center}
    \caption{The extended quantile--quantile plots for the GPA of the bipolar PNe in the North, South, East and West subsamples.}
    \label{QQsect1}
\end{figure*}

\begin{figure*}
    \begin{center}

$
    \begin{array}{cc}

    \includegraphics[width=8.1cm]{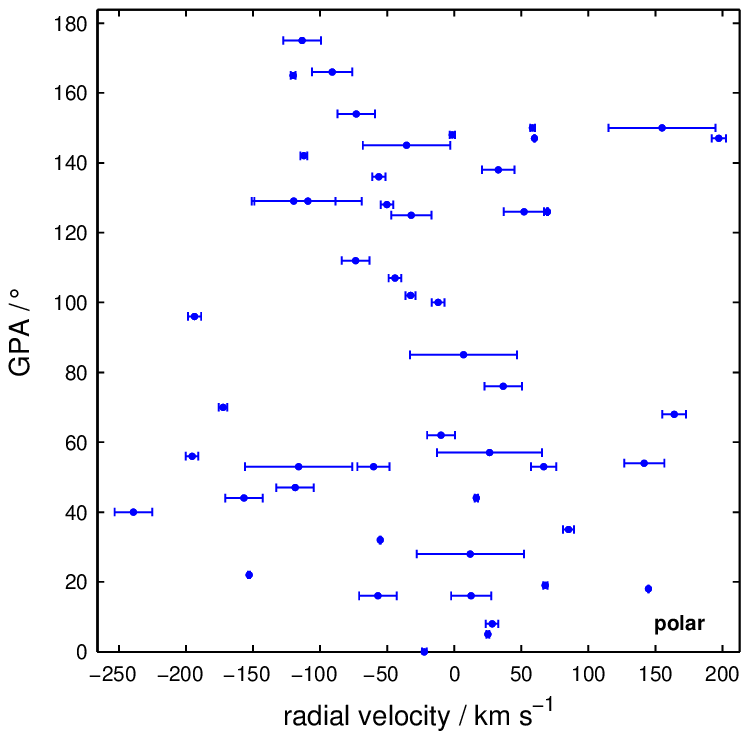} &
    \includegraphics[width=8.1cm]{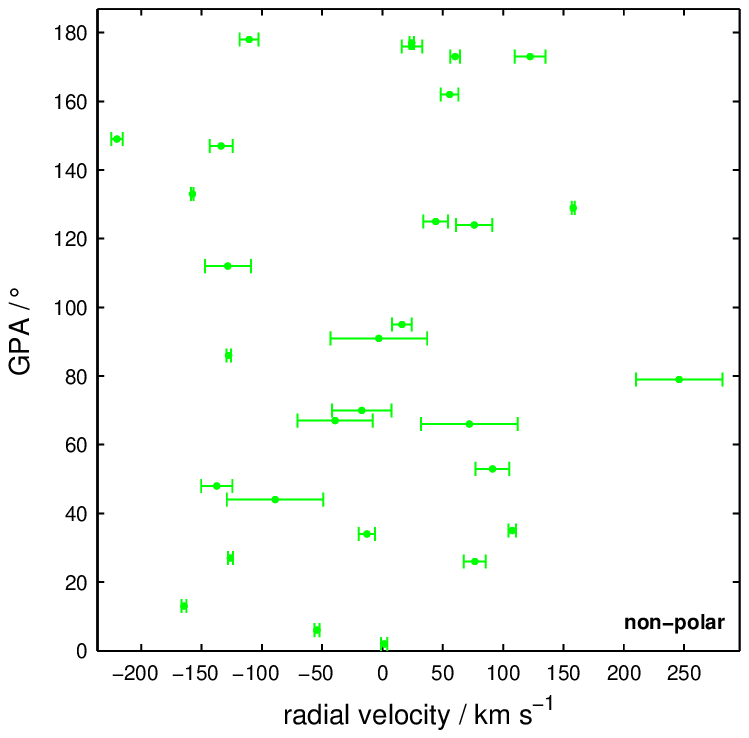}

    \end{array}
$
    \end{center}
    \caption{The GPA of the polar and non-polar PNe plotted against their Radial Velocities. The uncertainties forming the error bars for the radial velocities are taken from the catalogues.}
    \label{velang_pnp}
\end{figure*}

\begin{figure*}
    \begin{center}

$
    \begin{array}{cc}    
    \includegraphics{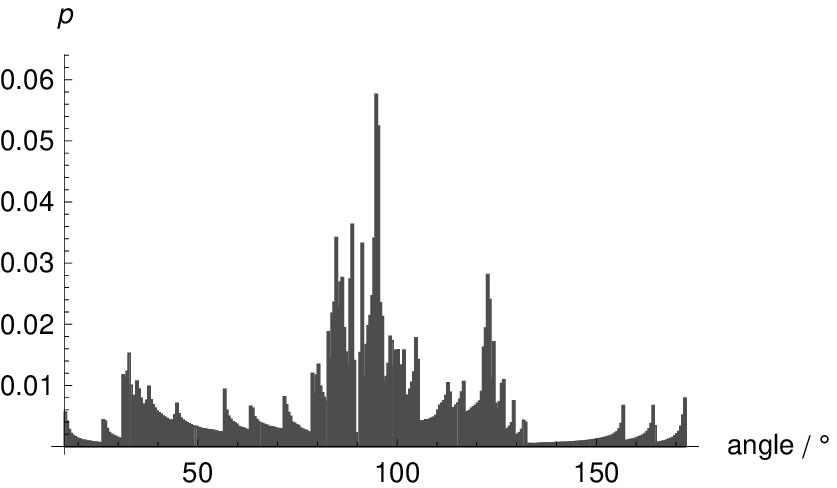} &
    \includegraphics{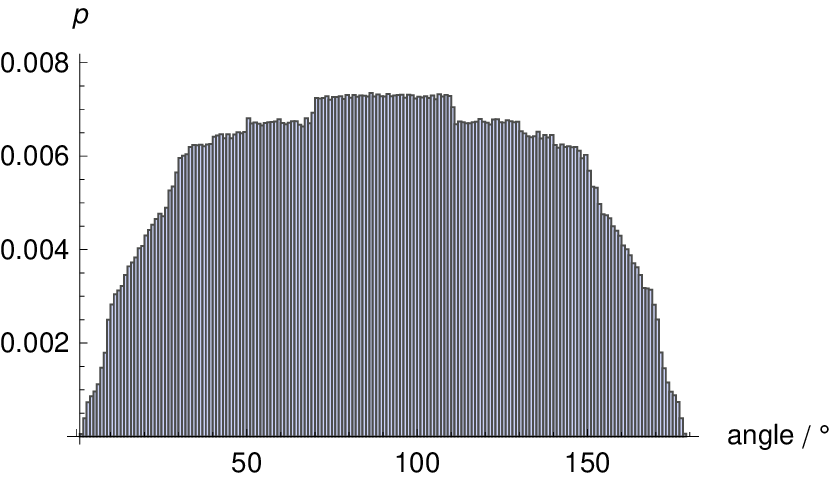}
       
    \end{array}
$
    \end{center}
    \caption{The histogram for the derived angles from the North Galactic Pole to the axes of the bipolar PNe is shown on the left. The peaks correspond to the GPA. The angles on the Galactic Plane used to derive the angles from the North are shown in the histogram on the right. Note the departure from uniformity in the distribution of those angles due to the constraints of the measured PN lengths, the maximum PN length and the GPA. Note also that only a subset of those angles will apply to each PN. There are 440\,000 entries contributing to each histogram and the area of each histogram is 1.}
    \label{3DHist}
\end{figure*}